\pdfoutput=1
\documentclass[usenatbib]{mn2e}
\usepackage{apjfonts}
\usepackage{amssymb}
\usepackage{amsmath}
\usepackage{ctable}
\usepackage{verbatim}
\usepackage{url}
\usepackage{fixltx2e} 
\usepackage[implicit=false,breaklinks,colorlinks,citecolor=blue]{hyperref}

\newcommand{\be}{\begin{equation}}
\newcommand{\ee}{\end{equation}}

\newcommand{\etal}{et al.}

\newcommand{\msun}{M_{\sun}}

\newcommand{\paperone}{Paper {\small I}}
\newcommand{\papertwo}{Paper {\small II}}

\newcommand{\Alf}{{Alfv\'en}}

\newcommand{\fref}[1]{Fig.~\ref{#1}}

\newcommand{\ICsurl}{\href{http://www.tapir.caltech.edu/~phopkins/publicICs}{\url{http://www.tapir.caltech.edu/~phopkins/publicICs}}}
\newcommand{\FIREurl}{\href{http://fire.northwestern.edu}{\url{http://fire.northwestern.edu}}}
\newcommand{\gizmourl}{\href{http://www.tapir.caltech.edu/~phopkins/Site/GIZMO.html}{\url{http://www.tapir.caltech.edu/~phopkins/Site/GIZMO.html}}}
\newcommand{\movieurl}{\href{http://www.tapir.caltech.edu/~phopkins/Site/animations/}{\url{http://www.tapir.caltech.edu/~phopkins/Site/animations/}}}

\newcommand\plotonesize[2]
 {\centering \leavevmode \includegraphics[width={#2\columnwidth}]{#1}}

\newcommand{\acknowledgments}{\begin{small}\section*{Acknowledgments}\end{small}}
\newcommand\altaffilmark[1]{$^{#1}$}
\newcommand\altaffiltext[1]{$^{#1}$}
\newcommand{\datastatement}[1]{\begin{small}\section*{Data Availability Statement}\end{small}{\noindent #1}\vspace{5pt}}
\title[Cosmic Rays on FIRE: Winds]{Cosmic-Ray Driven Outflows to Mpc Scales from $L_{\ast}$ Galaxies
\vspace{-0.5cm}}

\vspace{-0.2cm}
\author[Hopkins \etal]{
\parbox[t]{\textwidth}{ 
Philip F.~Hopkins\thanks{E-mail:phopkins@caltech.edu}\altaffilmark{1},
T.~K.\ Chan\altaffilmark{2,3},
Suoqing Ji\altaffilmark{1},
Cameron B.~Hummels\altaffilmark{1},
Du\v{s}an Kere\v{s}\altaffilmark{2}, 
Eliot Quataert\altaffilmark{4},
Claude-Andr{\'e} Faucher-Gigu{\`e}re\altaffilmark{5}
} 
\vspace*{6pt} \\
\altaffiltext{1}{TAPIR, Mailcode 350-17, California Institute of Technology, Pasadena, CA 91125, USA} \\
\altaffiltext{2}{Department of Physics, Center for Astrophysics and Space Science, University of California at San Diego, 9500 Gilman Drive, La Jolla, CA 92093} \\ 
\altaffiltext{3}{Institute for Computational Cosmology, Durham University, South Road, Durham, DH1 3LE, UK} \\ 
\altaffiltext{4}{Department of Astronomy and Theoretical Astrophysics Center, University of California Berkeley, Berkeley, CA 94720} \\ 
\altaffiltext{5}{Department of Physics and Astronomy and CIERA, Northwestern University, 2145 Sheridan Road, Evanston, IL 60208, USA} 
\vspace{-0.5cm}
}

\date{Working Document\vspace{-0.6cm}}
\begin{document}
\maketitle
\label{firstpage}

\vspace{-0.2cm}
\begin{abstract}
We study the effects of cosmic rays (CRs) on outflows from star-forming galaxies in the circum and inter-galactic medium (CGM/IGM), in high-resolution, fully-cosmological FIRE-2 simulations (accounting for mechanical and radiative stellar feedback, magnetic fields, anisotropic conduction/viscosity/CR diffusion and streaming, and CR losses). We showed previously that massive ($M_{\rm halo}\gtrsim 10^{11}\,M_{\sun}$), low-redshift ($z\lesssim 1-2$) halos can have CR pressure dominate over thermal CGM pressure and balance gravity, giving rise to a cooler CGM with an equilibrium density profile. This dramatically alters outflows. Absent CRs, high gas thermal pressure in massive halos ``traps'' galactic outflows near the disk, so they recycle. With CRs injected in supernovae as modeled here, the low-pressure halo allows ``escape'' and CR pressure gradients continuously accelerate this material well into the IGM in ``fast'' outflows, while lower-density gas at large radii is accelerated in-situ into ``slow'' outflows that extend to $>$Mpc scales. CGM/IGM outflow morphologies are radically altered: they become mostly volume-filling (with inflow in a thin mid-plane layer) and coherently biconical from the disk to $>$Mpc. The CR-driven outflows are primarily cool ($T\sim10^{5}\,$K) and low-velocity. All of these effects weaken and eventually vanish at lower halo masses ($\lesssim 10^{11}\,M_{\sun}$) or higher redshifts ($z\gtrsim 1-2$), reflecting the ratio of CR to thermal+gravitational pressure in the outer halo. We present a simple analytic model which explains all of the above phenomena. We caution that these predictions may depend on uncertain CR transport physics.
\end{abstract}

\begin{keywords}
galaxies: formation --- galaxies: evolution --- galaxies: active --- 
stars: formation --- galaxies: intergalactic medium --- cosmology: theory\vspace{-0.5cm}
\end{keywords}

\section{Introduction}
\label{sec:intro}

\begin{footnotesize}
\ctable[
  caption={{\normalsize Zoom-in simulation volumes run to $z=0$ (see \citealt{hopkins:fire2.methods} for details). All units are physical.}\label{tbl:sims}},center,star
  ]{lcccccr}{
\tnote[ ]{Halo/stellar properties listed refer only to the original ``target'' halo around which the high-resolution volume is centered: these volumes can reach up to $\sim (1-10\,{\rm Mpc})^{3}$ comoving, so there are actually several hundred resolved galaxies in total. 
{\bf (1)} Simulation Name: Designation used throughout this paper. 
{\bf (2)} $M_{\rm halo}^{\rm vir}$: Virial mass \citep[following][]{bryan.norman:1998.mvir.definition} of the ``target'' halo at $z=0$.
{\bf (3)} $M_{\ast}^{\rm MHD+}$: Stellar mass of the central galaxy at $z=0$, in our non-CR, but otherwise full-physics (``MHD+'') run. 
{\bf (4)} $M_{\ast}^{\rm CR+}$: Stellar mass of the central galaxy at $z=0$, in our ``default'' (observationally-favored) CR+ ($\kappa=3e29$) run.
{\bf (5)} $m_{i,\,1000}$: Mass resolution: the baryonic (gas or star) particle/element mass, in units of $1000\,\msun$. The DM particle mass is always larger by the universal ratio, a factor $\approx 5$. 
{\bf (6)} $\langle \epsilon_{\rm gas} \rangle^{\rm sf}$: Spatial resolution: the gravitational force softening (Plummer-equivalent) at the mean density of star formation (gas softenings are adaptive and match the hydrodynamic resolution, so this varies), in the MHD+ run. Typical time resolution reaches $\sim 100-100\,$yr, density resolution $\sim 10^{3}-10^{4}\,{\rm cm^{-3}}$.
{\bf (7)} Additional notes.\vspace{-0.5cm}
}
}{
\hline\hline
Simulation & $M_{\rm halo}^{\rm vir}$ & $M_{\ast}^{\rm MHD+}$ & $M_{\ast}^{\rm CR+}$ & $m_{i,\,1000}$ & $\langle \epsilon_{\rm gas} \rangle^{\rm sf}$ & Notes \\
Name \, & $[\msun]$ &  $[\msun]$  &   $[\msun]$  &$[1000\,\msun]$ & $[{\rm pc}]$ & \, \\ 
\hline 
{\bf m09} & 2.4e9 &  2e4 & 3e4 & 0.25 & 0.7 & early-forming, ultra-faint field dwarf \\
{\bf m10v} & 8.3e9 & 2e5 & 3e5 & 0.25 & 0.7 & isolated dwarf in a late-forming halo \\
{\bf m10q} & 8.0e9 & 2e6 & 2e6 & 0.25 & 0.8 & isolated dwarf in an early-forming halo \\
{\bf m10y} & 1.4e10 & 1e7 & 1e7 & 0.25 & 0.7 & early-forming dwarf, with a large dark matter ``core'' \\
{\bf m10z} & 3.4e10 & 4e7 & 3e7 & 0.25 & 0.8 & ultra-diffuse dwarf galaxy, with companions \\
\hline
{\bf m11a} & 3.5e10 & 6e7 & 5e7 & 2.1 & 1.6 & classical dwarf spheroidal \\
{\bf m11b} & 4.3e10 & 8e7 & 8e7 & 2.1 & 1.6 & disky (rapidly-rotating) dwarf \\
{\bf m11i} & 6.8e10 & 6e8 & 2e8 & 7.0 & 1.8 & dwarf with late mergers \&\ accretion \\
{\bf m11e} & 1.4e11 & 1e9 & 7e8 & 7.0 & 2.0 & low surface-brightness dwarf \\
{\bf m11c} & 1.4e11 & 1e9 & 9e8 & 2.1 & 1.3 & late-forming, LMC-mass halo \\
\hline
{\bf m11q} & 1.5e11 & 1e9 & 1e9 & 0.88 & 1.0 & early-forming, large-core diffuse galaxy\\
{\bf m11v} & 3.2e11 & 2e9 & 1e9 & 7.0 & 2.4 & has a multiple-merger ongoing at $z\sim0$ \\
{\bf m11h} & 2.0e11 & 4e9 & 3e9 & 7.0 & 1.9 & early-forming, compact halo \\
{\bf m11d} & 3.3e11 & 4e9 & 2e9 & 7.0 & 2.1 & late-forming, ``fluffy'' halo and galaxy \\
{\bf m11f} & 5.2e11 & 3e10 & 1e10 & 12 & 2.6 & early-forming, intermediate-mass halo \\
\hline
{\bf m11g} & 6.6e11 & 5e10 & 1e10  & 12 & 2.9 & late-forming, intermediate-mass halo \\
{\bf m12z} & 8.7e11 & 2e10 & 8e9 & 4.0 & 1.8 & disk with little bulge, ongoing merger at $z\sim0$ \\
{\bf m12r} & 8.9e11 & 2e10 & 9e9 & 7.0 & 2.0 & late-forming, barred thick-disk \\
{\bf m12w} & 1.0e12 & 6e10 & 2e10 & 7.0 & 2.1 & forms a low surface-brightness / diffuse disk \\
\hline
{\bf m12i} & 1.2e12 & 7e10 & 3e10 & 7.0 & 2.0 & ``Latte'' halo, later-forming MW-mass halo, massive disk\\  
{\bf m12b} & 1.3e12 & 9e10 & 4e10 & 7.0 & 2.2 & early-forming, compact bulge+thin disk \\
{\bf m12c} & 1.3e12 & 6e10 & 2e10 & 7.0 & 1.9 & MW-mass halo with $z\sim1$ major merger(s) \\
{\bf m12m} & 1.5e12 & 1e11 & 3e10 & 7.0 & 2.3 & earlier-forming halo, features strong bar at late times \\ 
{\bf m12f} & 1.6e12 & 8e10 & 4e10 & 7.0 & 1.9 & MW-like disk, merges with LMC-like companion \\ 
\hline\hline
}
\end{footnotesize}

Galactic outflows are ubiquitous in star-forming galaxies. Spectral observations of galaxies directly indicate outflows from the galactic interstellar medium (ISM) at a range of velocities, across a wide range of galaxy stellar masses and redshifts \citep{martin99:outflow.vs.m,heckman:superwind.abs.kinematics,weiner:z1.outflows,martin:2010.metal.enriched.regions,sato:2009.ulirg.outflows,steidel:2010.outflow.kinematics}. Observations of the circum and inter-galactic medium (CGM/IGM) also indicate that outflows must be ubiquitous in order to explain the pollution of these regions by heavy elements \citep{pettini:2003.igm.metal.evol,songaila:2005.igm.metal.evol}. Moreover, it has long been recognized that outflows must occur in essentially all star-forming galaxies in order to explain their relatively low stellar masses (compared to the Universal baryon fraction) and the existence of the mass-metallicity relation \citep[see e.g.][]{katz:treesph,somerville99:sam,cole:durham.sam.initial,springel:lcdm.sfh,tremonti:mass.metallicity.relation,keres:fb.constraints.from.cosmo.sims}. These outflows (primarily) stem from ``feedback'' from massive stars, which can act in a variety of forms including radiative (photo-heating and radiation pressure) and mechanical (thermal and kinetic energy from supernovae [SNe] explosions and outflows/jets), injection of magnetic fields and cosmic rays (CRs). In more massive galaxies, outflows from super-massive black holes (BHs) and active galactic nuclei (AGN) are almost certainly important as well \citep{croton:sam,hopkins:qso.all}, but these are sub-dominant in lower-mass, star-forming galaxies (owing to the very small BHs and low duty cycle of high-accretion rate activity, among other factors; see \citealt{krongold:seyfert.outflow,hopkins:seyferts,hopkins:seyfert.limits,kormendy:2011.bh.nodisk.corr,greene:2011.nlr.outflows,daa:BHs.on.FIRE}).

The existence of galactic outflows in star-forming galaxies, their significance for galaxy formation and CGM/IGM evolution, and their generic attribution to ``stellar feedback'' processes are well-established. However almost everything else remains controversial at some level, including e.g.\ the actual physical state[s] of outflowing gas (the phases/densities/temperatures/velocities which carry most of the mass/momentum/energy, and which of these if any is ``most important''), the acceleration sites (within the disk, near massive stars, or ``above the midplane'' or in the CGM), the ultimate fate of outflows (whether they are unbound, or halt and are efficiently recycled, and if so over what time and spatial scales this occurs), their morphologies (bi-conical or spherical or filamentary or clumpy), and the physical feedback mechanisms that accelerate the winds (e.g.\ SNe vs.\ radiation pressure vs.\ cosmic rays, which may act on different spatial and timescales with different efficiencies in galaxies of different types). 

In recent years, numerical simulations have begun to directly resolve the relevant scales of some of these acceleration processes in global galaxy-wide simulations, making it possible to {\em self-consistently} predict the {\em generation} of galactic winds and therefore some of the properties above (e.g.\ their phases and velocities), as opposed to inserting assumptions about wind properties ``by hand'' \citep[see e.g.][]{tasker:2011.photoion.heating.gmc.evol,hopkins:rad.pressure.sf.fb,hopkins:fb.ism.prop,wise:2012.rad.pressure.effects,kannan:2013.early.fb.gives.good.highz.mgal.mhalo,agertz:2013.new.stellar.fb.model,roskar:2014.stellar.rad.fx.approx.model}. One such effort is the ``Feedback In Realistic Environments'' (FIRE)\footnote{\label{foot:movie}See the {\small FIRE} project website:\\
\FIREurl \\
For additional movies and images of FIRE simulations, see:\\
\movieurl} project \citep{hopkins:2013.fire}, which attempts to explicitly incorporate and at least begin to resolve mechanical feedback from individual SNe (Types Ia \&\ II) as well as stellar mass-loss (O/B and AGB), following \citet{hopkins:sne.methods}, and multi-band radiation-hydrodynamics to follow photo-electric and photo-ionization heating and radiation pressure \citep{hopkins:radiation.methods}, in fully-cosmological simulations. These simulations have been used to explore the generation and properties of multi-phase galactic outflows \citep{muratov:2015.fire.winds,muratov:2016.fire.metal.outflow.loading} as well as their consequences for galactic abundances \citep{ma:2015.fire.mass.metallicity,escala:turbulent.metal.diffusion.fire}, dark matter profiles \citep{onorbe:2015.fire.cores,chan:fire.dwarf.cusps}, CGM absorbers around galaxies \citep{faucher.2016:high.mass.qso.halo.covering.fraction.neutral.gas.fire,hafen:2016.lyman.limit.absorbers,2018arXiv181111753H,2019arXiv191001123H}, stellar halos \citep{sanderson:stellar.halo.mass.vs.fire.comparison,elbadry:most.ancient.mw.halo.stars.kicked}, gas-phase kinematics of galaxies \citep{wheeler.2015:dwarfs.isolated.not.rotating,elbadry:fire.morph.momentum,elbadry:HI.obs.gal.kinematics,ma:radial.gradients,ma:2016.disk.structure,bonaca:gaia.structure.vs.fire} and galaxy star formation histories and stellar masses \citep{sparre.2015:bursty.star.formation.main.sequence.fire,garrisonkimmel:local.group.fire.tbtf.missing.satellites,ma:fire2.reion.gal.lfs}. 

Although the simulations above directly treat many of the important stellar feedback processes (e.g.\ SNe Types Ia \&\ II, O/B and AGB mass-loss, photo-ionization and photo-electric heating, multi-wavelength radiation pressure), they neglect (among other things) CRs. In the ISM, the CRs which dominate their pressure/energy density ($\sim$\,GeV protons) are distributed smoothly with a $\gtrsim 1\,$kpc scale-height above the disk, with order-of-magnitude similar energy densities to thermal and magnetic pressure \citep{Ginz85,Boul90}. The idea that this smooth ``additional pressure'' term could contribute to galactic outflows by accelerating material down the CR pressure gradient if it was ``lofted'' above the midplane (by e.g.\ galactic fountains) has existed for decades \citep{Ipav75,Brei91,Brei93,Zira96,Socr08,Ever08,Dorf12,Mao18}, and in the last several years there has been a flurry of activity exploring this in numerical simulations \citep{jubelgas:2008.cosmic.ray.outflows,Boot13,Wien13,Sale14,Chen16,Simp16,Rusz17,Buts18,Farb18}. This work has shown not only that this is viable, but potentially consistent with a variety of observational constraints; moreover it has also argued that this produces more ``cool'' material in outflows, which can potentially enhance wind mass-loading and observable CGM absorption in certain species.

However (as is always the case), this work has limitations. Most (although not all) of the studies focused on CR winds have focused on ``idealized'' simulations: either ``slabs'' of a galactic disk or ISM, or isolated (non-cosmological) galaxies. This means that one cannot make meaningful predictions for the large-scale acceleration or propagation of winds beyond the immediate vicinity of the galactic disk, let alone their interaction with e.g.\ cosmological inflows and the high thermal pressure, virialized gaseous halo. On the other hand, the global and/or cosmological simulations that have been run have (largely owing to resolution limitations) generally treated the multi-phase ISM/CGM, star formation, and mechanical/radiative stellar feedback in a highly approximate fashion (in some cases ignoring these effects entirely, or not including cooling below $\sim 10^{4}\,$K, or putting in galactic outflows ``by hand,'' or simply adding SNe mechanical energy as a thermal energy component which is rapidly radiated away). In those cases, it is difficult if not impossible to self-consistently assess the impact of CR-driven outflows on different phases of gas, or their interplay with outflows driven by mechanical feedback (SNe kinetic or thermal feedback), their influence on the thermal instability, etc. 

Moreover, some of the simulations above did not incorporate potentially-critical CR physics: ignoring CR losses (so CRs have essentially ``infinite'' energy/cooling times), treating only CR streaming or diffusion, ignoring magnetic fields (which can confine the CRs and regulate their transport), or (for numerical time-step reasons) using artificially low CR diffusion coefficients\footnote{Throughout, we will use $\kappa \equiv \kappa_{\|}$ to refer to the {\em parallel} diffusivity of CRs, specifically at the energies (a few GeV) which dominate the CR pressure. In many analyses of e.g.\ galactic CR propagation, magnetic field structure is not included so the typical value $\tilde{\kappa}_{\rm iso}$ quoted is the effective isotropic-averaged diffusivity $\tilde{\kappa}_{\rm iso}  = \langle |\hat{\bf B}\cdot \hat{\nabla} e_{\rm cr}|^{2}\,\kappa \rangle \sim \kappa/3$ for random fields. Moreover note that older ``leaky-box'' models of the galaxy which assume CRs escape if $\gtrsim 200\,$pc above the disk derive order-of-magnitude lower $\tilde{\kappa}_{\rm iso}$ compared to modern (favored) models than allow for the existence of a diffuse gaseous halo extending $\sim 5-10\,$kpc above the disk, which require $\kappa \sim 3\,\tilde{\kappa}_{\rm iso} \gtrsim 3\times 10^{29}\,{\rm cm^{2}\,s^{-1}}$ \citep[see e.g.][]{blasi:cr.propagation.constraints,vladimirov:cr.highegy.diff,gaggero:2015.cr.diffusion.coefficient,2016ApJ...831...18C,2016ApJ...824...16J,2016PhRvD..94l3019K,evoli:dragon2.cr.prop,2018AdSpR..62.2731A}.} $\kappa \lesssim 10^{29}\,{\rm cm^{2}\,s^{-1}}$, which artificially confines CRs near galaxies (generating stronger effects there) but violates observational constraints from spallation in the MW and $\gamma$-ray emission in nearby galaxies \citep[see][]{lacki:2011.cosmic.ray.sub.calorimetric,2016ApJ...831...18C,2016ApJ...824...16J,2016PhRvD..94l3019K,2016ApJ...819...54G,fu:2017.m33.revised.cr.upper.limit,lopez:2018.smc.below.calorimetric.crs,chan:2018.cosmicray.fire.gammaray,2018JCAP...07..051G}. It is therefore critical to study the effects of CRs in fully-cosmological simulations, which attempt to directly treat the multi-phase ISM and mechanical/radiative (i.e.\ non-CR) stellar feedback processes, at least marginally resolve ISM structure and wind generation, and incorporate magnetic fields and anisotropic CR diffusion and streaming with transport coefficients that have been shown to reproduce observational constraints. 

Working towards this goal, \citet{chan:2018.cosmicray.fire.gammaray} performed and presented the first simulations combining the specific physics from the FIRE simulations, described above, with explicit CR injection and transport, accounting for advection and fully-anisotropic streaming and diffusion, as well as hadronic and Coulomb collisional and streaming (\Alf) losses, and showed that for reasonable parameter choices (e.g.\ $\kappa \sim 10^{29-30}\,{\rm cm^{2}\,s^{-1}}$) these simulations were consistent with empirical constraints on CR propagation in the MW and nearby galaxies (both dwarf and starburst systems). In \citet{hopkins:2018.cosmicray.fire.masses} (hereafter \paperone), we presented a new large suite of $>100$ fully-cosmological FIRE simulations incorporating these physics and the detailed physics of cooling, star formation, and stellar feedback described above, in halos from ultra-faint to $>$\,MW masses, with resolution reaching $\sim$\,pc scales. We showed that CRs produce weak effects in dwarfs and very high-redshift galaxies, but in massive ($M_{\rm halo} \gtrsim 10^{11}\,M_{\sun}$) halos at $z\lesssim 2$, they can substantially suppress SFRs and stellar masses. Moreover we showed that this was primarily via their interaction with the CGM, rather than their {\em direct} action deep within the ISM. But in this mass and redshift range, the galaxies develop ``CR-dominated'' halos, where CRs form the dominant source of pressure support over e.g.\ gas thermal pressure. In \paperone\ and \citet{ji:fire.cr.cgm} we followed this up and showed that a simple analytic model can predict where CRs should dominate and the ensuing equilibrium pressure and gas density profiles in CR-dominated halos; we also showed that where CR-dominated halos exist, they have a dramatic impact on CGM absorption statistics, and gas phase/temperature distributions. Given this and the motivation above, in this paper, we explore the effects on galactic outflows, primarily in the CGM and IGM, of these CR-dominated halos. Indeed, we will argue that the most dramatic impact of CRs on galactic outflows occurs in the CGM and IGM, where cosmological simulations are required.

Importantly, because the micro-physics of CR transport remain highly uncertain, we assume a very simple \Alf{ic} streaming plus constant-parallel-diffusivity model for the CR transport parameters. In a pair of companion papers \citep{hopkins:cr.transport.constraints.from.galaxies,2020arXiv200402897H}, we explore more complicated models for the CR transport parameters, and show these can lead to significant differences in CGM CR pressure profiles. However, the conclusions here are generally robust {\em where CRs dominate the pressure}, insofar as these models are representative of reality.

In \S~\ref{sec:methods} we briefly review the numerical methods and simulation suite from \paperone. \S~\ref{sec:theory} develops and presents an analytic model for the effects of CR-dominated halos on galactic winds. \S~\ref{sec:results} presents our simulation results and compares them to these theoretical expectations. We review and conclude in \S~\ref{sec:conclusions}. 

\vspace{-0.5cm}
\section{Methods}
\label{sec:methods}

\begin{figure*}
\begin{centering}
\includegraphics[width=0.49\textwidth]{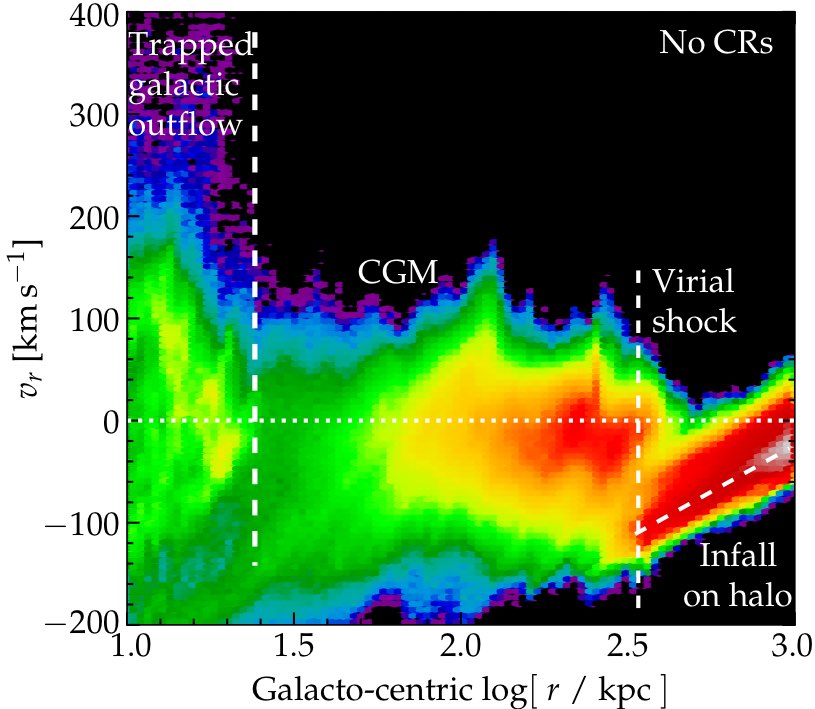}
\includegraphics[width=0.49\textwidth]{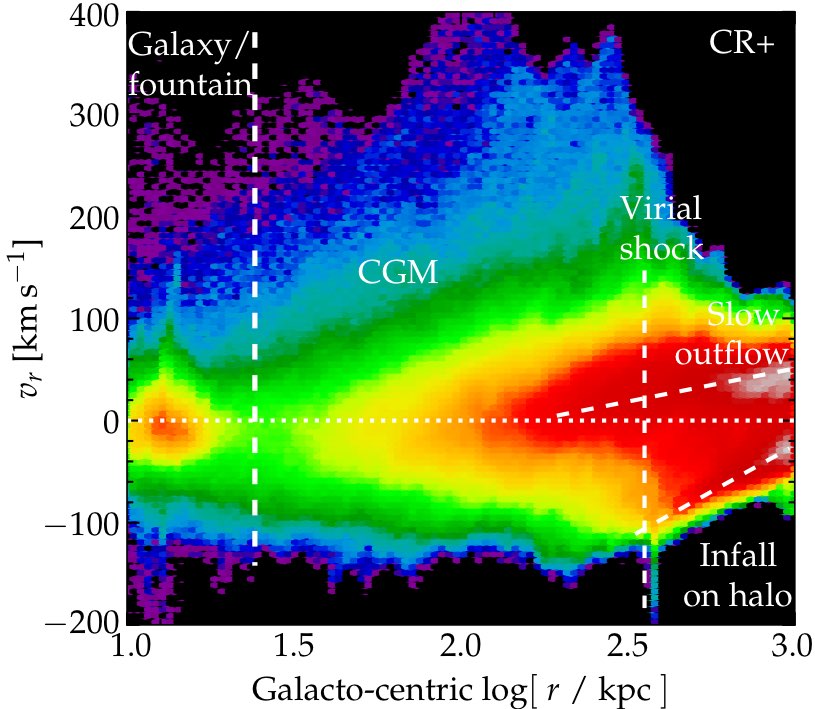}
\end{centering}
\caption{Distribution of gas radial inflow (negative) or outflow (positive) velocities $v_{r}$ in our halo {\bf m12i} (at $z=0$) -- a case study where the effects of CRs are most dramatic. 
We compare runs without CRs (``No CRs'' or ``MHD+''; {\em left}) and with CRs (``CR+'' or ``CR+($\kappa=3e29$)''; {\em right}), as a function of galactocentric distance $r$. 
Colors show (logarithmically-scaled) mass-weighted density in the plot (increasing purple-to-white). 
We label some components: 
(a) gas infalling from $\gtrsim 1\,$Mpc to $\sim R_{\rm vir}$ at (roughly) free-fall velocities; 
(b) the virial radius, where a strong shock is evident in the ``No CRs'' run ($v_{r}$ ``jumping'' to $\sim 0$); 
(c) the CGM; 
(d) the ``outer galaxy'' or ``fountain'' regime ($\lesssim 30\,$kpc). 
Absent CRs (still including MHD, conduction, radiation-hydrodynamics, stellar feedback, etc.), outflows are trapped by large halo thermal pressure, stirring large velocities in the disk. 
With CRs (for the parameters here, which have a near-maximal effect), CR pressure continuously accelerates material past $\gtrsim 10\,$kpc, and the low-thermal-pressure halo allows it to escape easily, producing fast outflows to $\sim R_{\rm vir}$, and ``slow'' ($< 100\,{\rm km\,s^{-1}}$) outflows accelerated in-situ by CRs at $\sim 0.5-5\,R_{\rm vir}$. 
\label{fig:demo.vr.vs.r}}
\end{figure*}

\begin{figure*}
\begin{centering}
\includegraphics[width={0.47\textwidth}]{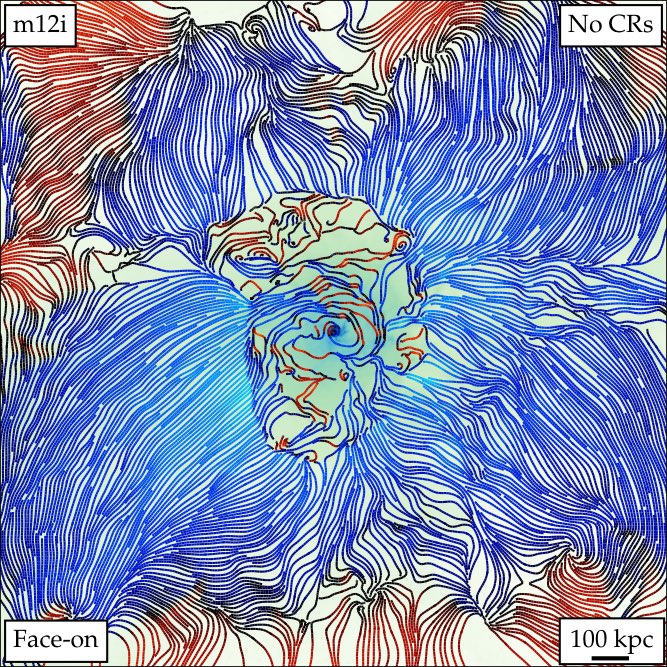}
\hspace{0.3cm}
\includegraphics[width={0.47\textwidth}]{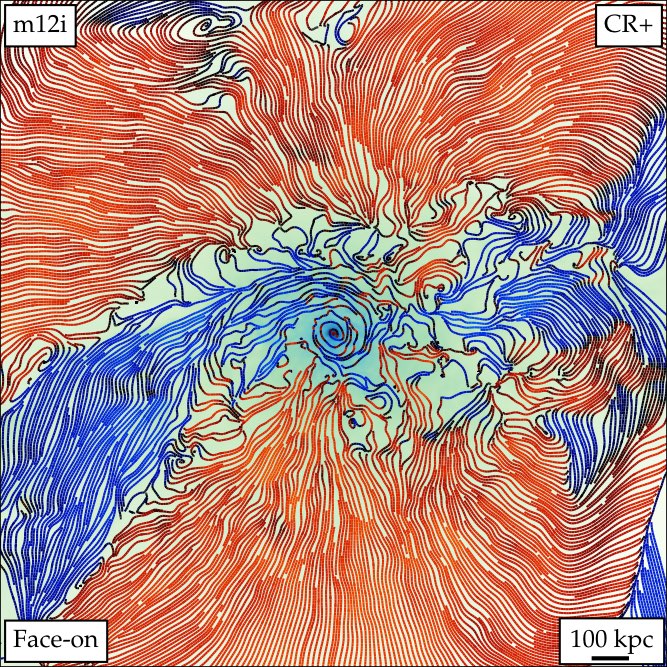} \\
\includegraphics[width={0.47\textwidth}]{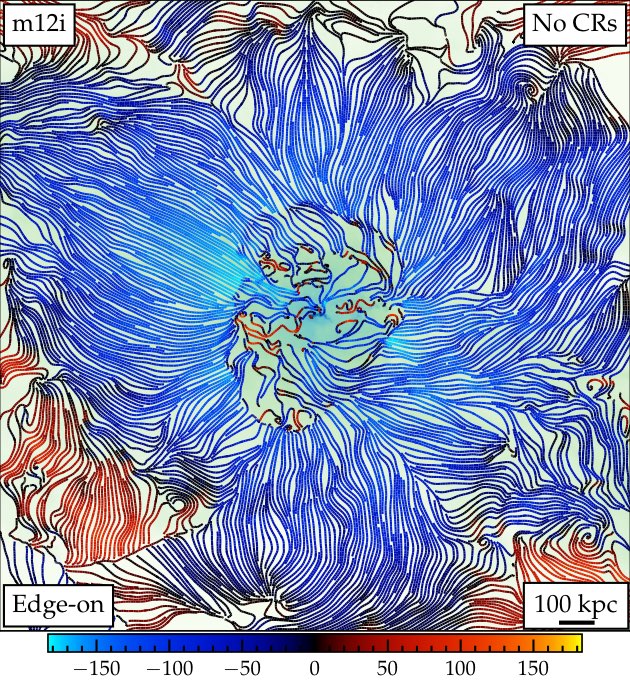}
\hspace{0.3cm}
\includegraphics[width={0.47\textwidth}]{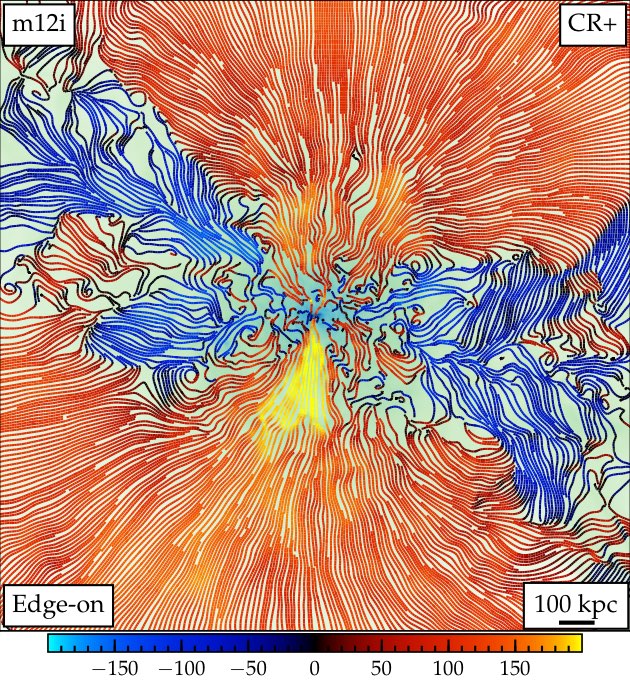} \\
\end{centering}
\caption{Inflow/outflow structure in {\bf m12i}, runs without CRs ({\em left}) and with CRs ({\em right}). We plot gas velocity (${\bf v}$) streamlines, in a 2D slice (background color shows gas density, to indicate the disk location in cyan). Lines are colored by radial velocity $v_{r}$ in ${\rm km\,s^{-1}}$ (see colorbar: red is outflow, blue is inflow). 
We compare face-on ({\em top}) and edge-on ({\em bottom}) projections (with respect to the galactic disk plane), in a box with extending to $\pm 1\,$Mpc ($\approx 8\,R_{\rm vir}$ across) away from the galaxy center in both directions (see scale bar). 
The CR run exhibits qualitatively different structure: ``No CRs'' (MHD+) shows inflow in all directions from the cosmic web onto a very obvious/sharp virial shock at $\sim 250\,$kpc, with a turbulent, inflow-dominated halo interior to this. ``CR+'' shows inflow penetrating in the midplane and filament feeding the disk, with strong bipolar outflow filling almost all the large-scale volume to $>$\,Mpc scales.
\label{fig:inflow.outflow.m12i}}
\end{figure*}

\begin{figure*}
\begin{centering}
\includegraphics[width={0.47\textwidth}]{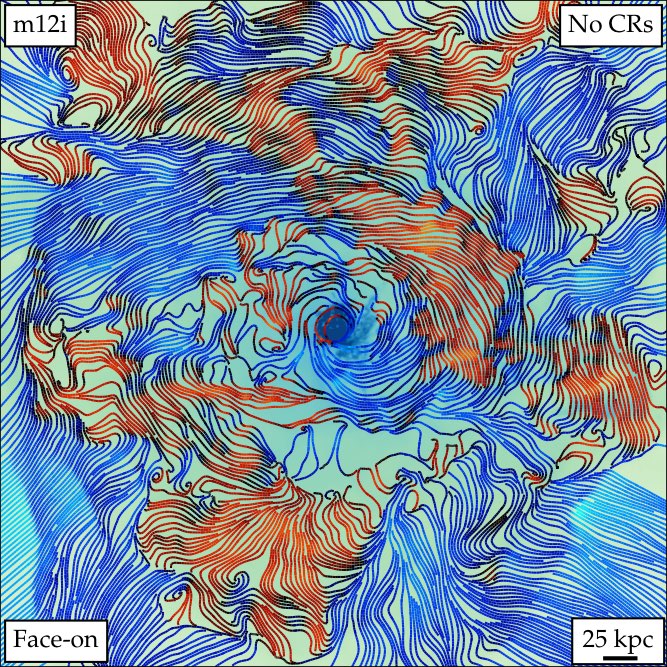}
\hspace{0.3cm}
\includegraphics[width={0.47\textwidth}]{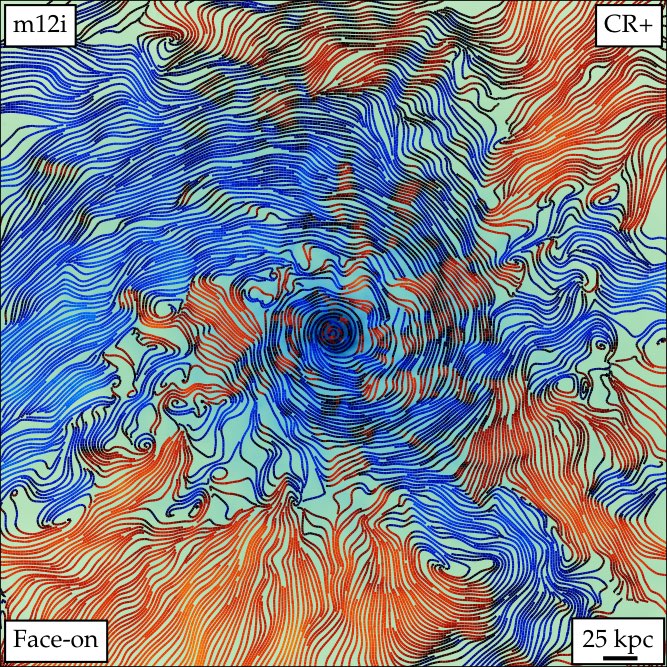} \\
\includegraphics[width={0.47\textwidth}]{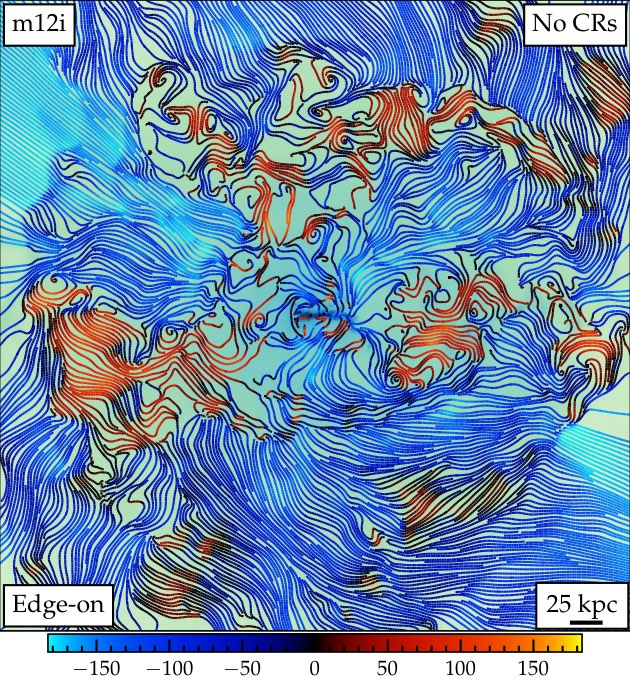}
\hspace{0.3cm}
\includegraphics[width={0.47\textwidth}]{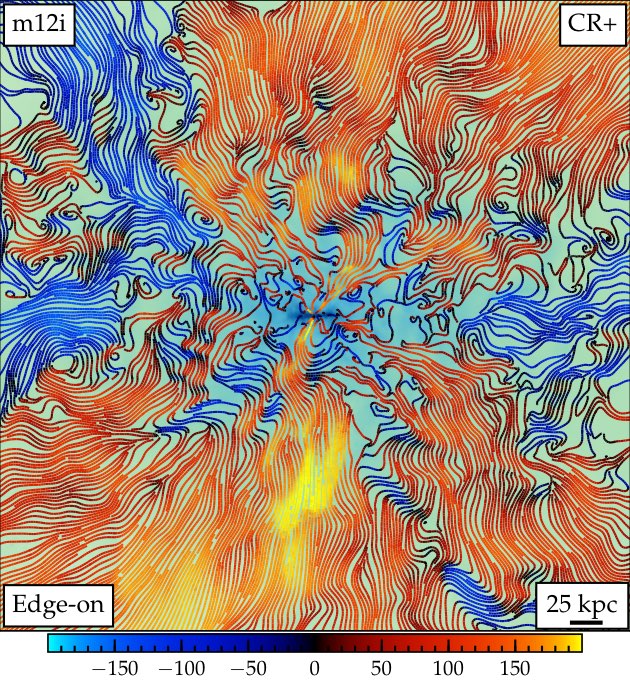} \\
\end{centering}
\caption{Inflow/outflow as Fig.~\ref{fig:inflow.outflow.m12i}, but zooming in to a smaller box at about the radius of the virial shock in the ``No CRs'' run (side-length $\pm1\,R_{\rm vir}$; see scale bar). The qualitative difference between CR and non-CR runs persists, but is less dramatic. We still see substantial turbulence in the inflowing gas for the non-CR run, and more similar mid-plane structure of inflowing gas in both runs (the primary difference is the bipolar outflows, which extend down to the disk). 
\label{fig:inflow.outflow.m12i.zoomed}}
\end{figure*}

\begin{figure*}
\begin{centering}
\includegraphics[width={0.48\textwidth}]{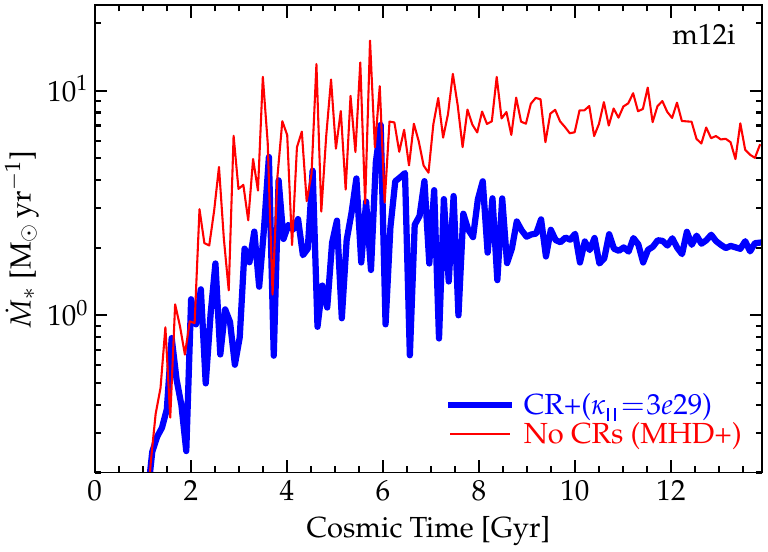} \includegraphics[width={0.48\textwidth}]{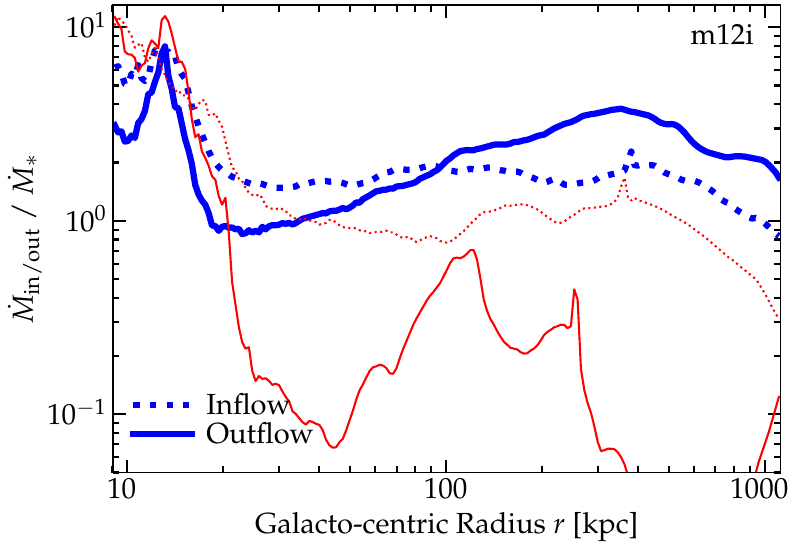} \\
\includegraphics[width={0.48\textwidth}]{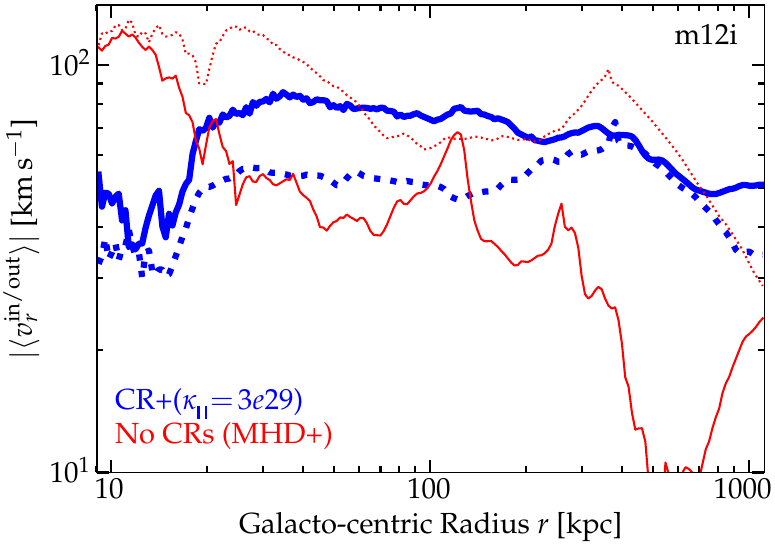} \includegraphics[width={0.48\textwidth}]{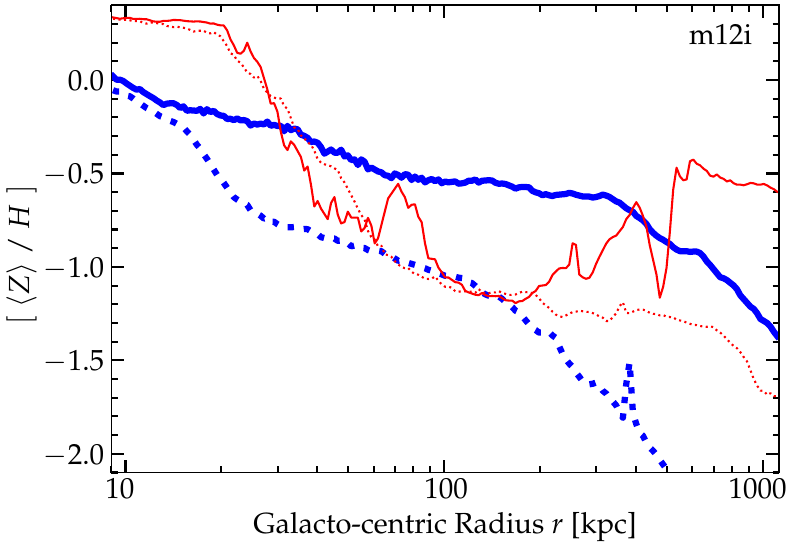}\\     \vspace{-0.25cm}
    \end{centering}
        \caption{Basic star formation (SF) \&\ inflow/outflow properties in {\bf m12i}. 
        {\em Top Left:} SFR ($\dot{M}_{\ast}$) vs.\ cosmic time. Including CRs from SNe, SFRs in MW-mass halos are significantly suppressed below $z\sim 1-2$ (see \paperone). 
        {\em Top Right:} Inflow ({\em dotted}) and outflow ({\em solid}) rates through each radial annulus at $z=0$ (normalized to the SFR). Note inflow+outflow co-exist because the flow is not spherically-symmetric. 
        {\em Bottom Left:} Inflow and outflow $\dot{M}_{\rm in/out}$-weighted mean radial velocities $\langle v_{r} \rangle$  in or out, versus radius at $z=0$.
        {\em Bottom Right:} Inflow and outflow $\dot{M}_{\rm in/out}$-weighted metallicities [Z/H]$=\log_{10}(\langle Z \rangle/Z_{\odot})$. 
        Gross inflow rates $\dot{M}_{\rm in}$ onto the halo (at $\gtrsim R_{\rm vir}$) are similar (more so in absolute units), indicating most of the inflow comes in the dense planar structures that remain in the ``CR+'' run (\fref{fig:inflow.outflow.m12i}); but inflows accelerate to larger $\langle v_{r} \rangle$ (by a factor $\sim 2$) absent CRs. 
        Outflow rates are similar near the disk ($r \lesssim 30\,$kpc): runs without CRs actually have larger $\dot{M}_{\rm out}$, with $\sim 2-3\times$ ``faster'' mean $\langle v_{r} \rangle$ ($\gtrsim 100\,{\rm km\,s^{-1}}$). 
        But in the CGM ($\gtrsim 30\,$kpc), absent CRs the outflow is dramatically ``halted,'' while with CRs it actually accelerates and $\dot{M}_{\rm out}$ increases again, to give $\dot{M}_{\rm out} > \dot{M}_{\rm in}$ at essentially all radii $\gtrsim 100\,$kpc. The CR outflows have intermediate $|\langle v_{r} \rangle^{\rm out}| \sim |\langle v_{r} \rangle^{\rm in}| \sim 50-100\,{\rm km\,s^{-1}}$. Without or without CRs, outflow metallicities decrease with $r$, indicating continuing entrainment, but with CRs the trend is monotonic at all $r$ and the outflows have higher metallicity (vs.\ inflow) at all $r$ (while outflows absent CRs mix within $\sim R_{\rm vir}$, giving no inflow/outflow difference). 
    \label{fig:mdot.inflow.outflow.m12i}}
\end{figure*}

\begin{figure*}
\begin{centering}
\includegraphics[width=0.32\textwidth]{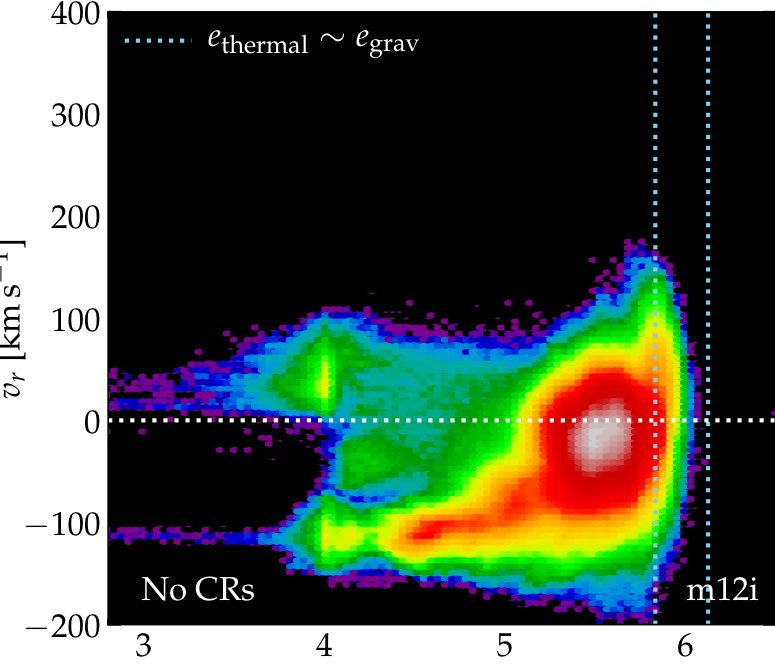}
\includegraphics[width=0.33\textwidth]{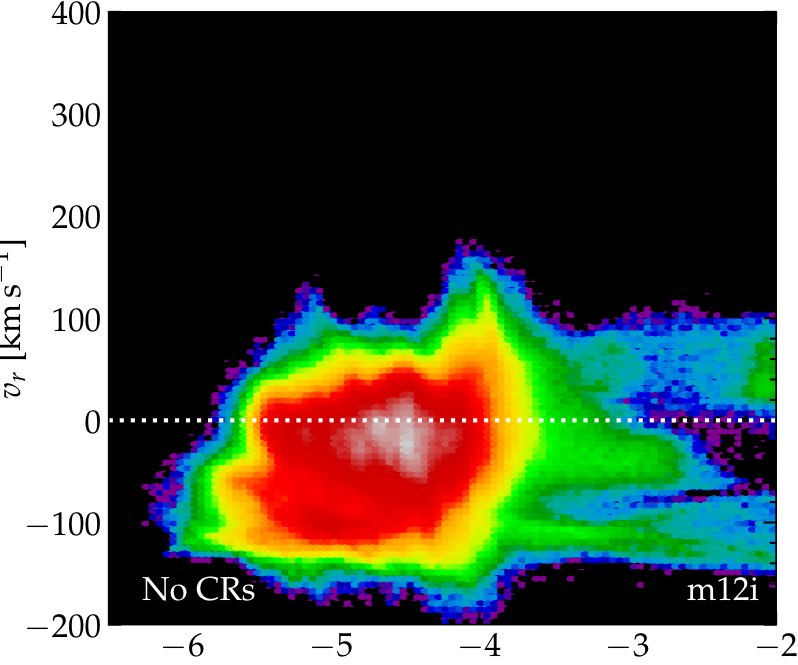}
\includegraphics[width=0.32\textwidth]{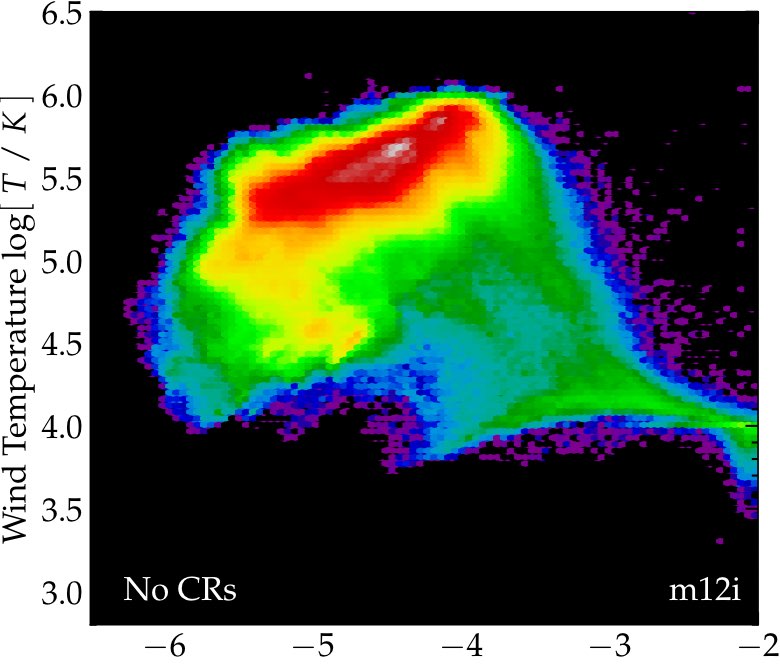}\\
\includegraphics[width=0.32\textwidth]{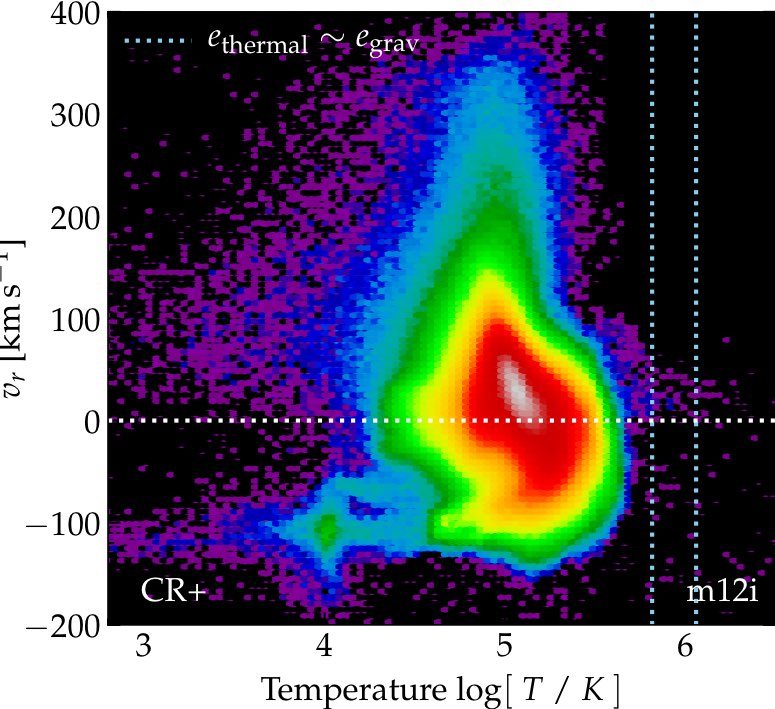}
\includegraphics[width=0.33\textwidth]{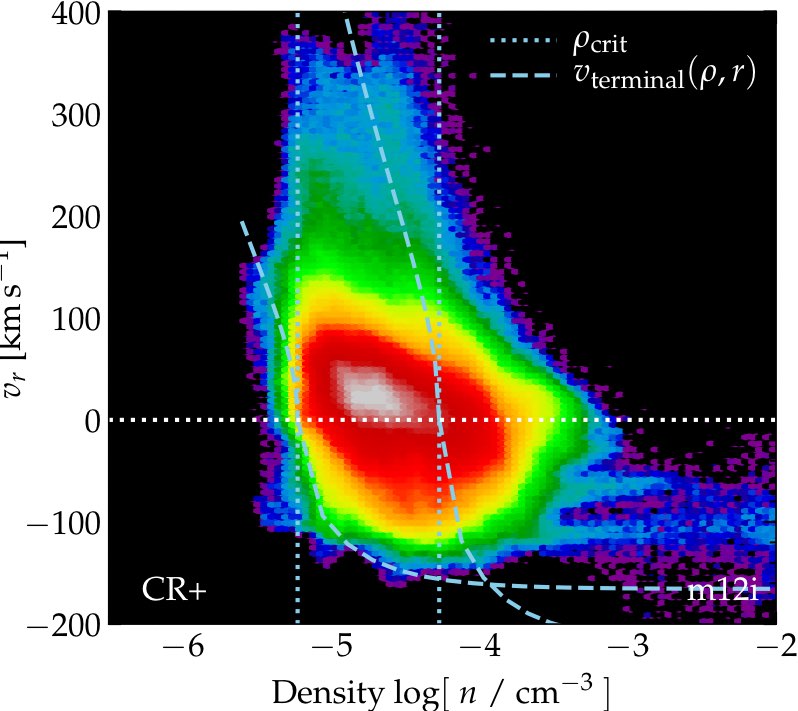}
\includegraphics[width=0.32\textwidth]{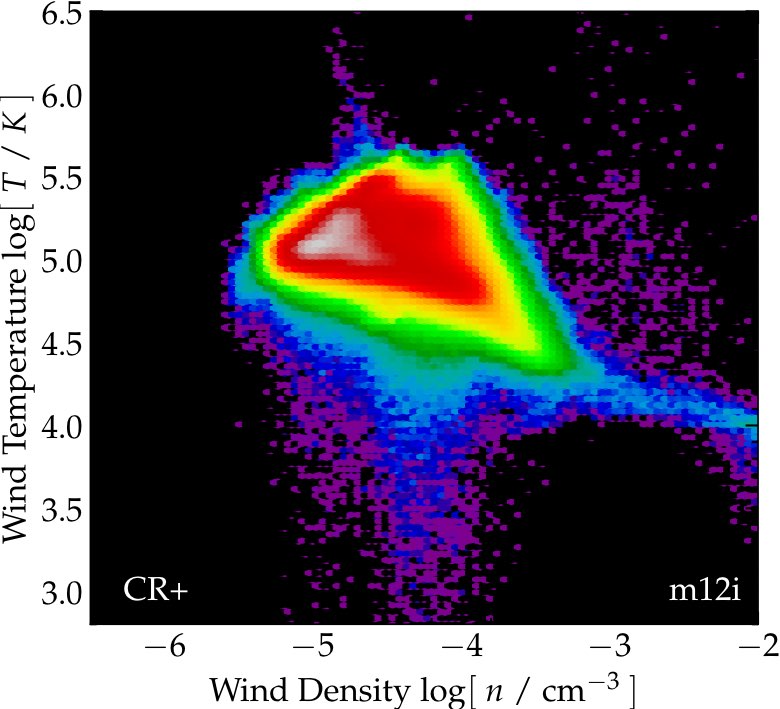}\\
\end{centering}
\caption{Distribution of gas outflow properties (as \fref{fig:demo.vr.vs.r}), comparing our default {\bf m12i} ``No CRs'' ({\em top}) and ``CR+'' ({\em bottom}) runs. 
{\em Left:} Gas outflow velocity ($v_{r}$) versus temperature $T$, for gas selected at galacto-centric radii $0.5 < r/R_{\rm vir} < 1.5$. Vertical lines show the value of $T$ at the inner/outer radii in the ``slice'' where the gas thermal energy density would equal the gravitational potential. In the ``No CRs'' run fast outflows preferentially appear in ``hot'' gas which nears this virial-like value. In the ``CR+'' run the outflows are ``cool'' ($T\sim 10^{5}\,$K, well below this virial value). 
{\em Middle:} Outflow $v_{r}$ vs.\ density $n$, in the same slice in $r$. Vertical dotted lines label $\rho_{\rm crit}$, the critical density (\S~\ref{sec:theory}) where CR pressure balances gravity on the gas, at the inner/outer slice $r$. Dashed curves label the analytic expected ``terminal velocity'' for gas which deviates from $\rho_{\rm crit}$ (denser gas falls in under gravity, less dense gas accelerates out under CR pressure; see \S~\ref{sec:theory}). Without CRs, there is no density-$v_{r}$ relation; with CRs, most gas resides near $\rho_{\rm crit}$, and lower density gas preferentially flows out while higher-density gas almost exclusively flows in.
{\em Right:} Phase ($n-T$) diagram of outflow material (selected to have $v_{r} > V_{\rm vir}/2$): the ``CR+'' outflows are cooler and lower-density, on average.
\label{fig:demo.nh.t.wind}}
\end{figure*}

\begin{figure*}
\begin{centering}
\includegraphics[width=0.328\textwidth]{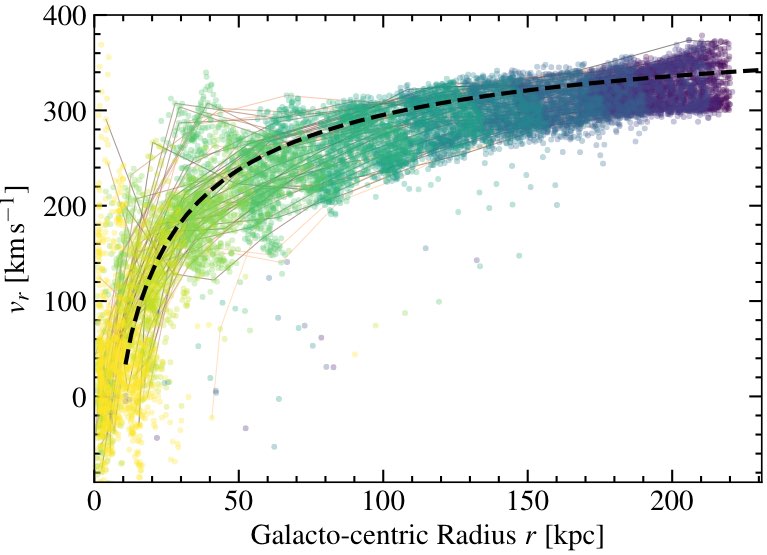}
\includegraphics[width=0.33\textwidth]{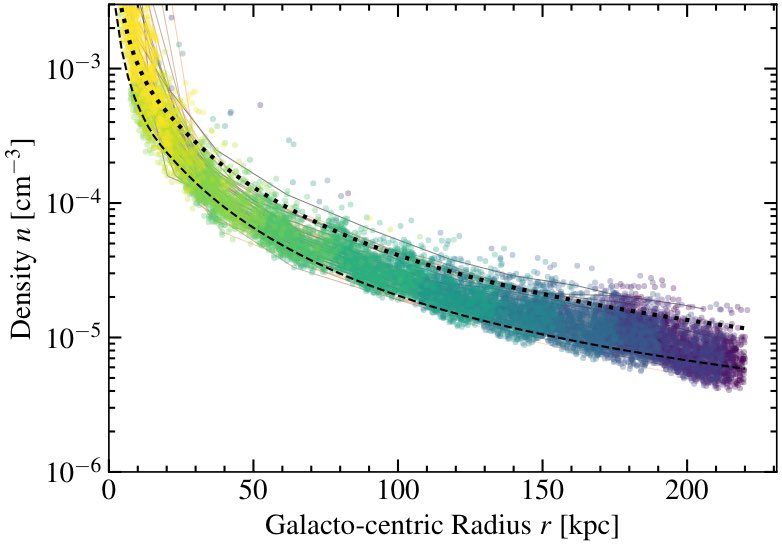}
\includegraphics[width=0.322\textwidth]{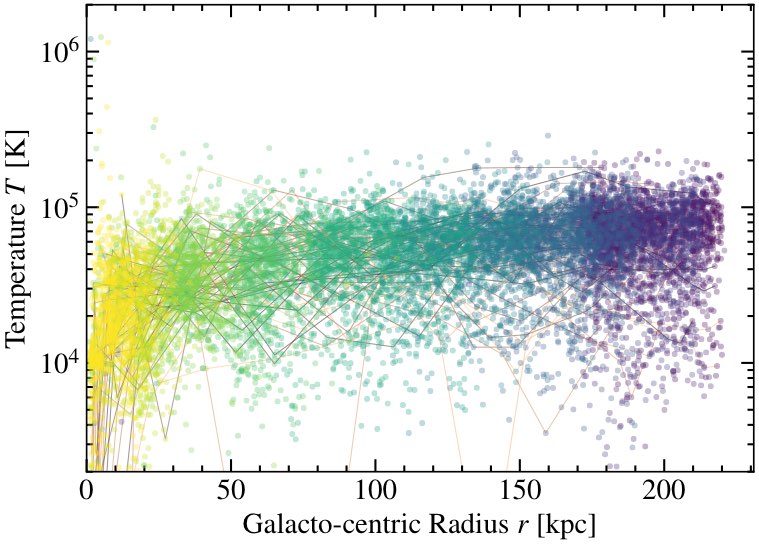}
\end{centering}\vspace{-0.25cm}
\caption{Time-evolution of the ``fast'' outflows in the CR-dominated {\bf m12i} halo (\fref{fig:demo.vr.vs.r}; ``CR+''). To define this, we select all gas with $175<r/{\rm kpc}<220$ and $300 < v_{r}/{\rm km\,s^{-1}} < 400$ at $z=0$, and follow the Lagrangian histories of those fluid elements back to redshift $z\approx 0.07$ ($\sim 1\,$Gyr lookback time). Points show different Lagrangian fluid elements at each snapshot in time, with colors/brightness indicating increasing lookback time (black-to-yellow, in $\sim 50\,$Myr increments); for illustrative purposes we select a random $\sim50$ elements and show lines connecting their trajectories. Note this is only the selected ``fast'' component -- not the average of all gas (which is much slower) shown in Fig.~\ref{fig:mdot.inflow.outflow.m12i}.
{\em Left:} Radial velocity and galacto-centric $r$, as a function of time. Thick (dashed) line shows the analytic prediction from \S~\ref{sec:theory} (as \fref{fig:demo.nh.t.wind}) for a CR pressure-accelerated outflow originating at $\sim 10\,$kpc with $\rho \sim (2/3)\,\rho_{\rm crit}$. 
{\em Middle:} Density and $r$ vs.\ time. Dotted line shows the equilibrium $\rho_{\rm crit}$ (\fref{fig:demo.nh.t.wind}), dashed line shows $(1/2)$ of this: the outflow ``launches'' where the gas falls below $\rho_{\rm crit}$ and is accelerated continuously while just modestly below this density at each $r$.
{\em Right:} Temperature and $r$ vs.\ time. The outflows ``begin'' cold in the disk, and are never heated to sustained very high temperatures before or during acceleration.
\label{fig:outflow.histories}}
\end{figure*}

The specific simulations studied here are the same as those presented and studied in \paperone, where the details of the numerical methods are described. We therefore only briefly summarize here. The simulations were run with {\small GIZMO}\footnote{A public version of {\small GIZMO} is available at \gizmourl} \citep{hopkins:gizmo}, in its meshless finite-mass MFM mode (a mesh-free finite-volume Lagrangian Godunov method). The simulations solve the equations of ideal magneto-hydrodynamics (MHD) as described and tested in \citet{hopkins:mhd.gizmo,hopkins:cg.mhd.gizmo}, with fully-anisotropic Spitzer-Braginskii conduction and viscosity as described in \citet{hopkins:gizmo.diffusion,su:2016.weak.mhd.cond.visc.turbdiff.fx} and \paperone\ (see Eqs.~1-3 therein). Gravity is solved with fully-adaptive Lagrangian force softening (so hydrodynamic and force resolutions are matched). 

All our simulations include magnetic fields, anisotropic Spitzer-Braginskii conduction and viscosity, and the physics of cooling, star formation, and stellar feedback from the FIRE-2 version of the Feedback in Realistic Environments (FIRE) project, described in detail in \citet{hopkins:fire2.methods}. Gas cooling is followed from $T=10-10^{10}\,$K (including a variety of processes, e.g.\ metal-line, molecular, fine-structure, dust, photo-electric, photo-ionization cooling/heating, and accounting for self-shielding and both local radiation sources and the meta-galactic background; see \citealt{hopkins:fire2.methods}). We follow 11 distinct abundances accounting for turbulent diffusion of metals and passive scalars as in \citet{colbrook:passive.scalar.scalings,escala:turbulent.metal.diffusion.fire}. Gas is converted to stars using a sink-particle prescription if and only if it is locally self-gravitating at the resolution scale \citep{hopkins:virial.sf}, self-shielded/molecular \citep{krumholz:2011.molecular.prescription}, Jeans-unstable, and denser than $>1000\,{\rm cm^{-3}}$. Each star particle is then evolved as a single stellar population with IMF-averaged feedback properties calculated following \citet{starburst99} for a \citet{kroupa:2001.imf.var} IMF and its age and abundances. We explicitly treat mechanical feedback from SNe (Ia \&\ II) and stellar mass loss (from O/B and AGB stars) as discussed in \citet{hopkins:sne.methods}, and radiative feedback including photo-electric and photo-ionization heating and UV/optical/IR radiation pressure with a five-band radiation-hydrodynamics scheme as discussed in \citet{hopkins:radiation.methods}. Conduction adds the parallel heat flux $\kappa_{\rm cond}\,\hat{\bf B}\,(\hat{\bf B}\cdot \nabla T)$, and viscosity the anisotropic stress tensor $\Pi \equiv -3\,\eta_{\rm visc}\,(\hat{\bf B}\otimes\hat{\bf B} - \mathbb{I}/3)\,(\hat{\bf B}\otimes\hat{\bf B} - \mathbb{I}/3) : (\nabla\otimes{\bf v})$ to the gas momentum and energy equations, where the parallel transport coefficients $\kappa_{\rm cond}$ and $\eta_{\rm visc}$ follow the usual \citet{spitzer:conductivity,braginskii:viscosity} form, accounting for saturation following \citet{cowie:1977.evaporation}, and accounting for plasma instabilities (e.g.\ Whistler, mirror, and firehose) limiting the heat flux and anisotropic stress at high plasma-$\beta$ following \citet{komarov:whistler.instability.limiting.transport,squire:2017.max.braginskii.scalings,squire:2017.kinetic.mhd.alfven,squire:2017.max.anisotropy.kinetic.mhd}. The numerical implementation follows \citet{hopkins:gizmo.diffusion} to ensure stability. The simulations are fully-cosmological ``zoom-in'' runs with a high-resolution region (of size ranging from $\sim 1-5$ Mpc on a side, increasing with $M_{\rm halo}$) surrounding a ``primary'' halo of interest \citep{onorbe:2013.zoom.methods}.\footnote{For the MUSIC \citep{hahn:2011.music.code.paper} files necessary to generate all ICs here, see:\\ \ICsurl} The properties of these primary halos (our main focus here, as these are the best-resolved in each box) are given in Table~\ref{tbl:sims}. Details of all of these numerical methods are in \citet{hopkins:fire2.methods}.

Our ``CRs'' or ``CR+'' simulations include all of the above, and add our ``full physics'' treatment of CRs as described in detail in \citet{chan:2018.cosmicray.fire.gammaray} and \paperone. We evolve a ``single bin'' ($\sim$\,GeV) or constant spectral distribution of CRs as an ultra-relativistic ($\gamma=4/3$) fluid, accounting for injection in SNe shocks (with a fixed fraction $\epsilon_{\rm cr}=0.1$ of the initial SNe ejecta kinetic energy in each time-resolved explosion injected into CRs), streaming and collisional (hadronic and Coulomb and ionization) losses from the CRs (accounting for local neutral fractions and composition, with a fraction of this loss thermalizing and heating gas) following \citet{Mann94,guo.oh:cosmic.rays}, advection and adiabatic work (in the local ``strong coupling'' approximation, so the CR pressure contributes to the total pressure in the Riemann problem for the gas equations-of-motion), and CR transport including fully-anisotropic diffusion and streaming \citep{mckenzie.voelk:1982.cr.equations}. We solve the transport equations using a two-moment approximation to the collisionless Boltzmann equation (with a ``reduced speed of light'' $\tilde{c}\sim1000\,{\rm km\,s^{-1}}$), with a constant parallel diffusivity $\kappa_{\|}$ (perpendicular $\kappa_{\bot}=0$). The streaming velocity is ${\bf v}_{\rm stream} = -v_{\rm stream}\,\hat{\bf B}\,(\hat{\bf B}\cdot \hat{\nabla} P_{\rm cr})$ with $v_{\rm stream} =3\, v_{A}$ ($v_{A}$ the \Alf\ speed) our default choice, motivated by models favoring trans or modestly super-\Alf{ic} streaming  \citep{skilling:1971.cr.diffusion,holman:1979.cr.streaming.speed,kulsrud:plasma.astro.book,yan.lazarian.2008:cr.propagation.with.streaming}, although varying this widely (from $<1\,v_{A}$ to  $\sim 3\,(c_{s}^{2} + v_{A}^{2})^{1/2} \gg v_{A}$) has almost no effect on our  conclusions (see  \paperone). The ``streaming loss'' term ${\bf v}_{A}\cdot \nabla P_{\rm cr}$ represents losses to plasma instabilities at the CR gyro scale  and is 
thermalized \citep{wentzel:1968.mhd.wave.cr.coupling,kulsrud.1969:streaming.instability}. 

Our ``baseline'' or ``no CRs'' simulations include all the physics above except CRs: these are the ``MHD+'' simulations in \paperone. Note there we also compared a set without magnetic fields, conduction, or viscosity (the ``Hydro+'' runs); but as shown therein and in \citet{su:2016.weak.mhd.cond.visc.turbdiff.fx}  the differences in these runs are largely negligible, and we confirm this here. 
Our default ``CR'' simulations adopt $\kappa_{\|}=3\times10^{29}\,{\rm cm^{2}\,s^{-1}}$, along with the full physics of anisotropic streaming, diffusion, collisional losses, etc., above: these are the ``CR+($\kappa=3e29$)'' simulations in \paperone. Although we considered variations to all of these CR physics and, in particular, the diffusivity (which is not known {\em a priori}) in \paperone, we showed that the observational constraints from e.g.\ spallation and more detailed measurements in the MW and $\gamma$-ray emission in local galaxies were all consistent with the default ($\kappa_{\|}=3\times10^{29}\,{\rm cm^{2}\,s^{-1}}$) model here, and ruled out models (within the context of the approximations here) with much lower/higher $\kappa_{\|}$.

\begin{figure}
\hspace{0.03cm}\plotonesize{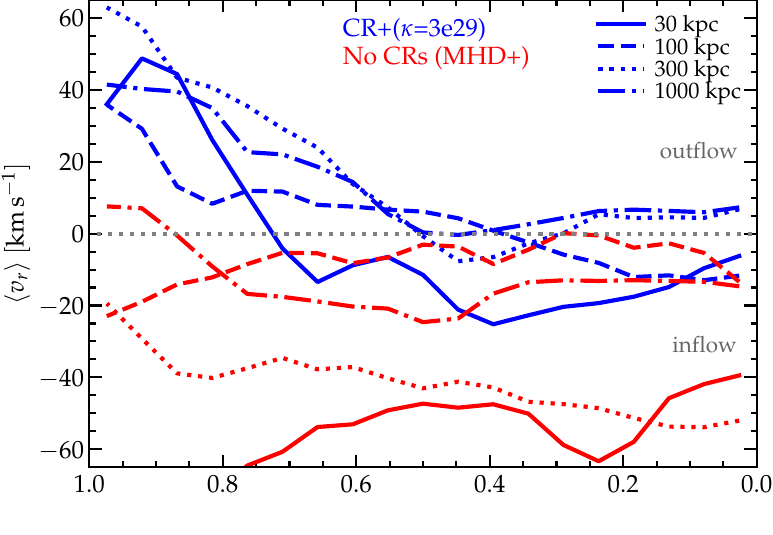}{0.91}\vspace{-0.3cm}\\
\plotonesize{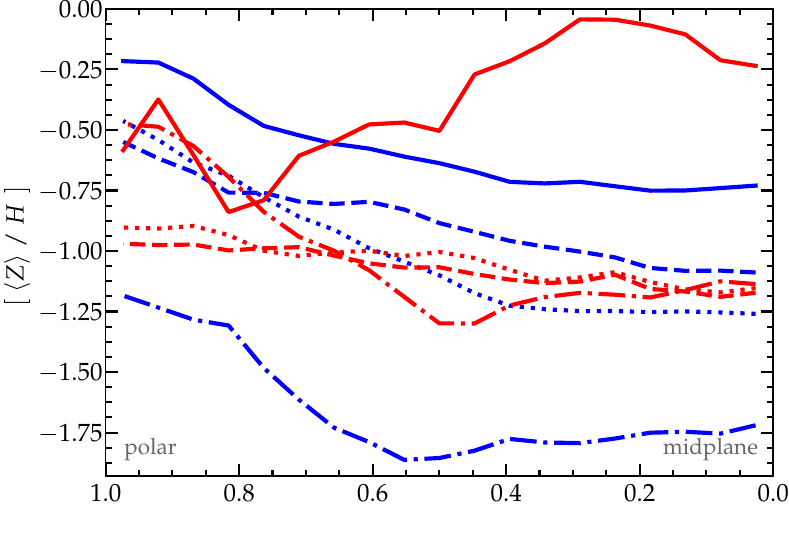}{0.93}\vspace{-0.3cm}\\
\plotonesize{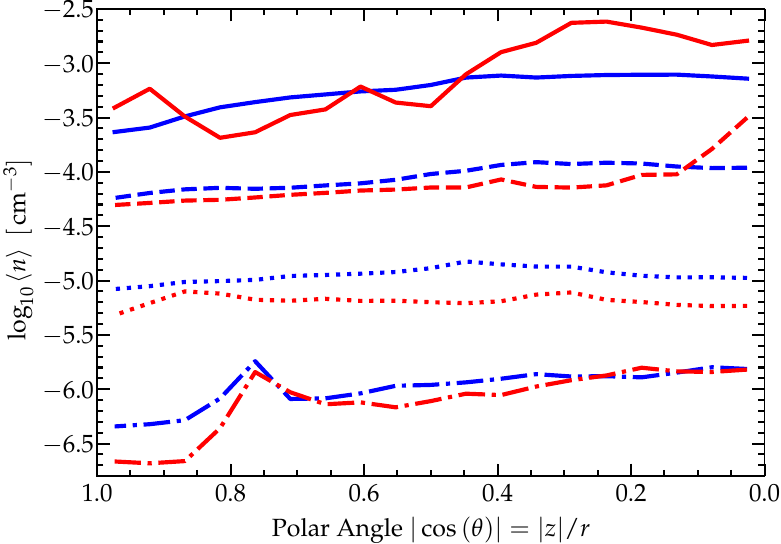}{0.92}\\
    \vspace{-0.25cm}
\caption{CGM gas properties versus polar angle $\cos{\theta} = z/r$, where $\hat{z}$ is the direction of the disk angular momentum axis within $10\,$kpc, in four spherical shells at radii $r=30,\,100,\,300,\,1000$\,kpc. We compare mass-weighted mean gas radial velocity ($\langle v_{r}\rangle$; {\em top}), metallicity ([Z/H]; {\em middle}), and density ($\langle n \rangle$; {\em bottom}), in runs with ({\em blue}) and without ({\em red}) CRs. Mean densities are similar with or without CRs: $\langle n \rangle$ decreases primarily with $r$ (as $\sim r^{-2}$, roughly), with a weaker trend with $\theta$ (polar $\langle \rho \rangle$ a factor $\sim 3$ lower than midplane). Absent CRs there is little coherent trend in $Z$, with CRs a monotonic radial trend is evident with again a weaker $\theta$-dependence (factor $\sim 3$ higher $Z$ at poles). In $\langle v_{r} \rangle$, we see a clear trend with CRs where polar angles are in outflow (accelerating away from the disk) with midplane inflow (accelerating near the disk); without CRs there is no polar-angle trend.
    \label{fig:angle.outflow}}
\end{figure}

\begin{figure}
\plotonesize{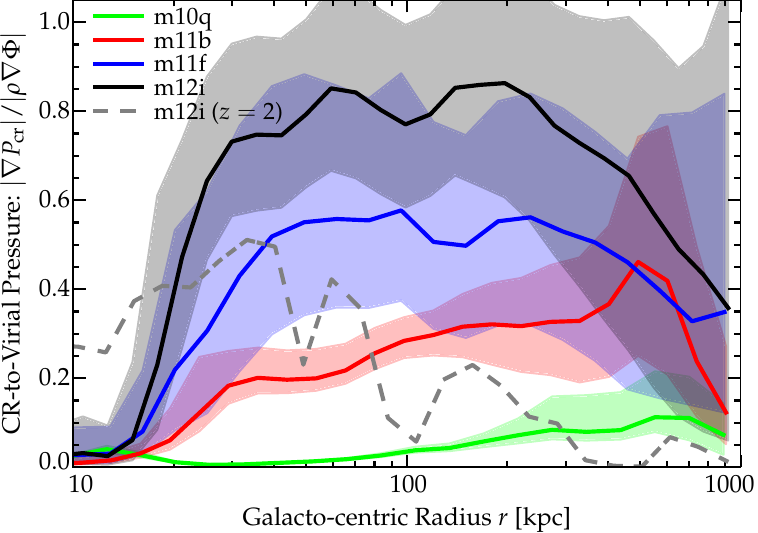}{0.9}
    \vspace{-0.25cm}
\caption{Ratio of the outward CR pressure force $\nabla P_{\rm cr}$ (averaged in spherical shells at a given radius $r$) to gravitational force $\rho\,\nabla \Phi$ at the same $r$, as a function of $r$ (solid line shows mean; shaded shows inter-quartile range), at $z=0$ (except for one line measured at $z=2$, labeled), for different halos from Table~\ref{tbl:sims}. As shown in \papertwo\ and \citet{ji:fire.cr.cgm}, in the simulations with CRs (with these particular diffusion coefficients), CRs dominate the CGM pressure for MW-mass galaxies at low redshifts $z\lesssim 1$. At lower masses (or higher redshifts) the CR pressure (relative to gravity) drops rapidly, becoming negligible for galaxies with halos $\lesssim 10^{11}\,M_{\odot}$ (or $z\gtrsim 1-2$). At $\lesssim 20\,$kpc, we see the effect of the gaseous disk (where rotation, not hydrostatic CR pressure, balances gravity); at $\gtrsim 200$\,kpc we see mean CR pressure falls but some sightlines with low densities still have $|\nabla P_{\rm cr}| \sim |\rho\,\nabla\Phi |$, so slow CR-pressure-driven outflows can continue to $\sim$\,Mpc.
    \label{fig:profile.pressure}}
\end{figure}

\begin{figure*}
\begin{centering}
\includegraphics[width={0.33\textwidth}]{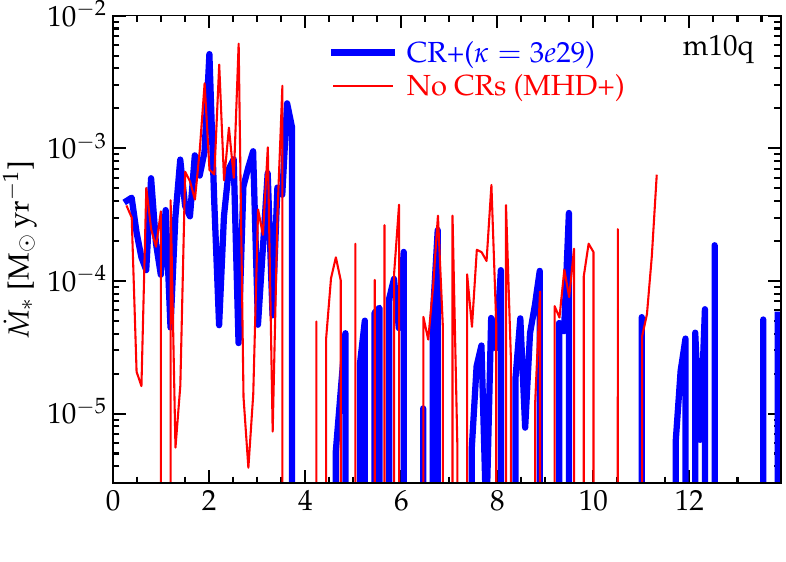} 
\includegraphics[width={0.33\textwidth}]{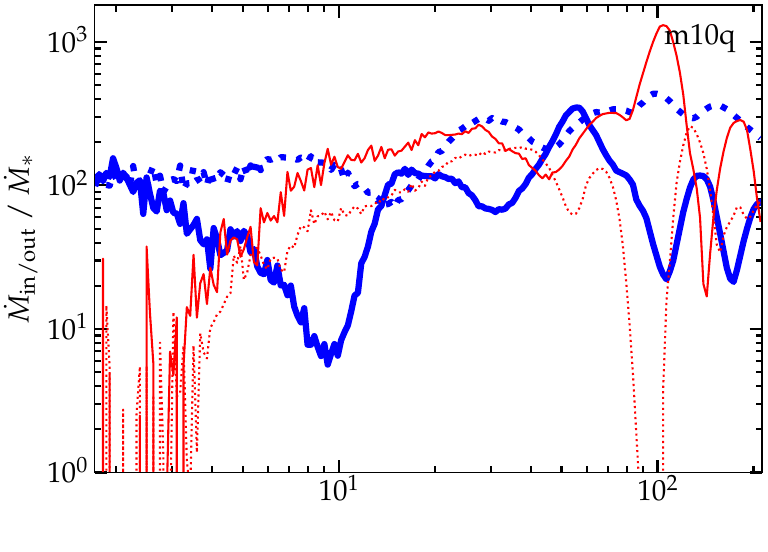} 
\includegraphics[width={0.33\textwidth}]{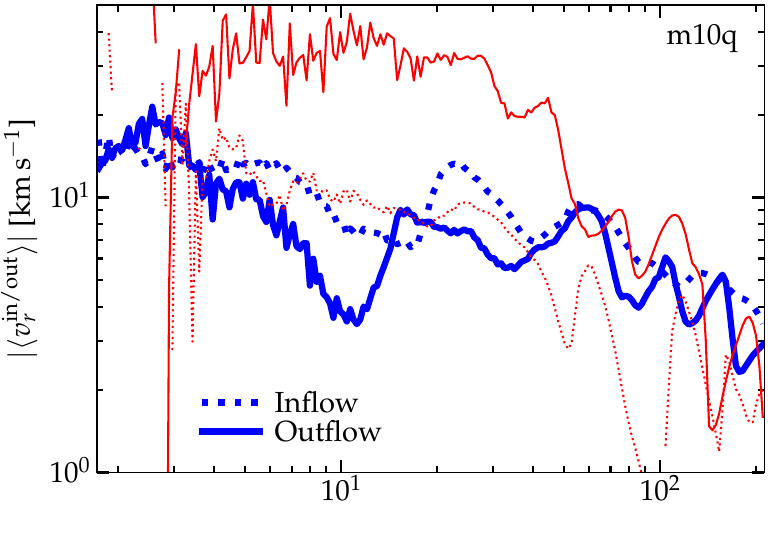} \\
\vspace{-0.3cm}
\includegraphics[width={0.33\textwidth}]{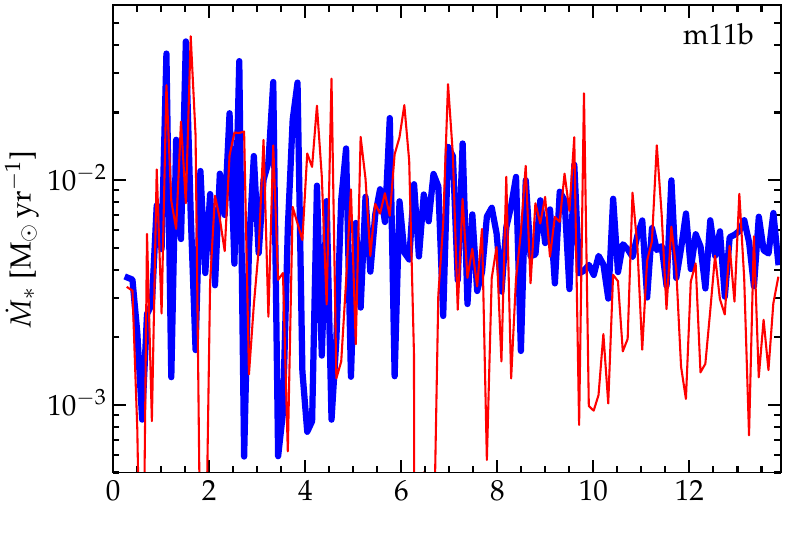} 
\includegraphics[width={0.33\textwidth}]{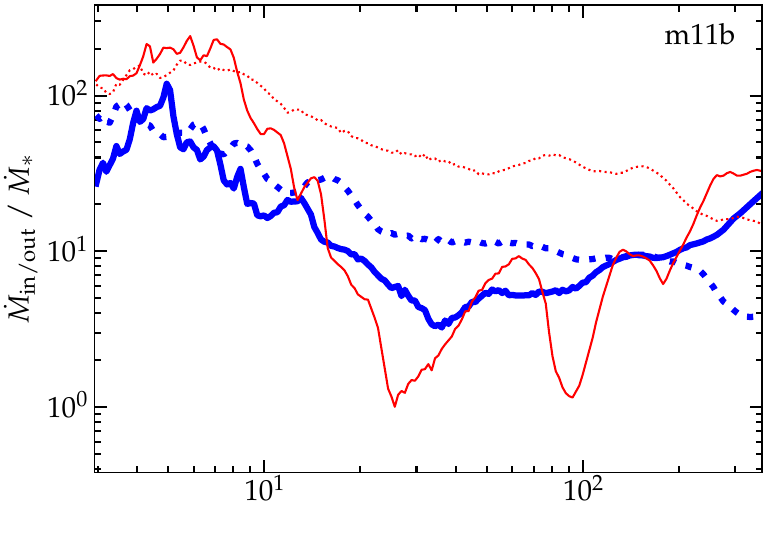} 
\includegraphics[width={0.33\textwidth}]{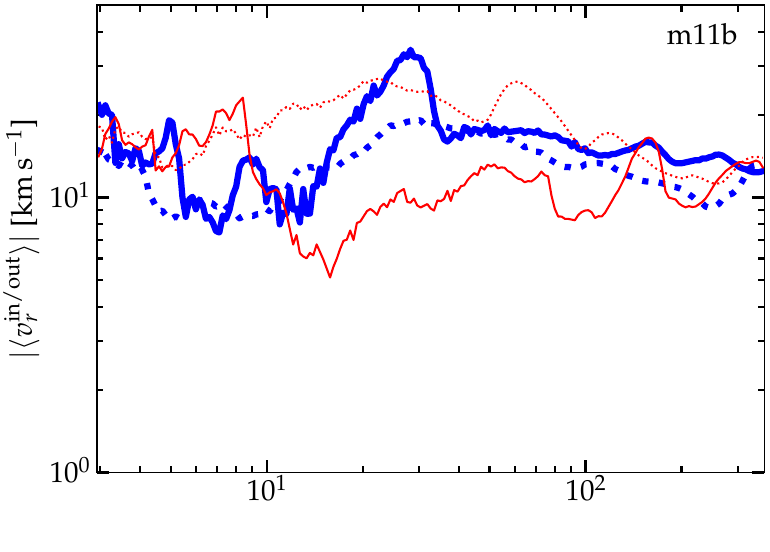} \\
\vspace{-0.3cm}
\includegraphics[width={0.33\textwidth}]{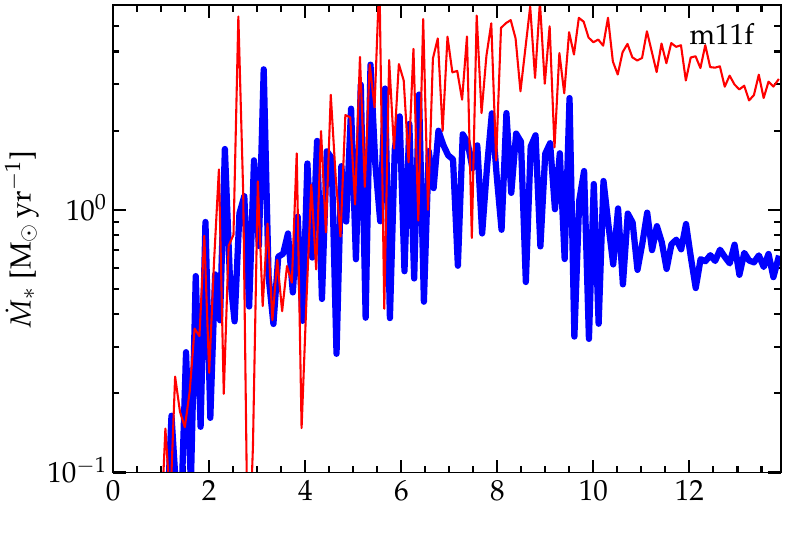} 
\includegraphics[width={0.33\textwidth}]{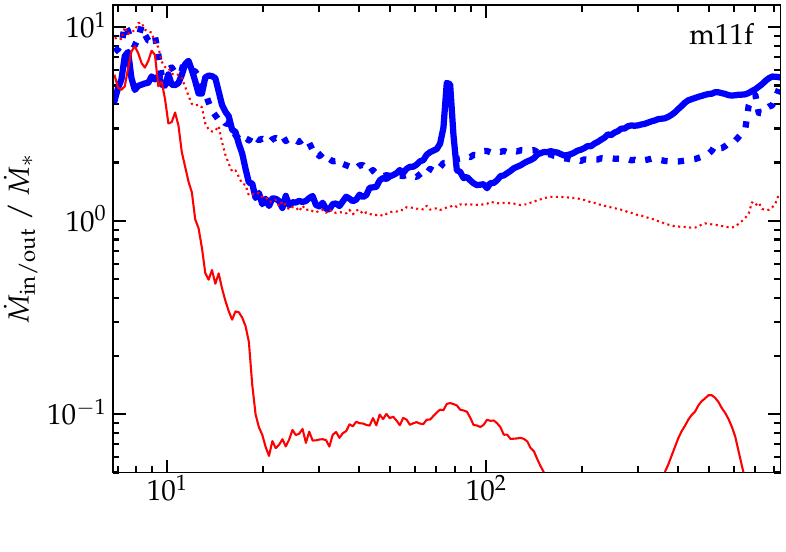} 
\includegraphics[width={0.33\textwidth}]{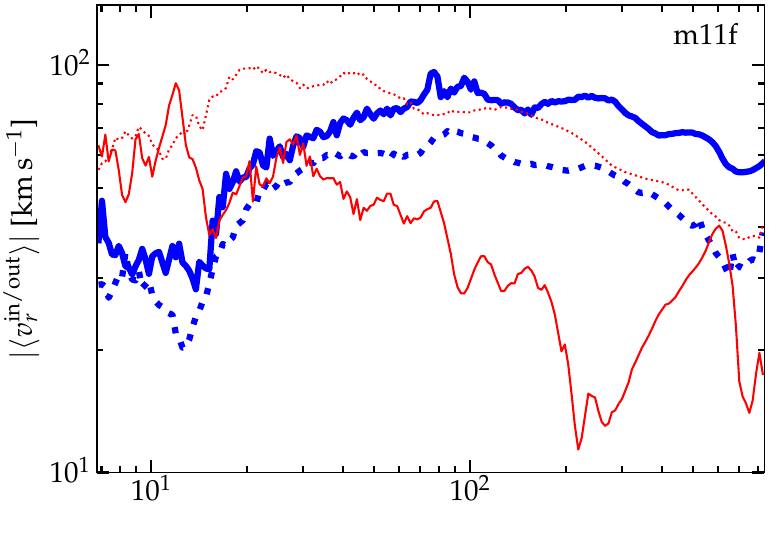} \\
\vspace{-0.3cm}
\hspace{0.05cm}\includegraphics[width={0.324\textwidth}]{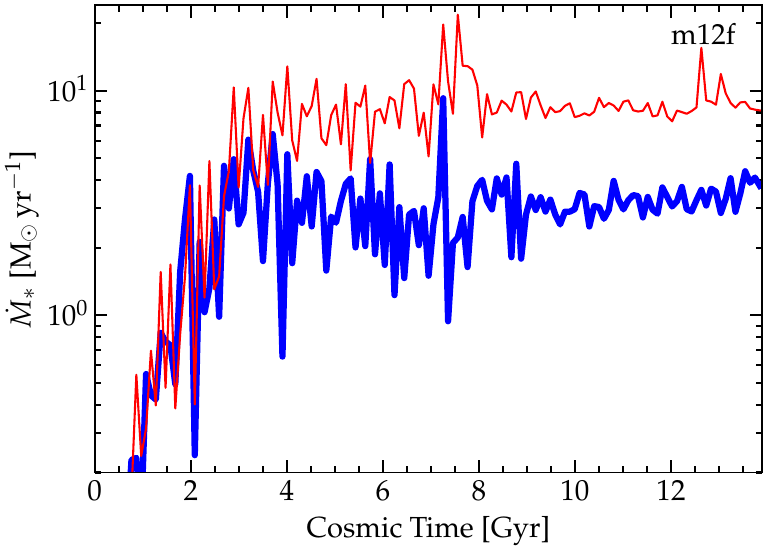} 
\includegraphics[width={0.33\textwidth}]{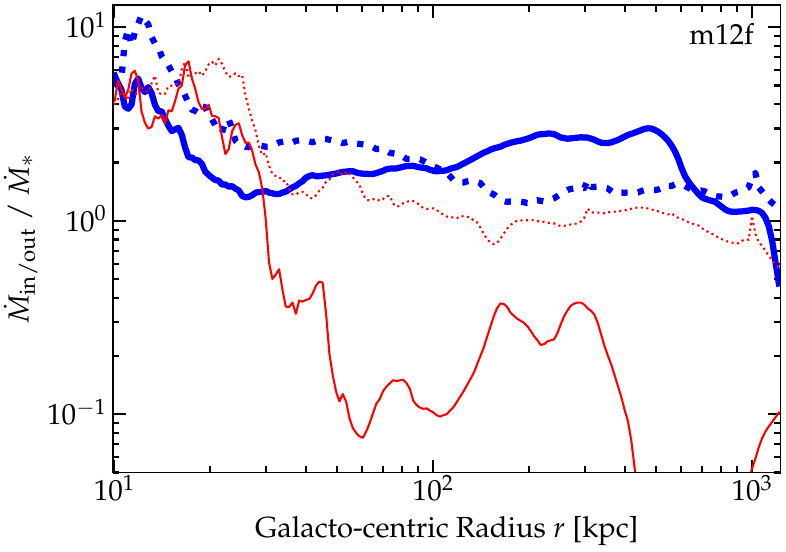} 
\includegraphics[width={0.33\textwidth}]{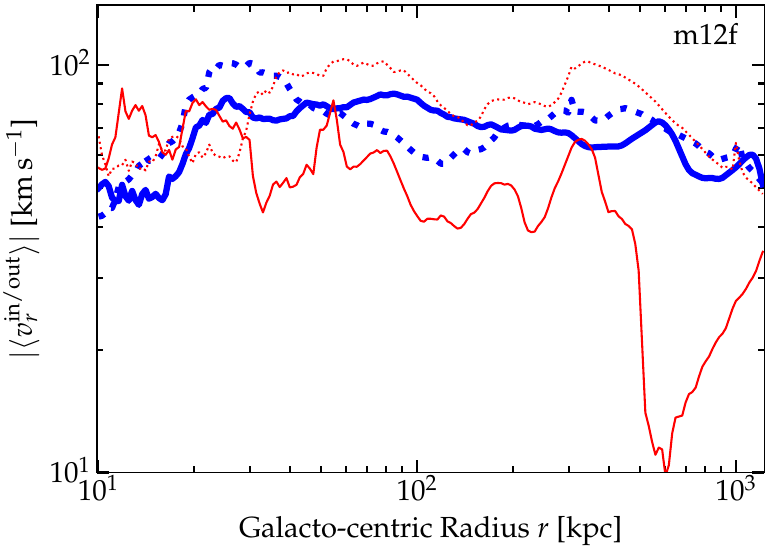} \\
    \end{centering}
    \vspace{-0.25cm}
        \caption{SFRs and inflow/outflow rates (as \fref{fig:mdot.inflow.outflow.m12i}) for {\bf m10q}, {\bf m11b}, {\bf m11f}, {\bf m12f} (top-to-bottom; see Table~\ref{tbl:sims}). As expected from where CRs dominate the pressure in the halo, the CRs have a weak effect on the SFR (see \paperone) or mass outflow rate at any annulus in runs with $M_{\rm halo} \lesssim 10^{11}\,M_{\sun}$ ({\bf m10q} and {\bf m11b}, here). There are some more subtle effects at these halo masses: note e.g.\ the somewhat less-bursty late-time SFR in {\bf m11b} (which is reflected in the outflow rate having less pronounced ``peaks'' at large $r$ from those previous ``bursts''). But in halos which reach $\gtrsim 10^{11}\,M_{\sun}$ at $z\sim 1$ ({\bf m11f}) or $z\sim 2$ ({\bf m12f}), the effects are similar to those for {\bf m12i} in \fref{fig:mdot.inflow.outflow.m12i}: inflow rates at large radii, and outflow mass-loading factors at small radii (``near the disk'') are relatively weakly modified, but CRs strongly suppress SF by accelerating material into outflow away from the disk in the CGM and IGM, maintaining net outflow with $\dot{M}_{\rm out}$  flat or rising to $\gtrsim\,$Mpc.
    \label{fig:mdot.inflow.outflow.others}}
\end{figure*}

\begin{figure*}
\begin{centering}
\includegraphics[width=0.245\textwidth]{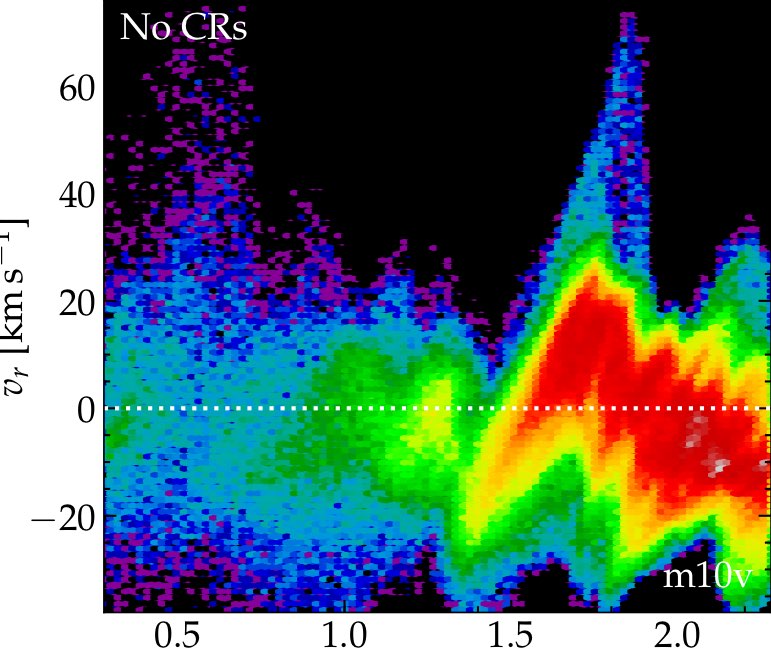}
\includegraphics[width=0.235\textwidth]{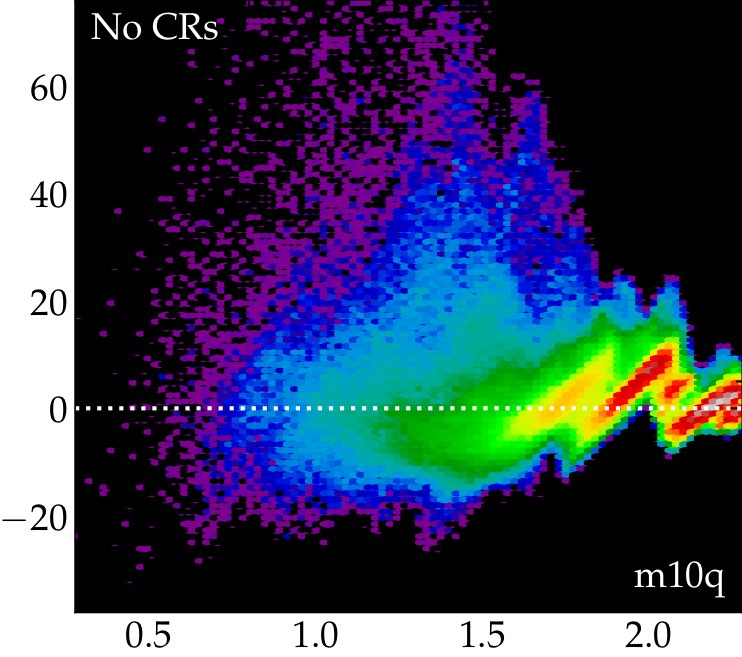}
\includegraphics[width=0.24\textwidth]{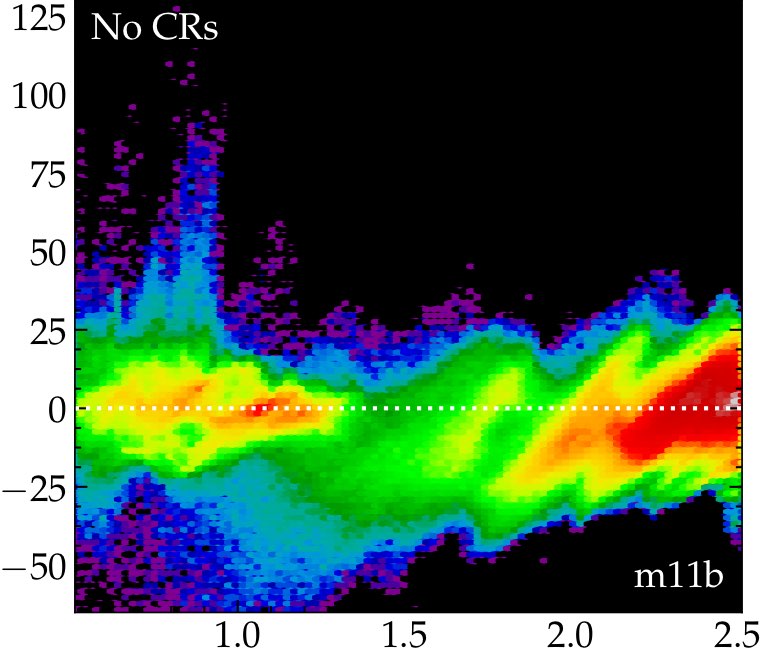}
\includegraphics[width=0.24\textwidth]{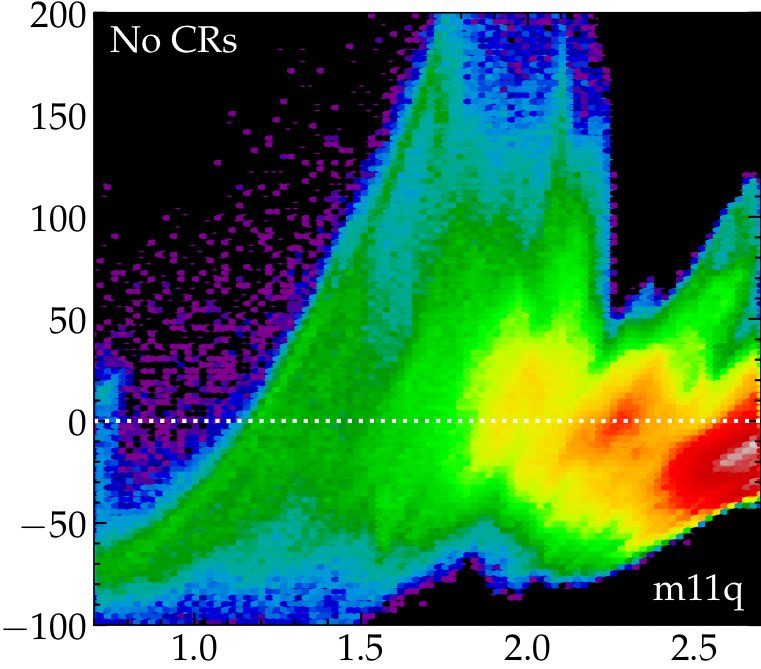}\\
\includegraphics[width=0.245\textwidth]{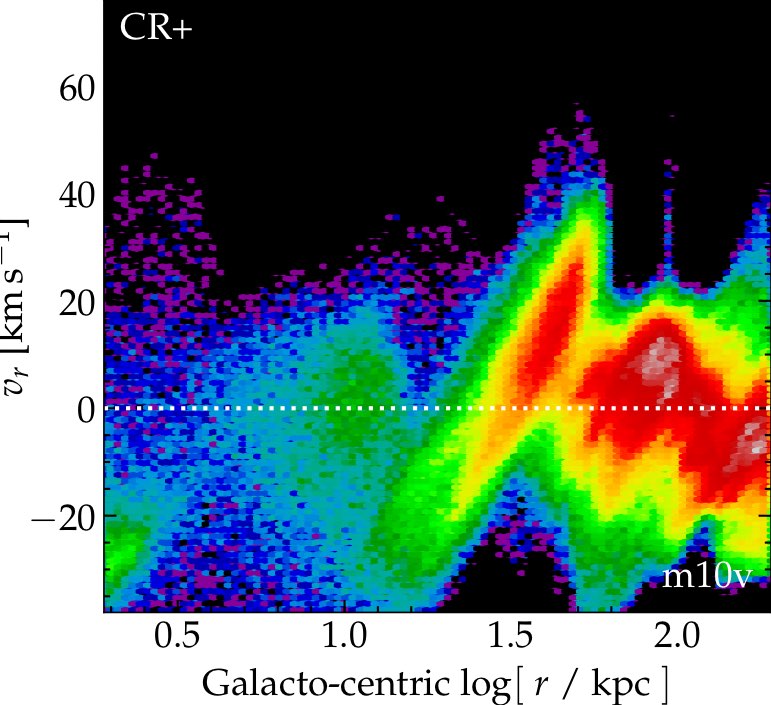}
\includegraphics[width=0.235\textwidth]{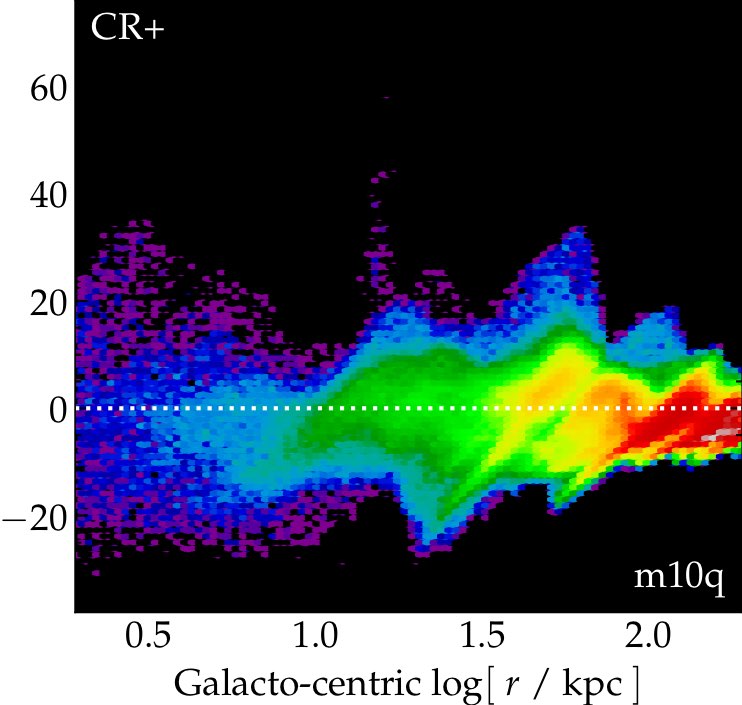}
\includegraphics[width=0.24\textwidth]{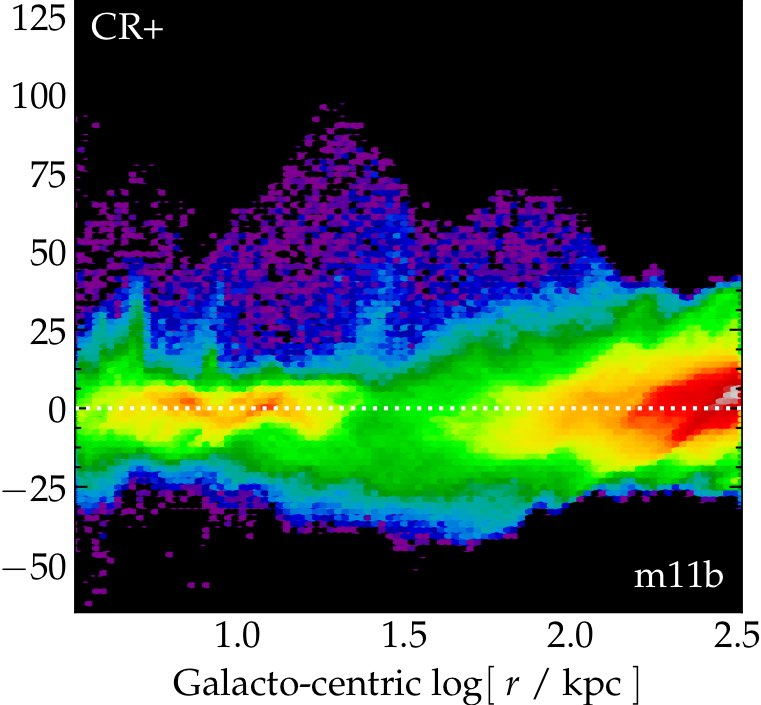}
\includegraphics[width=0.24\textwidth]{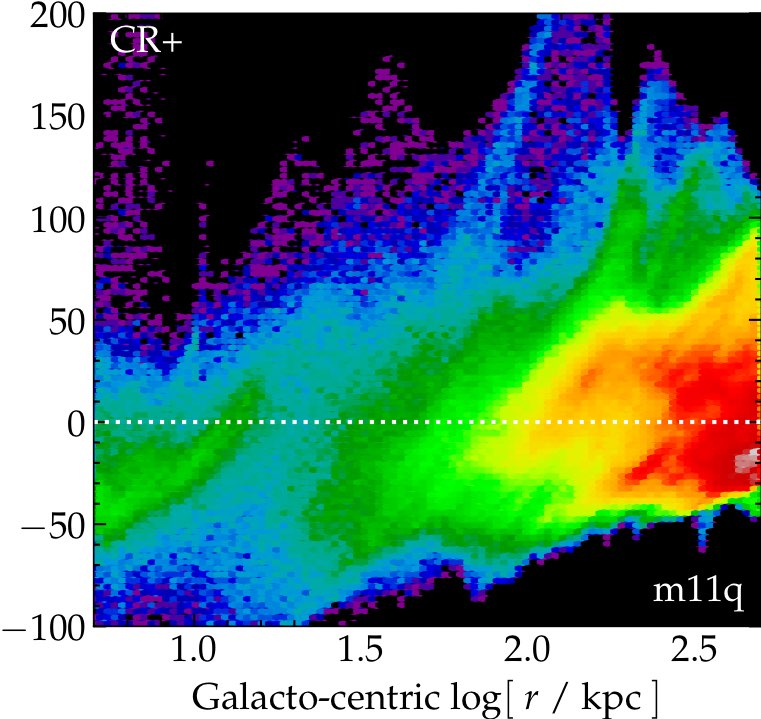}\\
\vspace{0.5cm}
\includegraphics[width=0.245\textwidth]{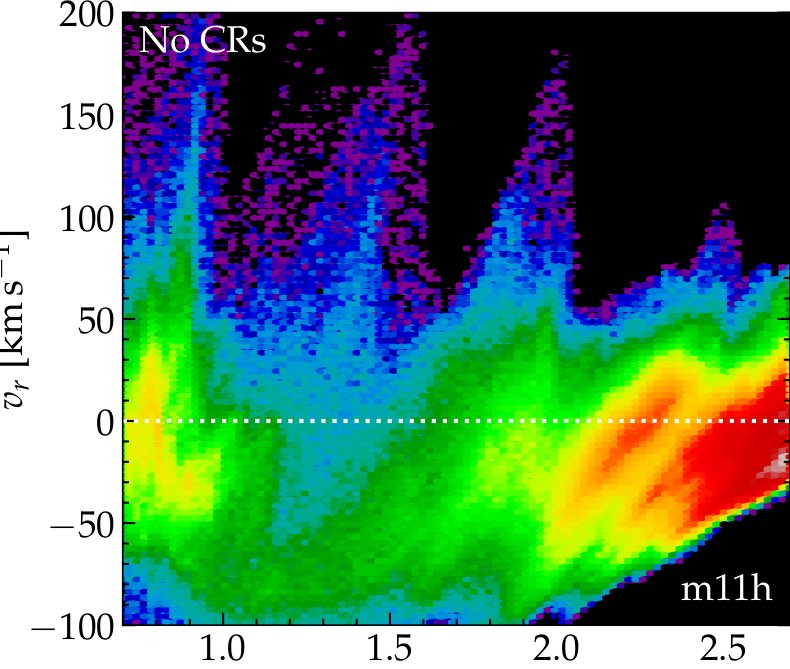}
\includegraphics[width=0.235\textwidth]{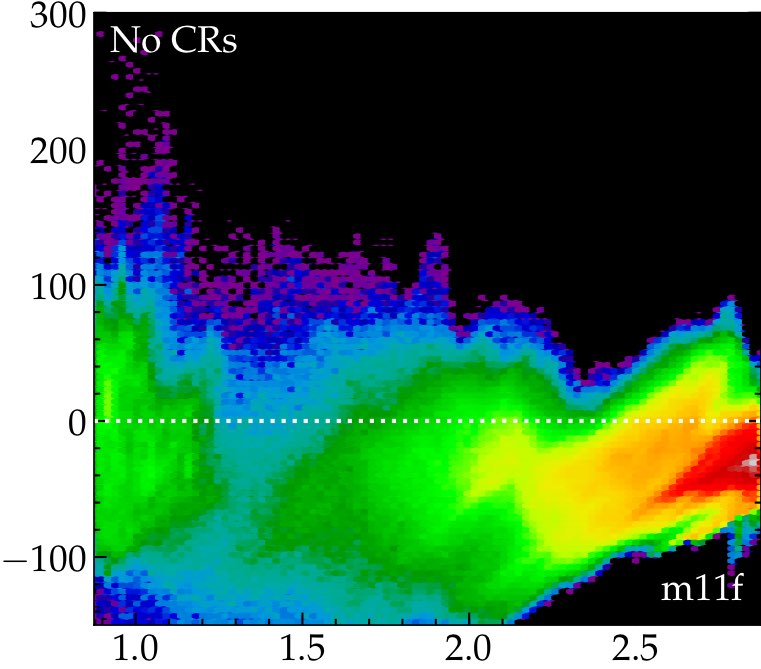}
\includegraphics[width=0.24\textwidth]{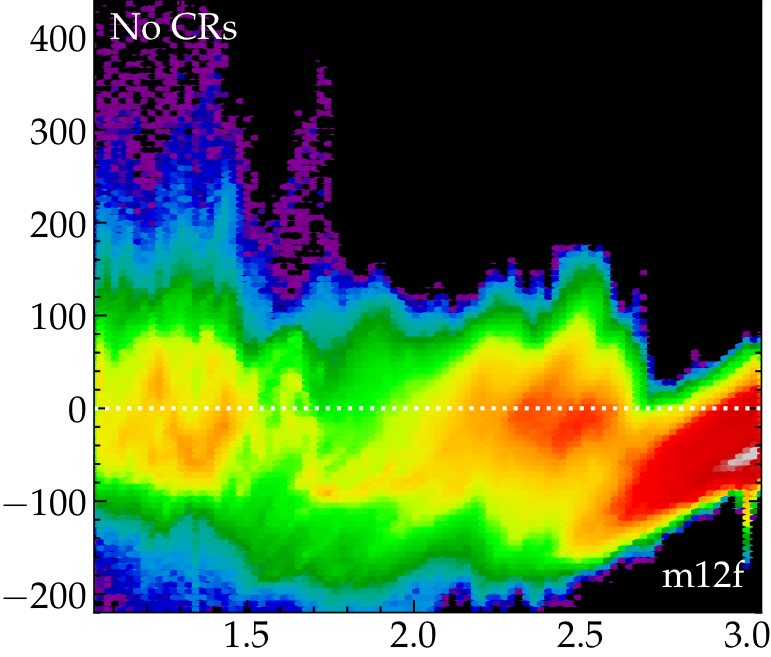}
\includegraphics[width=0.24\textwidth]{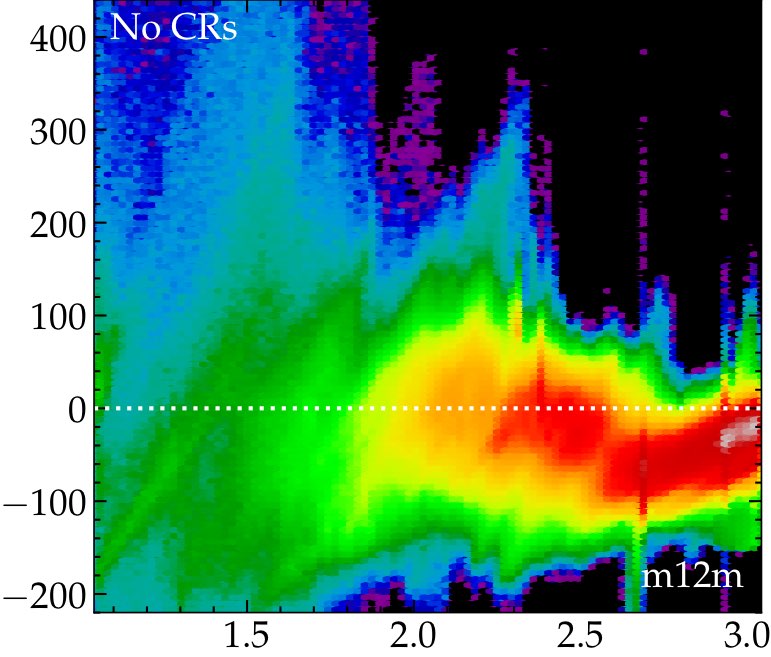}\\
\includegraphics[width=0.245\textwidth]{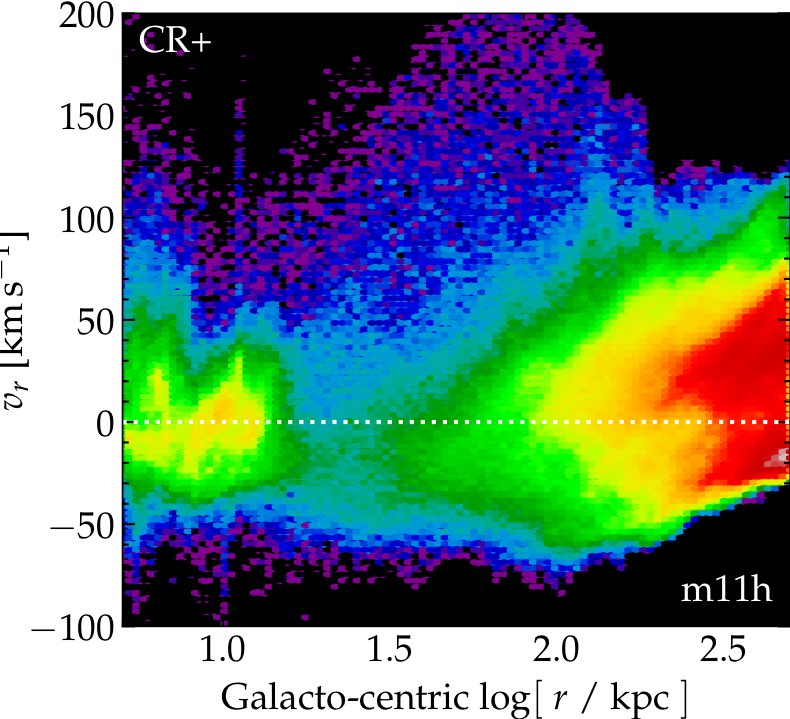}
\includegraphics[width=0.235\textwidth]{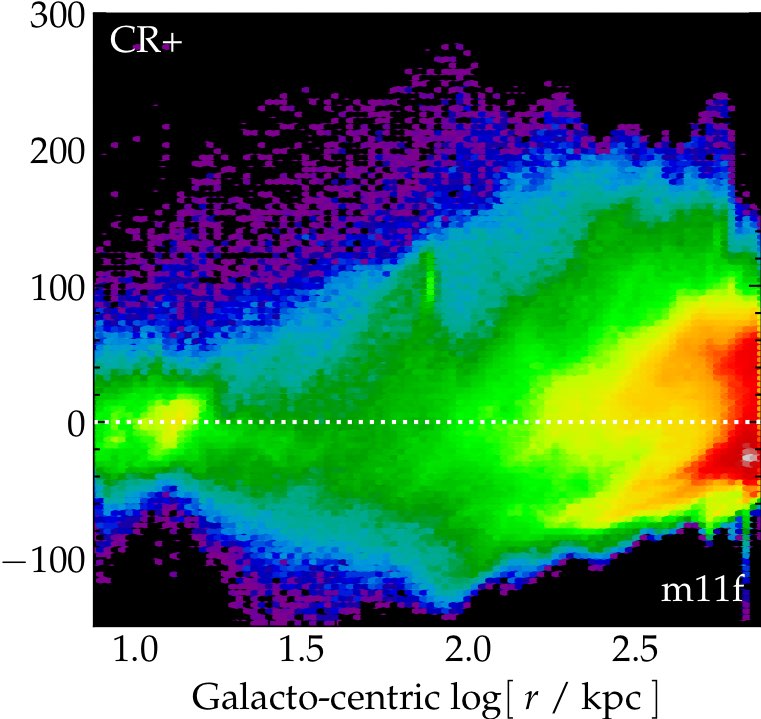}
\includegraphics[width=0.24\textwidth]{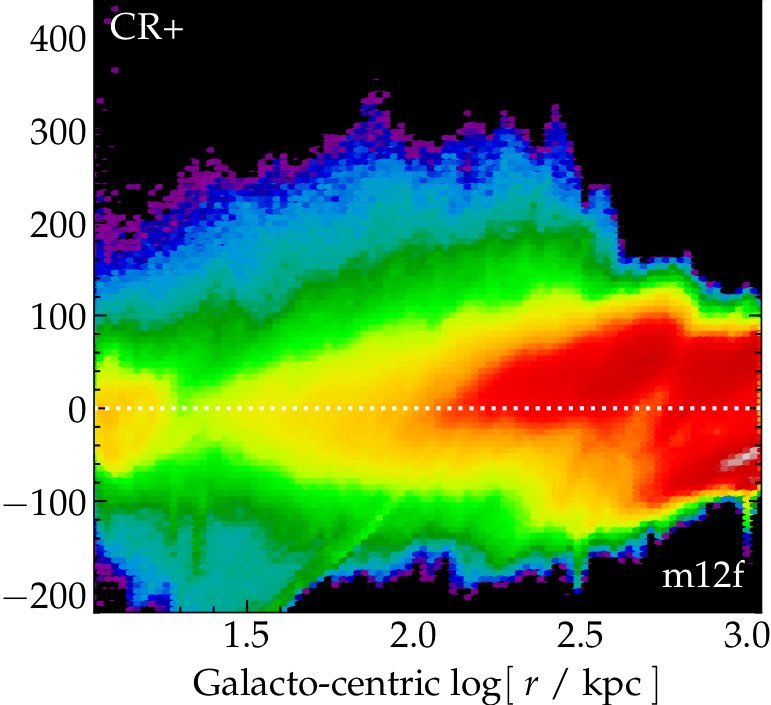}
\includegraphics[width=0.24\textwidth]{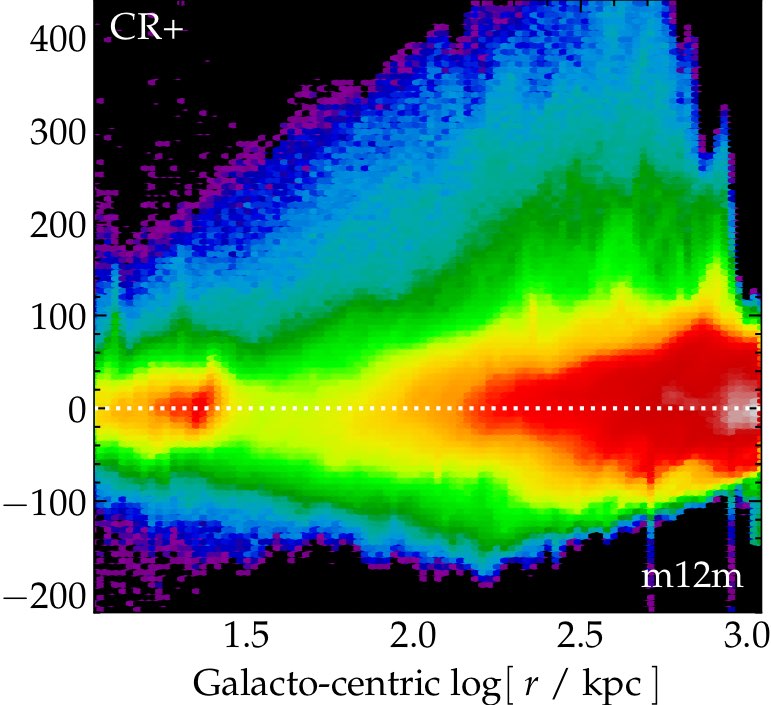}
\end{centering}
\caption{Gas outflow velocity versus radius, as \fref{fig:demo.vr.vs.r}, for a representative sub-sample of our simulated halos ordered by mass (increasing left-to-right, in the top and bottom ``groups''), in our ``No CRs'' ({\em top}) and ``CR+'' ({\em bottom}) runs. Below $M_{\rm halo} \lesssim 10^{11}\,M_{\sun}$ (top ``group''), the CRs have little obvious effect, consistent with their weak pressure relative to thermal (\fref{fig:profile.pressure}). At higher masses (bottom ``group'') the features in \fref{fig:demo.vr.vs.r} become progressively more prominent.
\label{fig:masssurvey.vr.vs.r}}
\end{figure*}

\begin{figure*}
\begin{centering}
\includegraphics[width=0.245\textwidth]{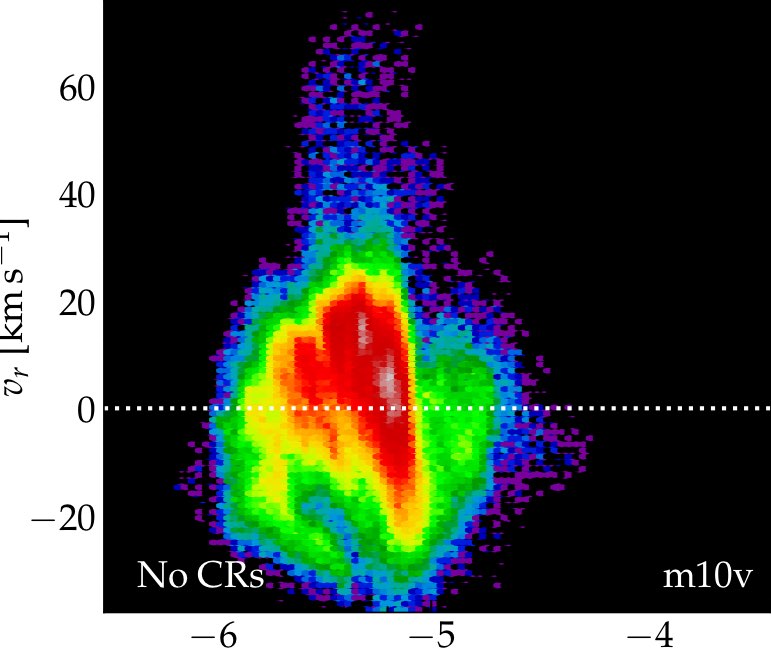}
\includegraphics[width=0.235\textwidth]{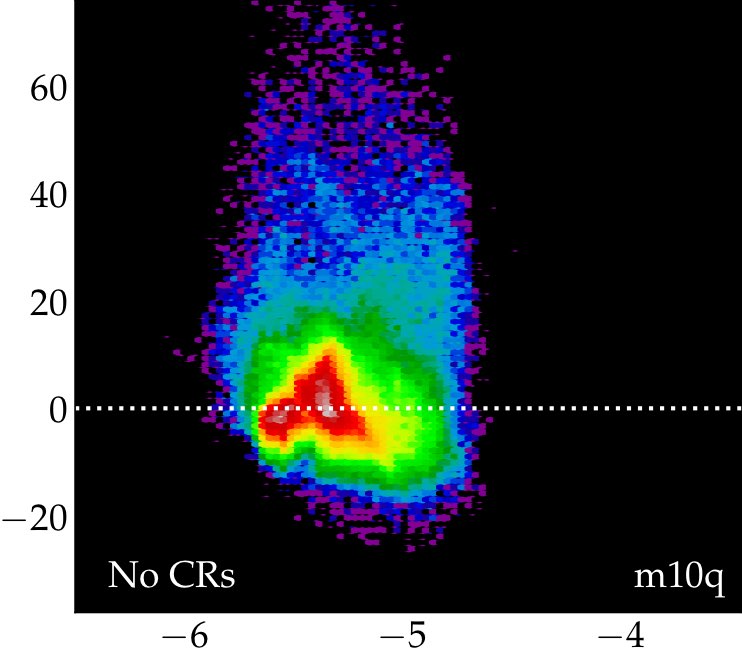}
\includegraphics[width=0.24\textwidth]{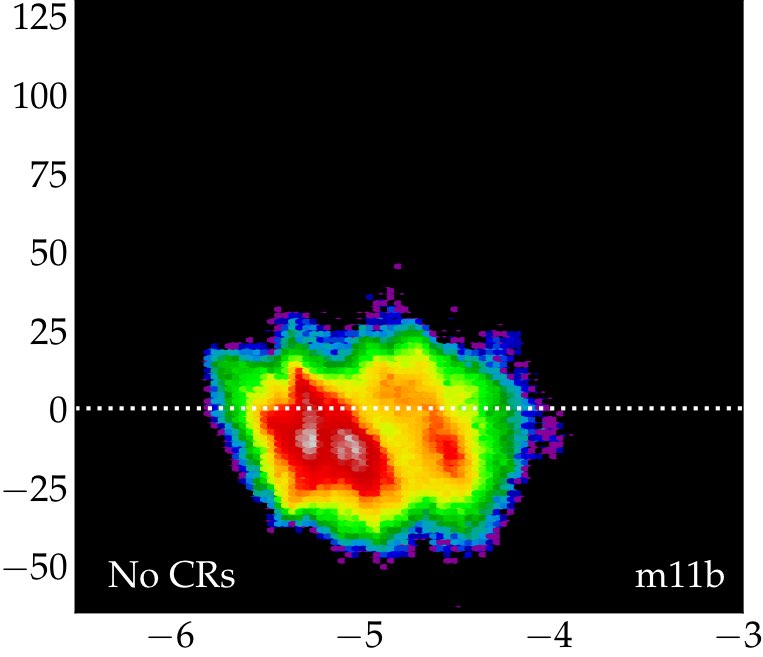}
\includegraphics[width=0.24\textwidth]{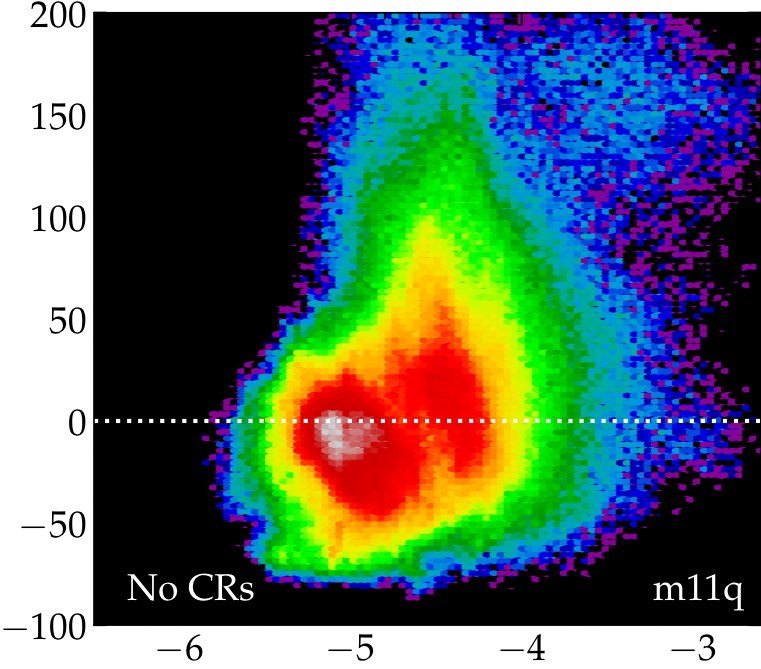}\\
\includegraphics[width=0.245\textwidth]{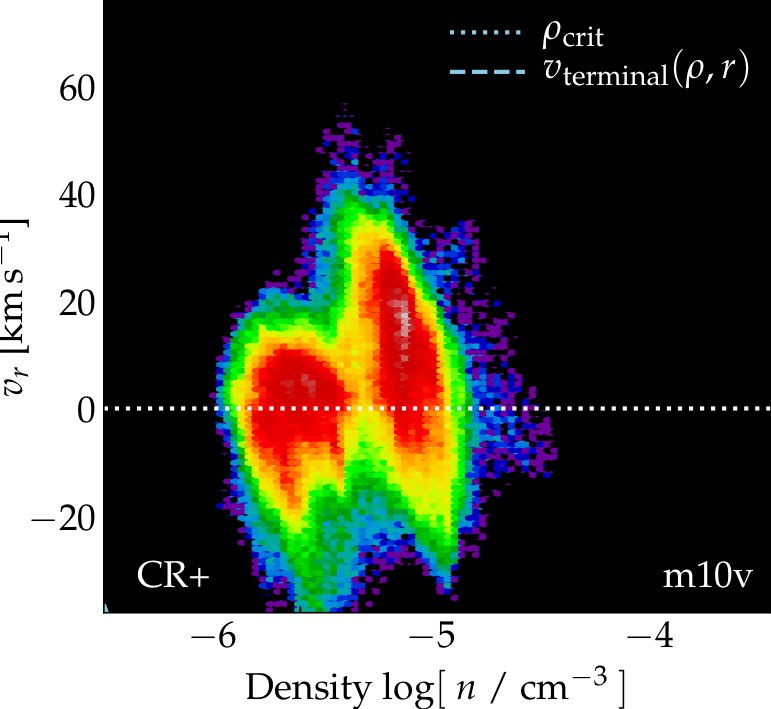}
\includegraphics[width=0.235\textwidth]{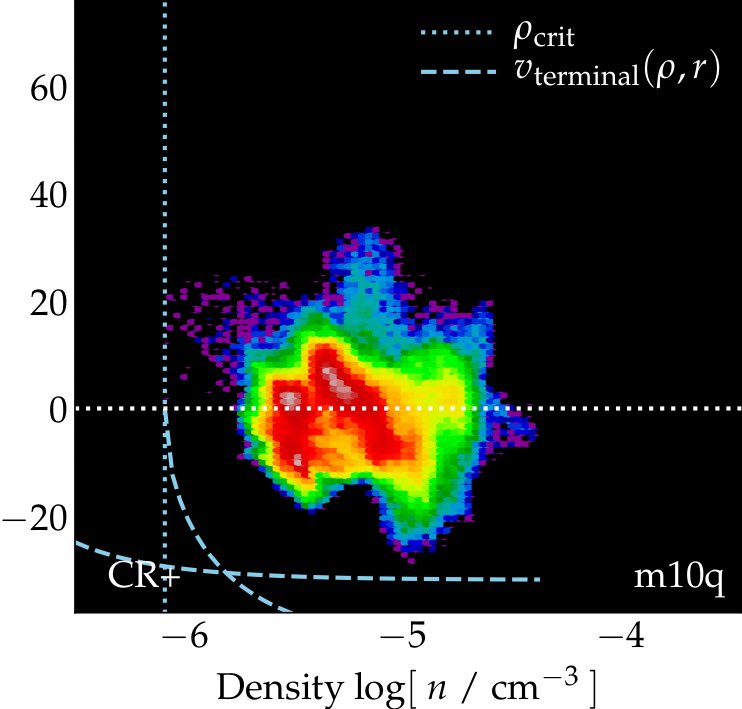}
\includegraphics[width=0.24\textwidth]{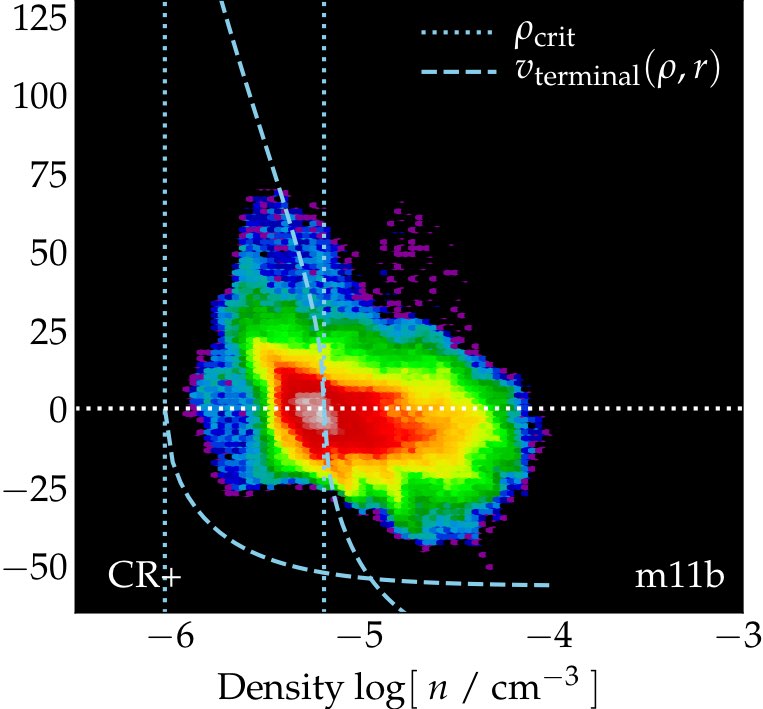}
\includegraphics[width=0.24\textwidth]{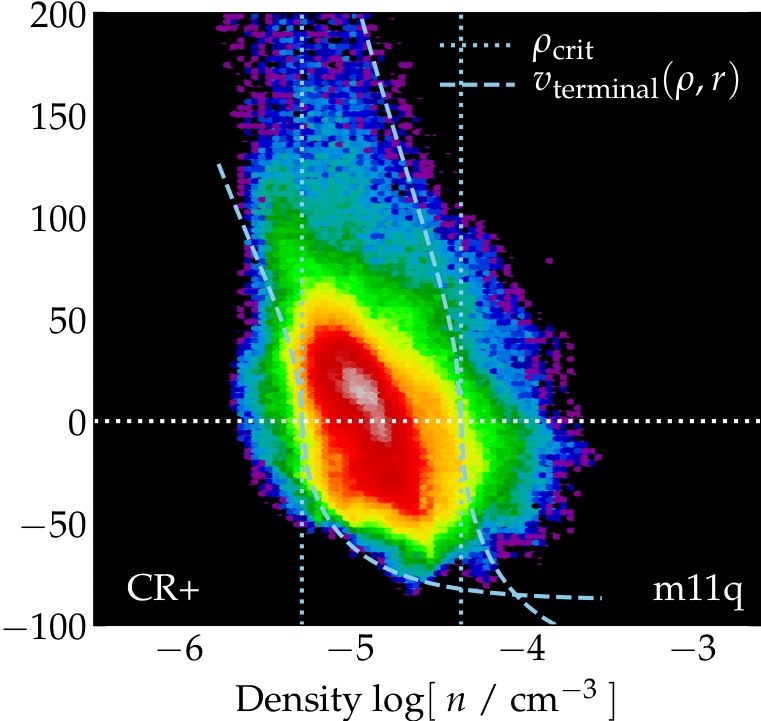}\\
\vspace{0.5cm}
\includegraphics[width=0.245\textwidth]{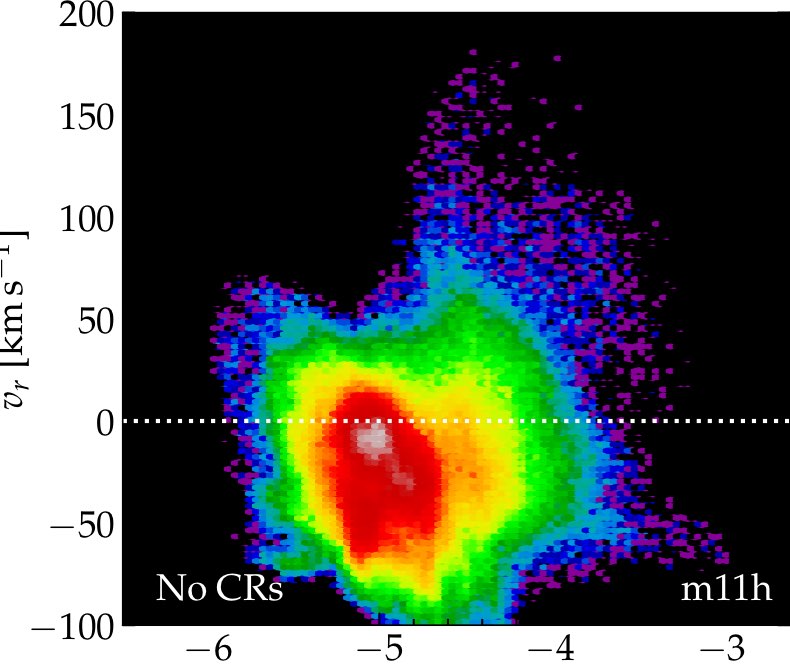}
\includegraphics[width=0.235\textwidth]{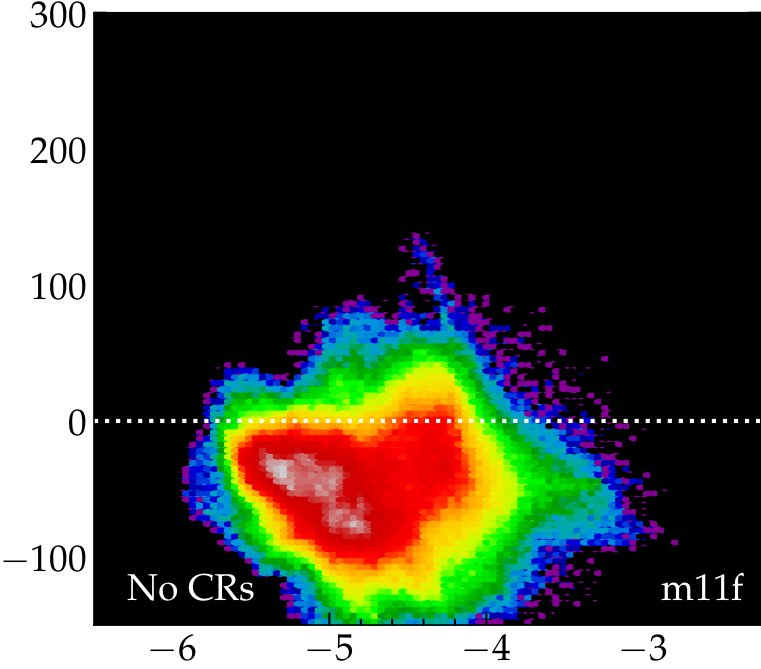}
\includegraphics[width=0.24\textwidth]{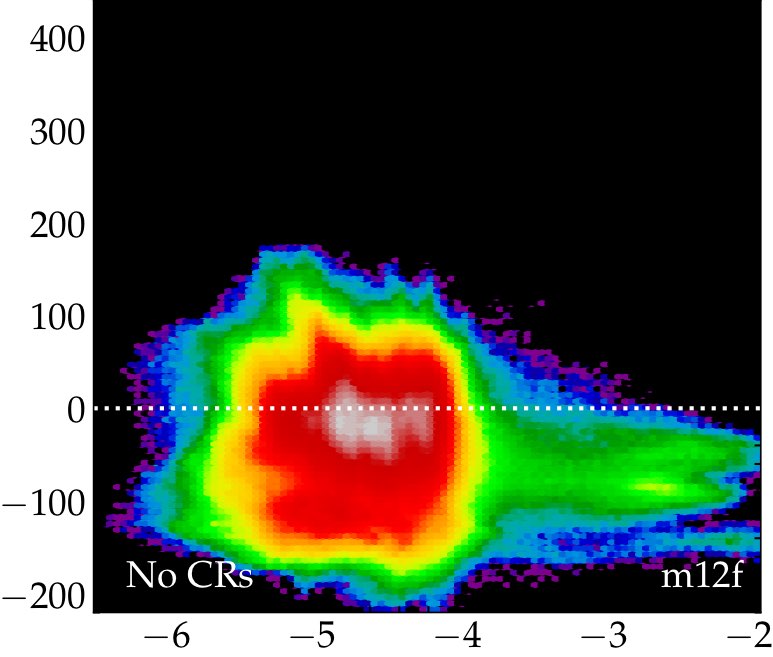}
\includegraphics[width=0.24\textwidth]{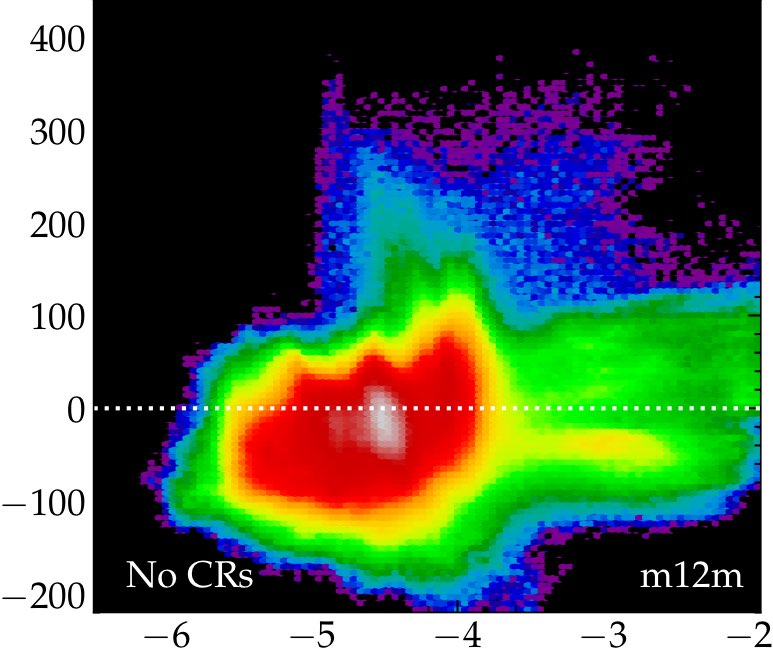}\\
\includegraphics[width=0.245\textwidth]{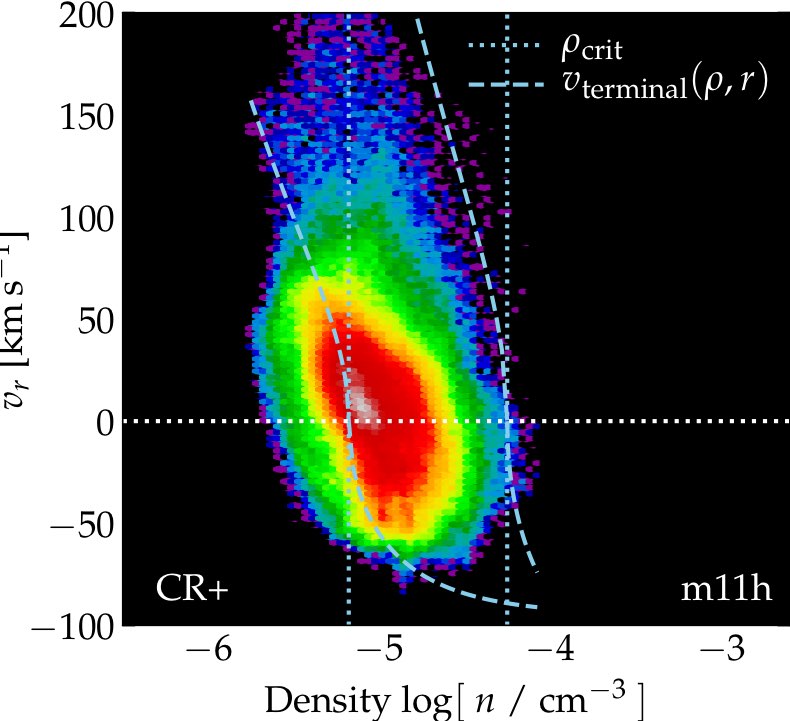}
\includegraphics[width=0.235\textwidth]{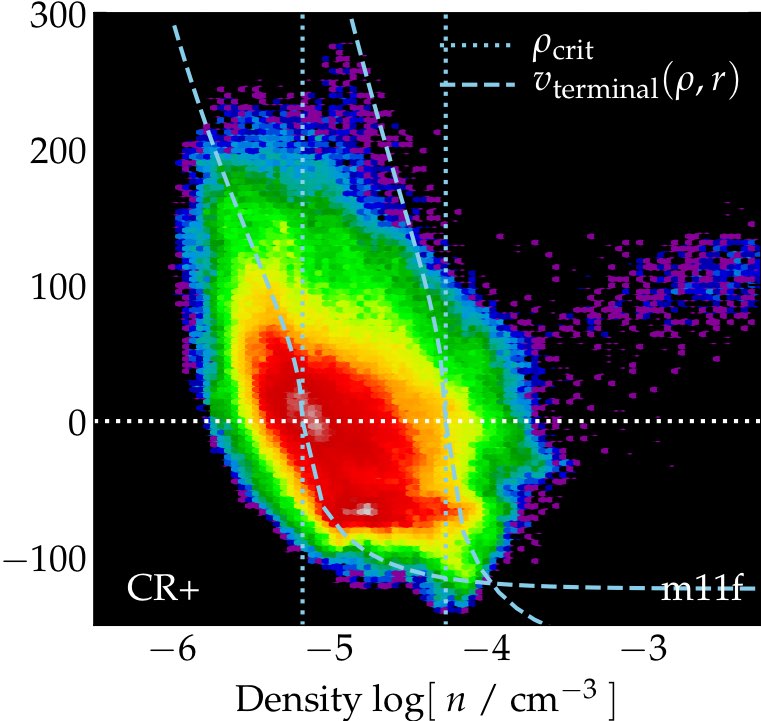}
\includegraphics[width=0.24\textwidth]{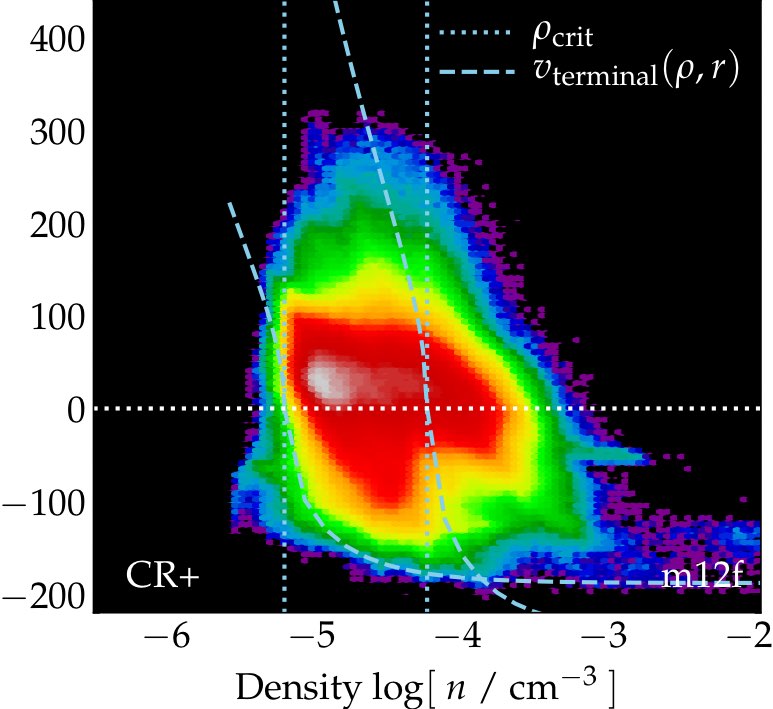}
\includegraphics[width=0.24\textwidth]{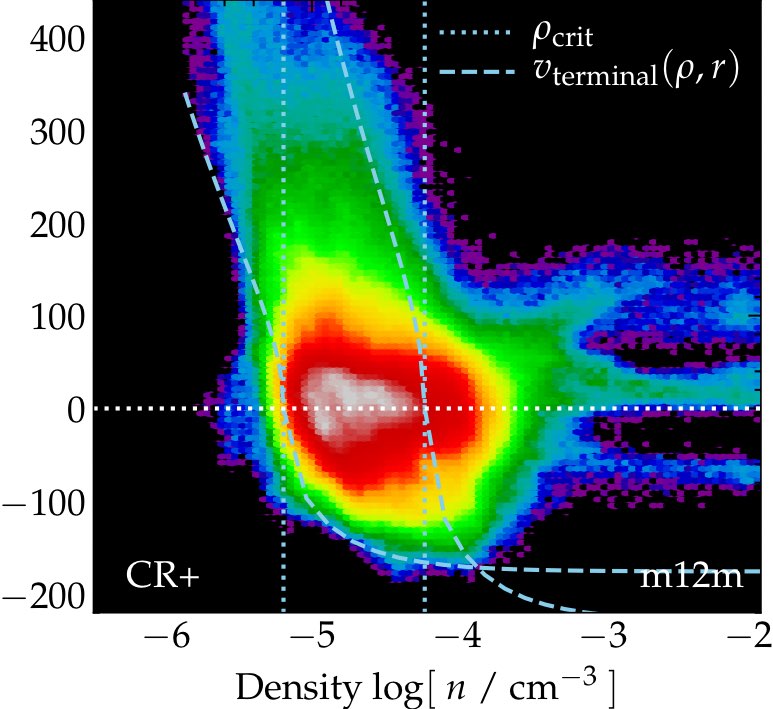}
\end{centering}
\caption{Gas outflow velocity versus density (as \fref{fig:demo.nh.t.wind}), for a sub-sample of our simulated halos ordered by mass (as \fref{fig:masssurvey.vr.vs.r}), in ``No CRs'' vs.\ ``CR+'' runs. For each we select gas with $0.5<r/R_{\rm vir}<1.5$, and compare the equilibrium $\rho_{\rm crit}$ where CR pressure balances gravity, and $v_{\rm terminal}$ for acceleration by CRs+gravity (see \S~\ref{sec:theory} \&\ \fref{fig:demo.nh.t.wind}). In low-mass ($\lesssim 10^{11}\,M_{\sun}$) halos, $\rho_{\rm crit} \ll \rho$ and $v_{\rm terminal} \ll v_{r}$, which simply reflects the fact that CRs make up a negligible contribution to the pressure (\fref{fig:profile.pressure}). Above this mass, the simple analytic scalings work remarkably well.
\label{fig:masssurvey.vr.vs.nh}}
\end{figure*}

\begin{figure*}
\begin{centering}
\includegraphics[width=0.245\textwidth]{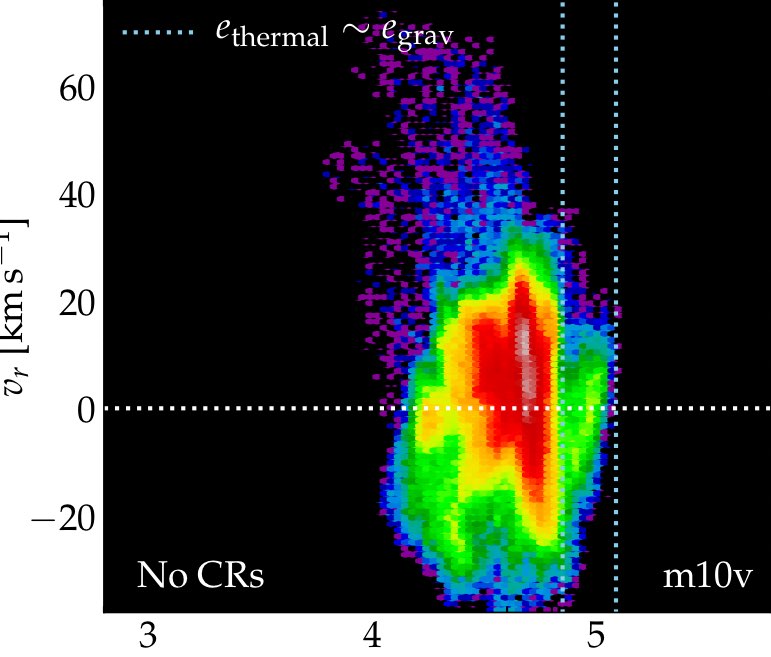}
\includegraphics[width=0.235\textwidth]{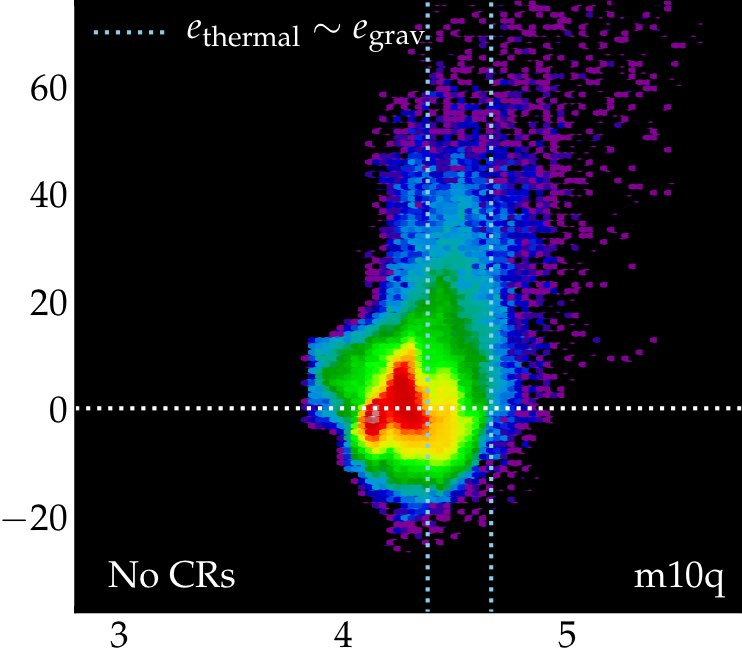}
\includegraphics[width=0.24\textwidth]{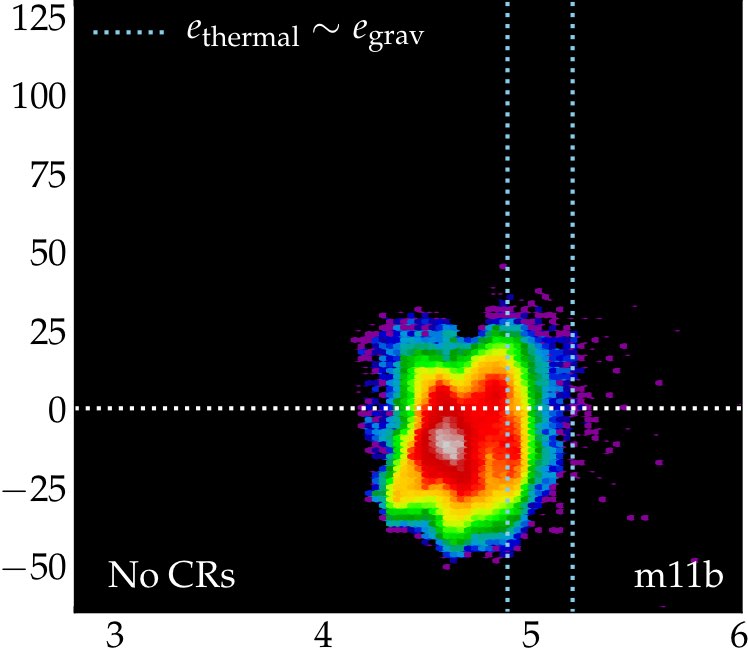}
\includegraphics[width=0.24\textwidth]{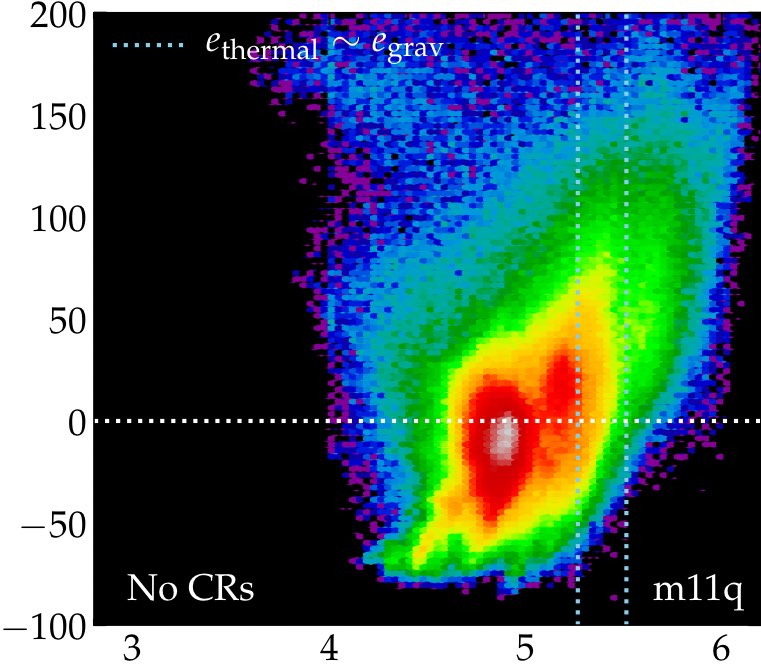}\\
\includegraphics[width=0.245\textwidth]{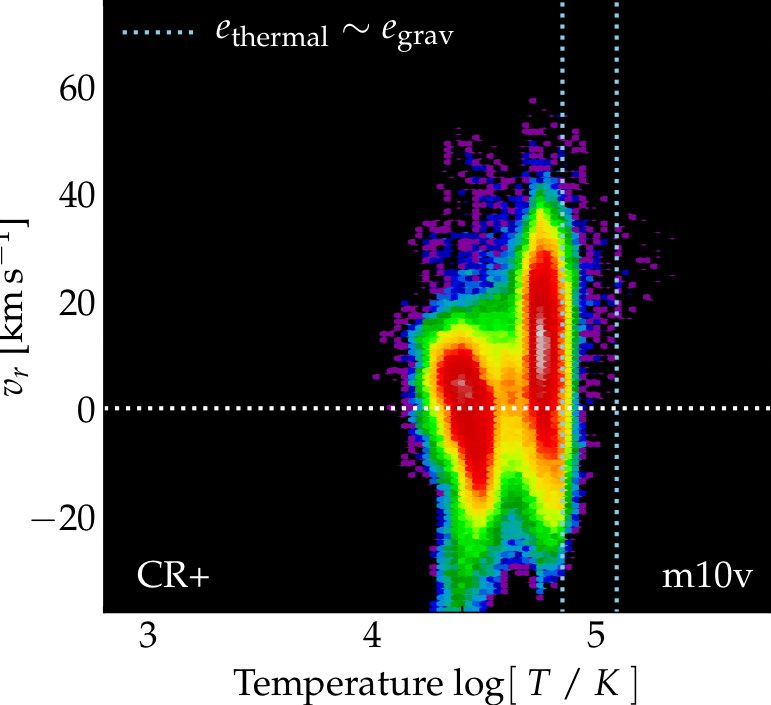}
\includegraphics[width=0.235\textwidth]{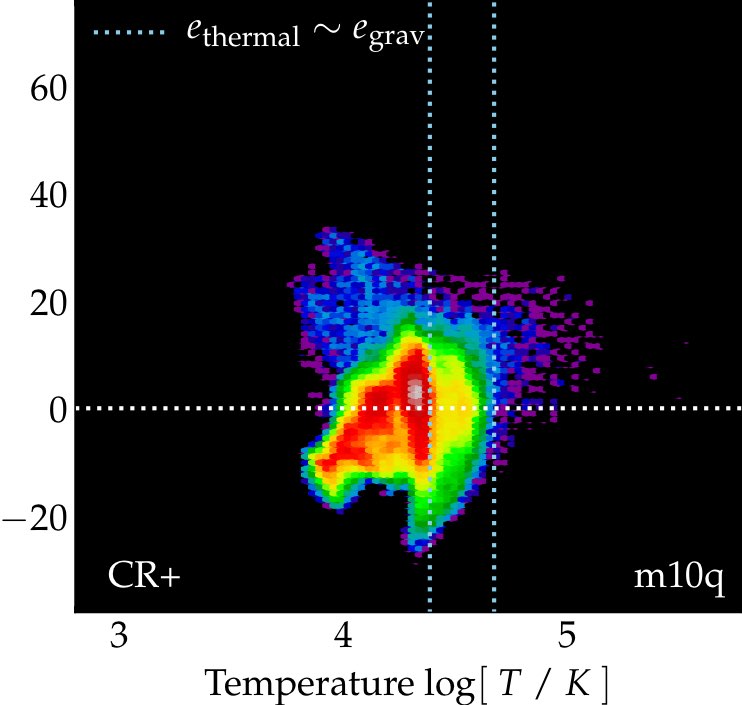}
\includegraphics[width=0.24\textwidth]{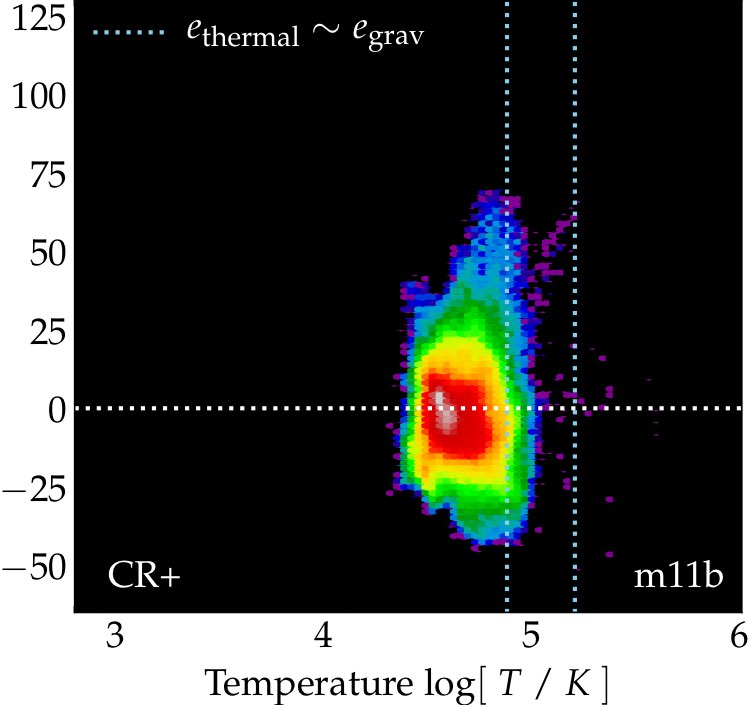}
\includegraphics[width=0.24\textwidth]{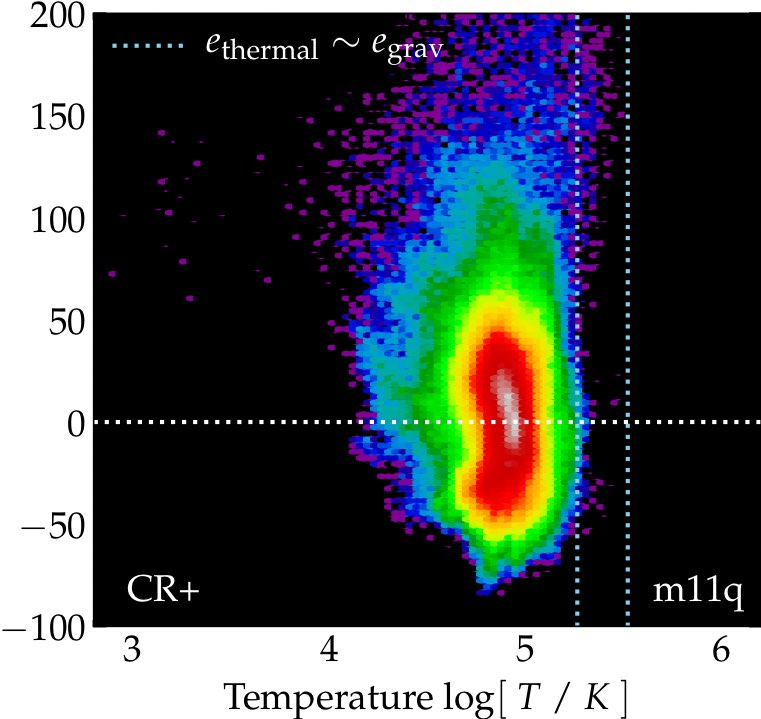}\\
\vspace{0.5cm}
\includegraphics[width=0.245\textwidth]{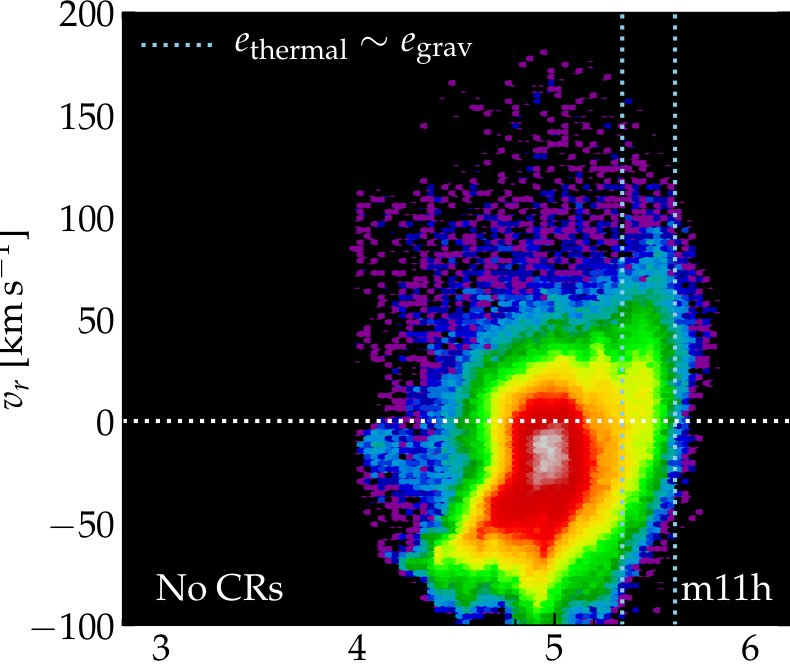}
\includegraphics[width=0.237\textwidth]{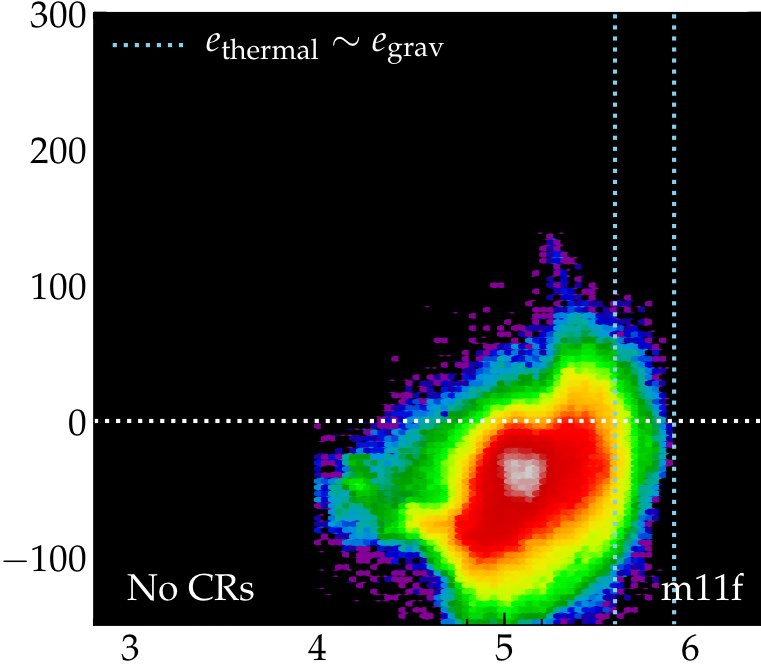}
\includegraphics[width=0.24\textwidth]{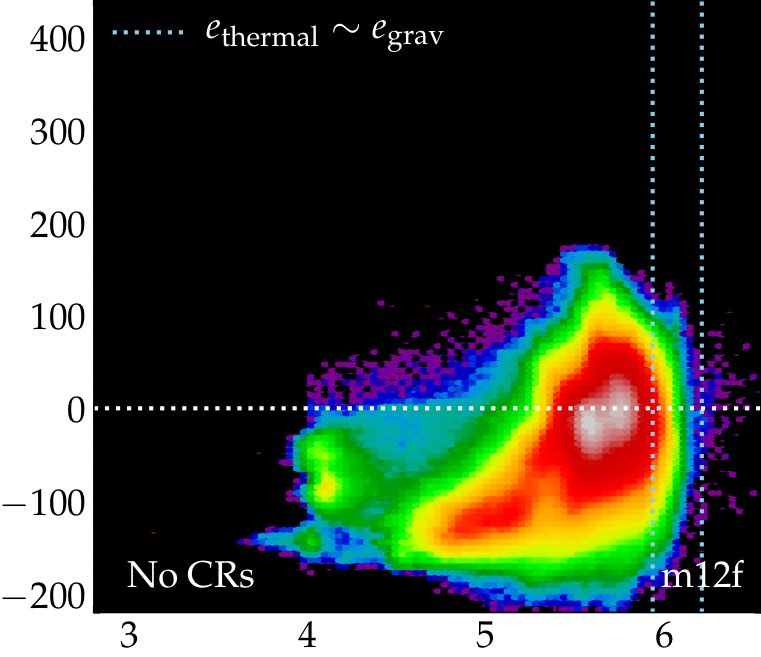}
\includegraphics[width=0.24\textwidth]{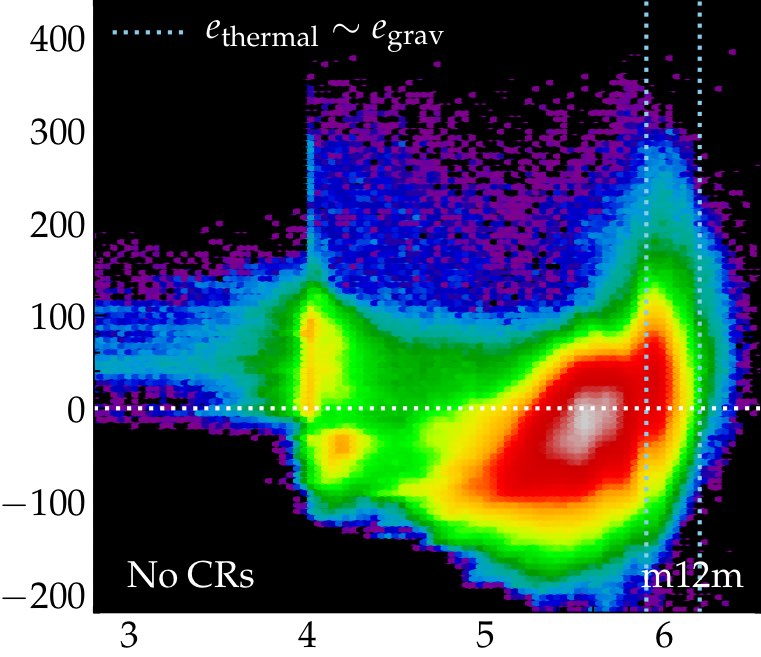}\\
\includegraphics[width=0.245\textwidth]{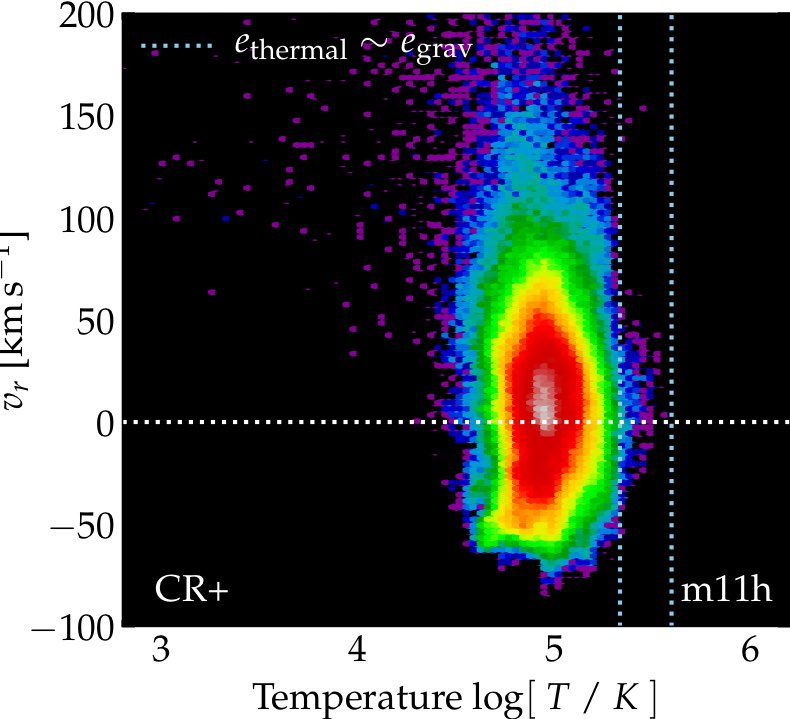}
\includegraphics[width=0.237\textwidth]{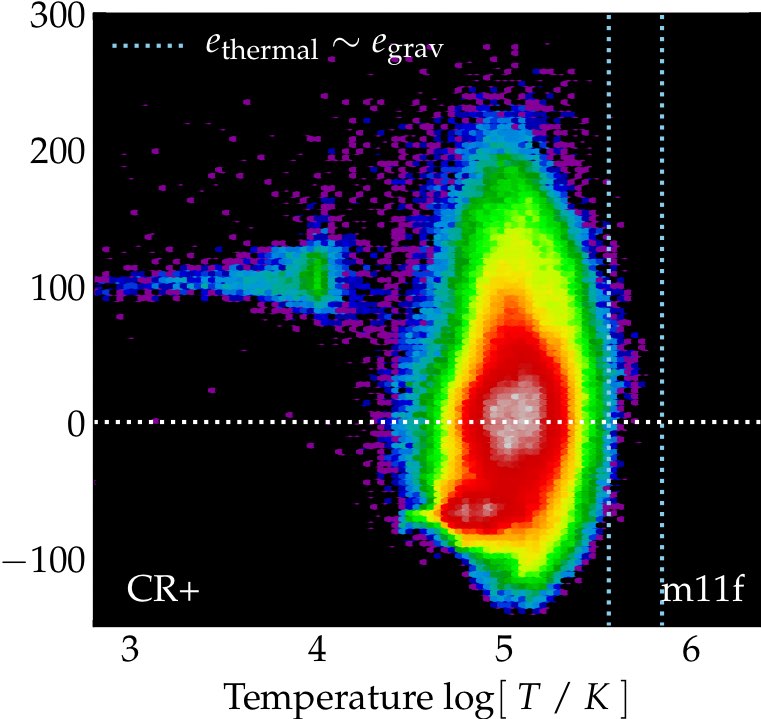}
\includegraphics[width=0.24\textwidth]{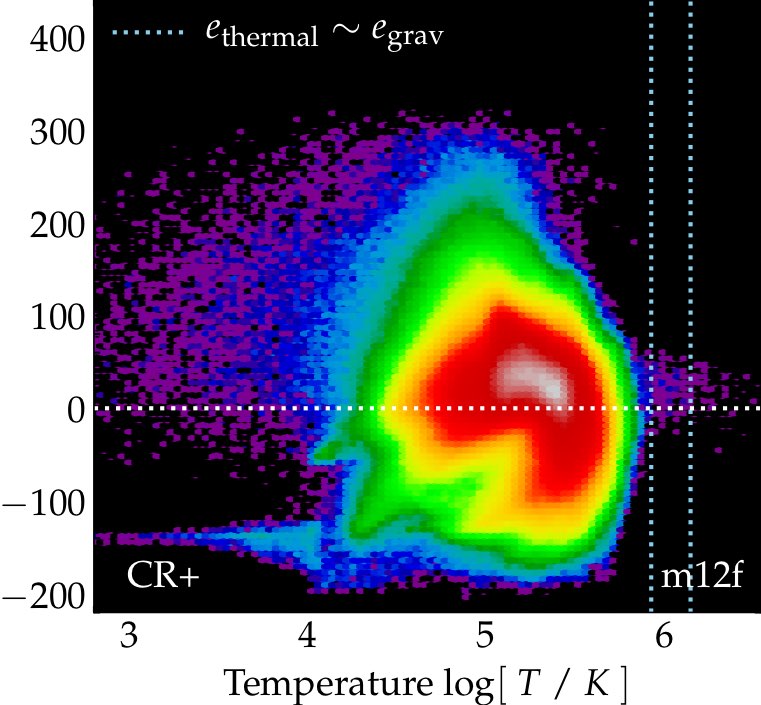}
\includegraphics[width=0.24\textwidth]{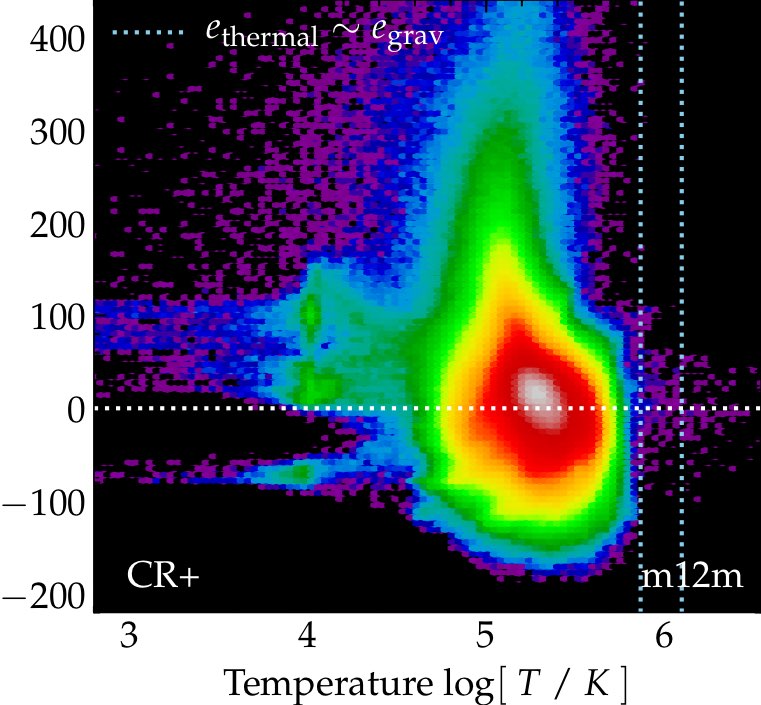}
\end{centering}
\caption{Gas outflow velocity versus temperature (as \fref{fig:demo.nh.t.wind}), for halos ordered by mass in ``No CRs'' vs.\ ``CR+'' runs (as \fref{fig:masssurvey.vr.vs.nh}, in the same $r$ ``slice''). There is little difference in low-mass halos but in high-mass halos, outflows in ``No CRs'' runs are preferentially ``hot'' gas ($e_{\rm thermal} \sim e_{\rm grav}$), while in ``CR+'' runs the outflows are primarily ``cool'' ($T\sim 10^{4-5}$\,K). 
\label{fig:masssurvey.vr.vs.t}}
\end{figure*}

\begin{figure*}
\begin{centering}
\includegraphics[width={0.24\textwidth}]{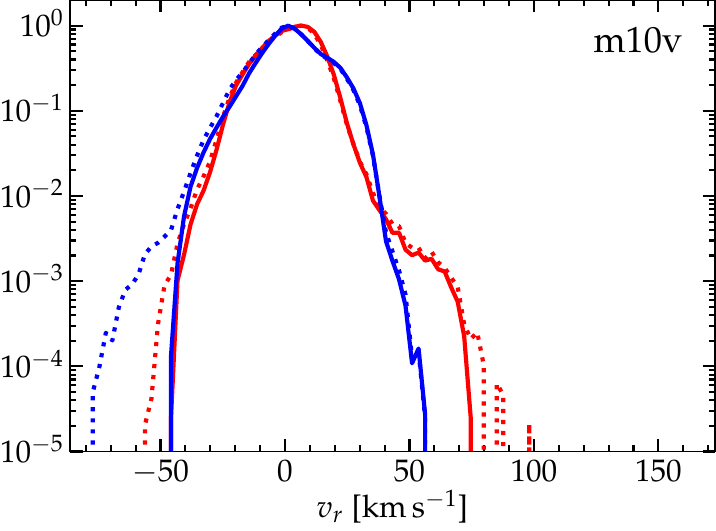}
\includegraphics[width={0.24\textwidth}]{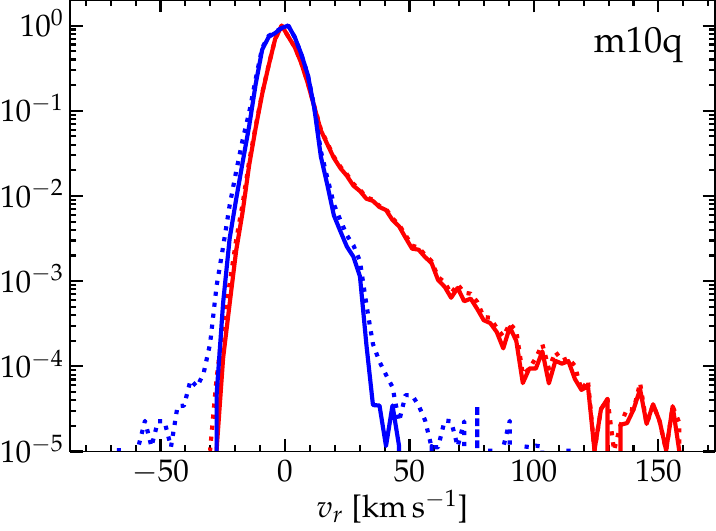}
\includegraphics[width={0.24\textwidth}]{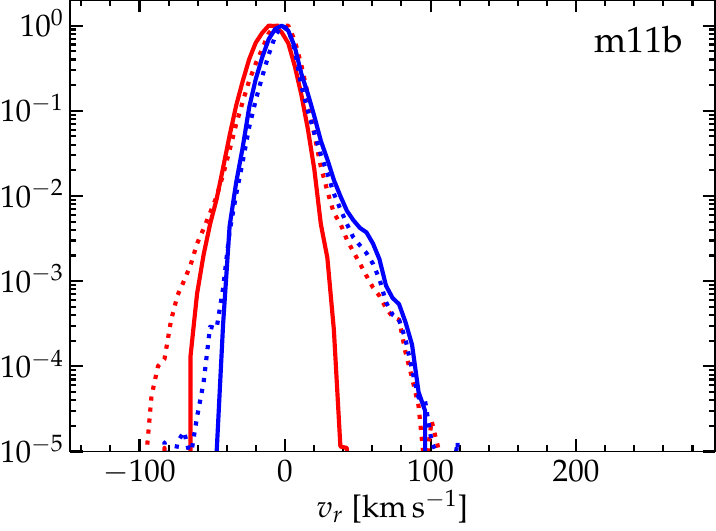}
\includegraphics[width={0.24\textwidth}]{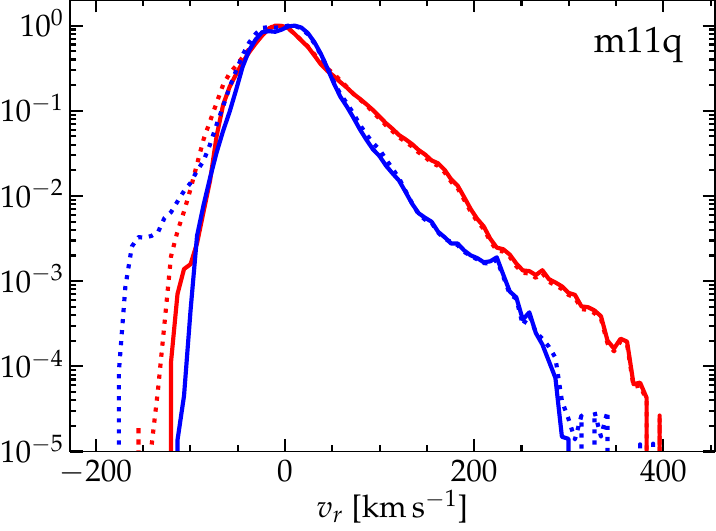} \\
\includegraphics[width={0.24\textwidth}]{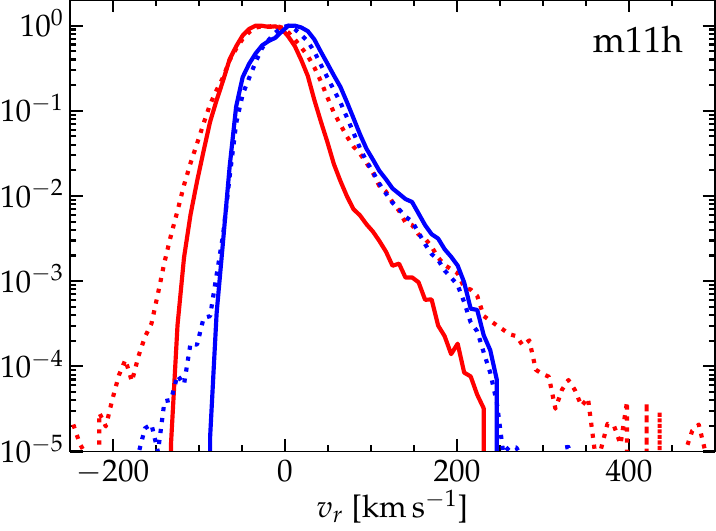}
\includegraphics[width={0.24\textwidth}]{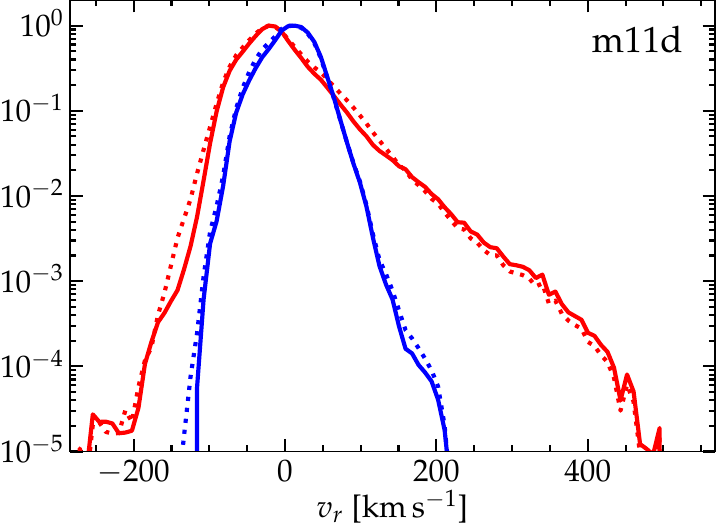}
\includegraphics[width={0.24\textwidth}]{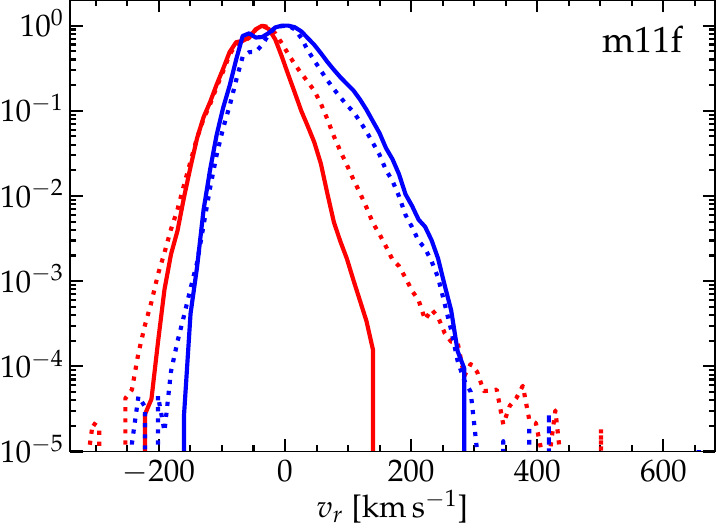}
\includegraphics[width={0.24\textwidth}]{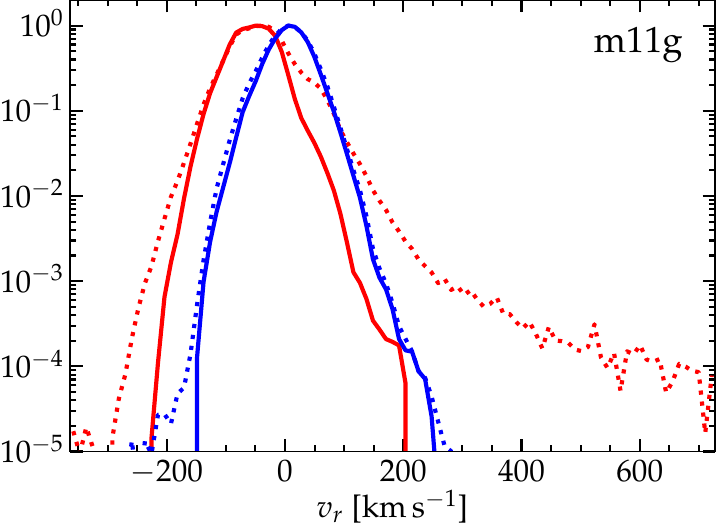} \\
\vspace{0.3mm}
\includegraphics[width={0.233\textwidth}]{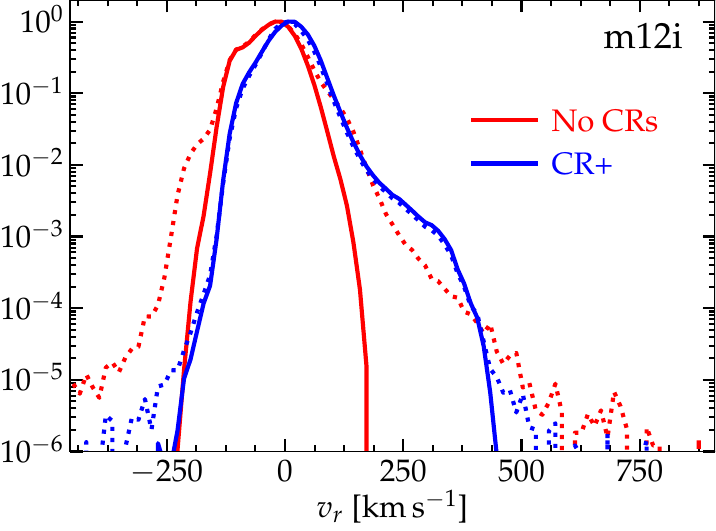}
\includegraphics[width={0.24\textwidth}]{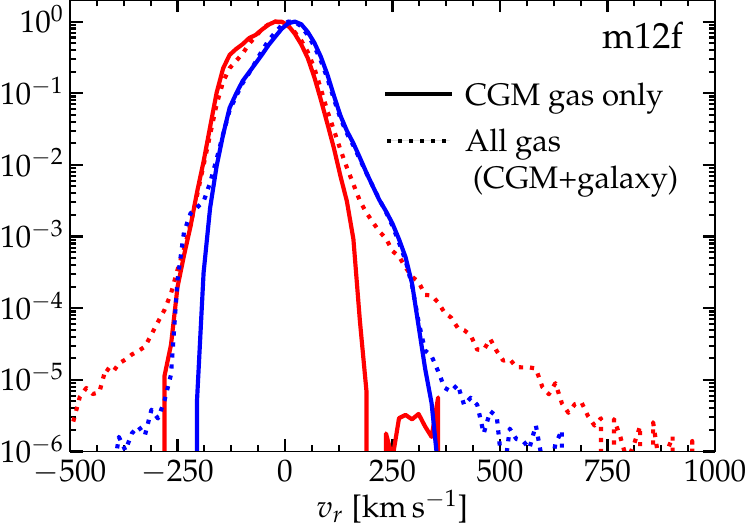}
\includegraphics[width={0.24\textwidth}]{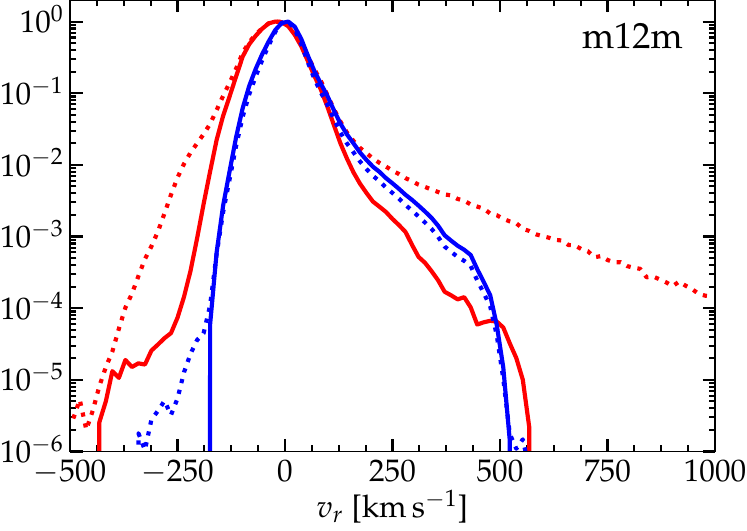} \\
    \end{centering}
    \vspace{-0.25cm}
        \caption{Radial velocity ($v_{r}$) distribution of gas in several simulations, ordered by mass. We compare the mass-weighted distribution of $dP_{M}/dv_{r}$ of gas at $z=0$ either (a) including all gas within $r<1.5\,R_{\rm vir}$, or (b) including just CGM gas at $0.2\,R_{\rm vir} < r < 1.5\,R_{\rm vir}$. In our low-mass dwarfs ({\em top}; {\bf m10v,q}, {\bf m11b,q}) there are differences but these are dominated by stochastic fluctuations in ``bursty'' star formation and outflow. In both ``No CRs'' and ``CR+'' runs at low masses, the tail of higher-velocity material is similar whether we consider ``all gas'' or ``CGM gas only,'' indicating that outflows are not strongly ``trapped.'' 
        In more massive galaxies, the ``No CRs'' runs commonly produce more high-velocity outflow when considering {\em all} gas, but this high-$v_{r}$ tail is strongly suppressed when we restrict to CGM gas, indicating outflow trapping/stalling in the inner halo ($R<0.2\,R_{\rm vir}$). As a result, the outflows {\em in the CGM} are stronger and reach higher velocities in the ``CR+'' runs.
    \label{fig:outflow.vdist}}
\end{figure*}

\begin{figure*}
\begin{centering}
\includegraphics[width={0.24\textwidth}]{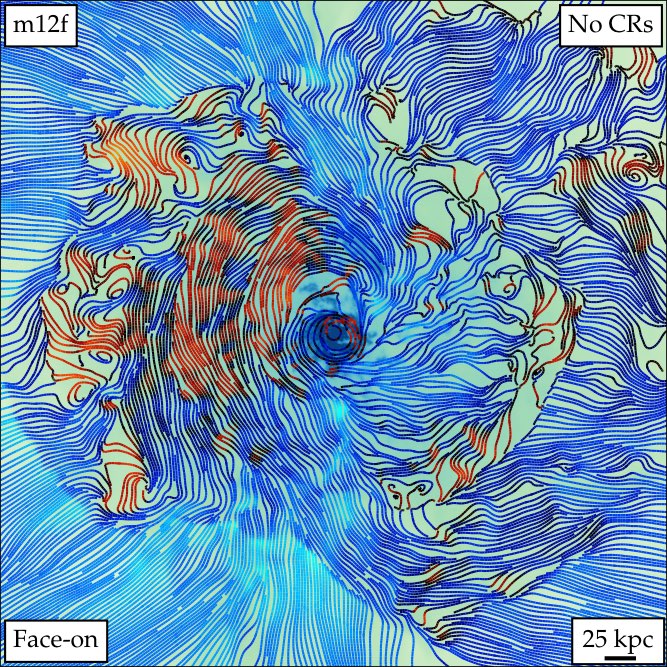}
\includegraphics[width={0.24\textwidth}]{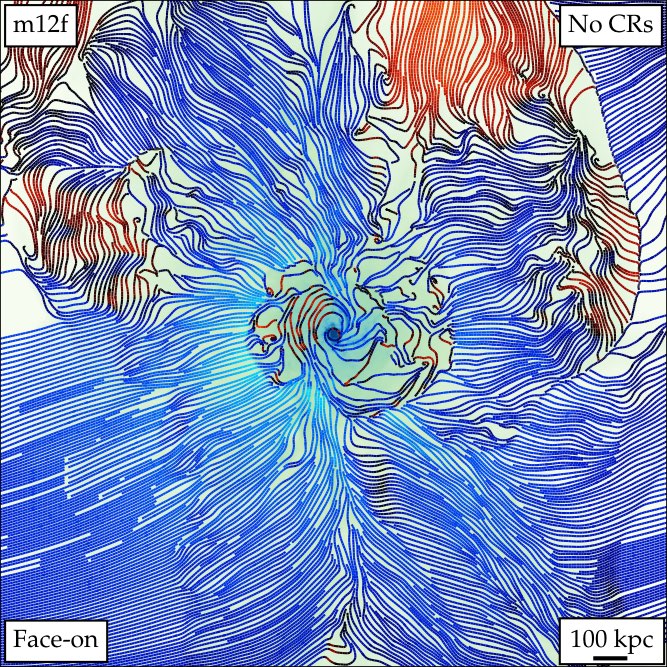}
\hspace{0.4cm}
\includegraphics[width={0.24\textwidth}]{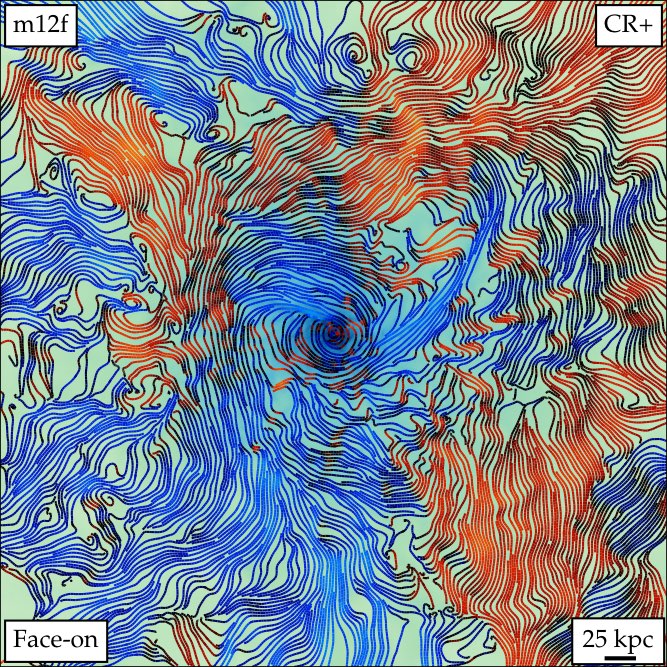}
\includegraphics[width={0.24\textwidth}]{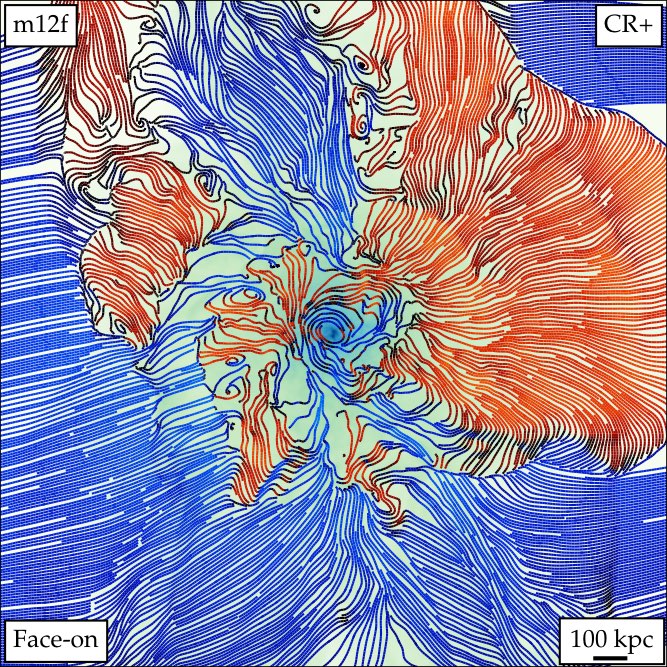} \\
\includegraphics[width={0.24\textwidth}]{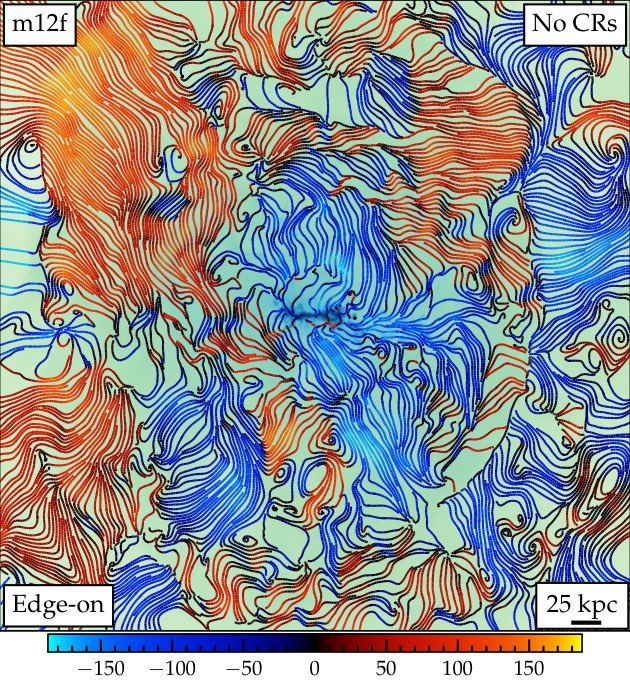}
\includegraphics[width={0.24\textwidth}]{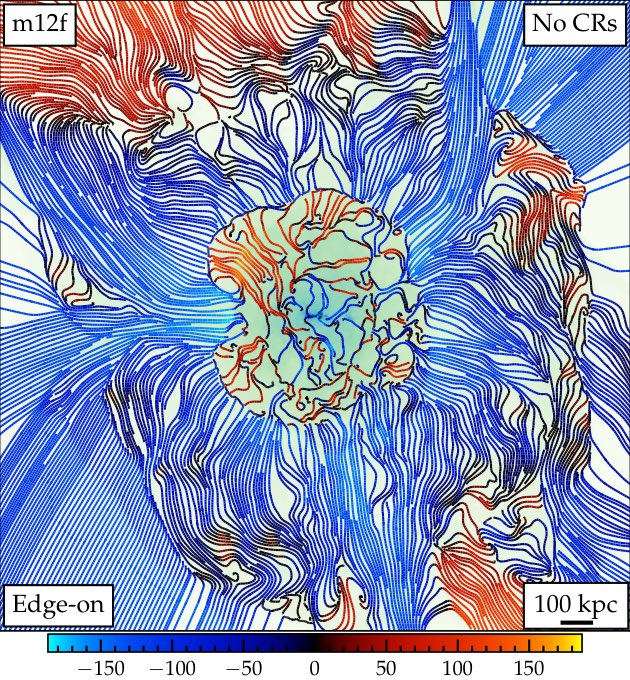}
\hspace{0.4cm}
\includegraphics[width={0.24\textwidth}]{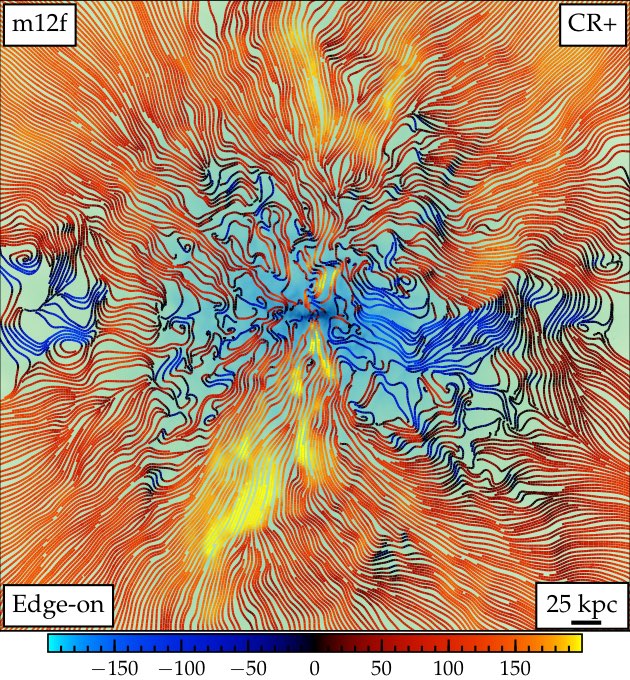} 
\includegraphics[width={0.24\textwidth}]{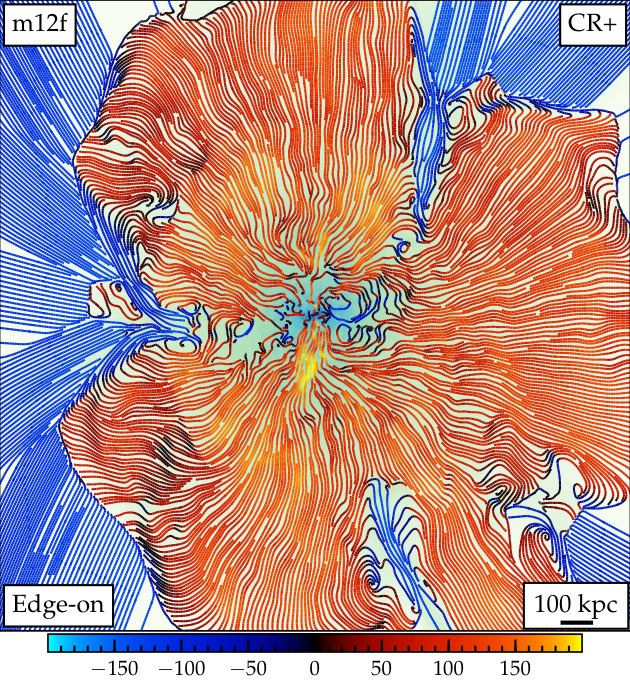} \\
\end{centering}
\caption{As Fig.~\ref{fig:inflow.outflow.m12i}, for halo {\bf m12f}, comparing the ``No CRs'' ({\em left} group) and ``CR+'' ({\em right} group) runs, projected face-on ({\em top}) and edge-on ({\em bottom}) to the disk, at two different spatial scales (within $\pm R_{\rm vir}$, {\em left}, or $\pm 4\,R_{\rm vir}$, {\em right}). Colorbar gives $v_{r}$ in ${\rm km\,s^{-1}}$. Like {\bf m12i}, the ``No CRs'' run exhibits inflows from $\gg R_{\rm vir}$ scales and a sharp virial shock with a turbulent halo; while the ``CR+'' run exhibits volume-filling outflows to $>$\,Mpc. These are broadly bipolar at $\sim R_{\rm vir}$ scales, but this becomes more volume-filling on larger scales with only narrow midplane channels continuing inflow.
\label{fig:inflow.outflow.m12f}}
\end{figure*}

\vspace{-0.5cm}
\section{Theoretical Expectations}
\label{sec:theory}

In \paperone, we developed a simple toy model for ``CR-dominated halos,'' and in \paperone\ and \citet{ji:fire.cr.cgm} we validated this as a surprisingly accurate description of the CR pressure and density profiles in the CGM  of our simulations. We therefore apply it here to outflows.

Although there has been significant study of CR-driven outflows ``within'' or ``just outside'' galaxies (i.e.\ within $\sim 1\,$kpc ``off'' the vertical surface of a thin disk, which we will study in detail in Chan et al., in prep.), we will argue below that many (not all) of the most dramatic differences owing to CRs occur on much larger scales in the CGM and IGM. Therefore on these scales, we can approximate the galaxy as small, so the injection of CRs is point-like, with quasi-steady rate $\dot{E}_{\rm cr} = \epsilon_{\rm cr}\,\dot{E}_{\rm SNe} = \epsilon_{\rm cr}\,u_{\rm SNe}\,\dot{M}_{\ast}$, averaged at a given radius over the CR diffusion time to that point ($\sim $\,Gyr). For the cases of interest, the CRs have some effective (isotropic-averaged) diffusivity $\tilde{\kappa}$ (which \citealt{ji:fire.cr.cgm} and \paperone\ argued should be $\sim \kappa_{\|}/3$), and escape the galaxy with negligible collisional losses (requiring $\tilde{\kappa} \gtrsim 10^{29}\,{\rm cm^{2}\,s^{-1}}$ in MW-like and dwarf galaxies; see \citealt{chan:2018.cosmicray.fire.gammaray} and \paperone). The CRs quickly form a spherically-symmetric radial pressure profile with $P_{\rm cr} \approx \dot{E}/(12\pi\,\tilde{\kappa}\,r)$ at $r < r_{\rm stream}$ and $P_{\rm cr} \approx \dot{E}/(12\pi\,v_{\rm stream}\,r^{2})$ at $r > r_{\rm stream}$, with $r_{\rm stream} \equiv \tilde{\kappa}/v_{\rm stream} \sim \tilde{\kappa}/v_{A}(r_{\rm stream})$. 

The case of particular interest is where this {\em dominates} over thermal pressure in the CGM. As discussed in \paperone, for this to be the case, it requires $P_{\rm cr} \gtrsim P_{\rm thermal,\,vir} \sim (3/16)\,200\,\bar{\rho}\,V_{\rm vir}^{2}$, or $\dot{E}_{\rm cr}/10^{41}\,{\rm erg\,s^{-1}} \gtrsim 1.7\,\tilde{\kappa}_{29}\,M_{\rm halo,\,12}\,(1+z)^{3}$. Assuming CRs from SNe with $\epsilon_{\rm cr}=0.1$ and our adopted IMF, with time-averaged SFRs $\dot{M}_{\ast} \equiv \alpha\,M_{\ast}(z)/t_{\rm Hubble}(z)$ ($\alpha\sim1$), this is equivalent to $P_{\rm cr}/P_{\rm vir} \approx 3\,(1+z)^{-3/2}\,(M_{\ast}/f_{\rm baryon}\,M_{\rm halo})\,\alpha\,\tilde{\kappa}_{29}^{-1}$, so we expect (as we showed in \paperone) the halos are only CR-dominated at redshifts $z\lesssim  1-2$, in the mass range $M_{\rm halo} \gtrsim 10^{11-13}\,M_{\sun}$ where $M_{\ast}/f_{\rm baryon}\,M_{\rm halo}$ is relatively large. 

In this regime, following \citet{ji:fire.cr.cgm}, if we take $P = P_{\rm total} \approx P_{\rm cr}$ (since, by definition, CRs dominate the pressure), we can compare the outward pressure gradient force to the gravitational force $\rho\,\partial \Phi/\partial r$, and we immediately see there is a critical density $\rho_{\rm crit}$ where the two are equal, $\rho_{\rm crit} \approx \dot{E}_{\rm cr} / (12\pi\,V_{c}^{2}\,\tilde{\kappa}\,r)$ at $r \ll r_{\rm stream}$ and $\rho_{\rm crit} \approx \dot{E}_{\rm cr} / (6\pi\,V_{c}^{2}\,v_{\rm stream}\,r^{2})$ at $r \gg r_{\rm stream}$. We can approximate $V_{c}$ with a \citet{hernquist:profile} profile ($\Phi \sim -G\,M_{\rm halo}/(r + 2\,r_{s})$) or NFW profile  where $r_{s}$ is the usual NFW scale radius $= R_{\rm vir}/c$ with $c\approx 10$ at $z=0$.\footnote{We obtain very similar results assuming a \citet{hernquist:profile} or NFW profile (at all $r \lesssim 100\,r_{s}$), but some of the expressions below must be evaluated numerically for NFW (or have weakly-varying logarithmic corrections) so we default to the expressions for a \citet{hernquist:profile} for simplicity.} If we define $V_{0} = \sqrt{G\,M_{\rm halo}/r_{s}}$, this gives  $\rho_{\rm crit} \sim 0.07\,\dot{E}_{\rm cr}\,r_{s} / (V_{0}^{2}\,\tilde{\kappa}\,r^{2})$ at small $r$ and  $\rho_{\rm crit} \sim 0.04\, \dot{E}_{\rm cr} / (V_{0}^{2}\,v_{\rm stream}\,r_{s}\,r)$ at large $r$. Regardless of $V_{c}$, with no other forces, at any $r$, gas with $\rho < \rho_{\rm crit}$ will rise (move outwards) while gas with $\rho > \rho_{\rm crit}$ will sink (infall).

\begin{figure}
\begin{centering}
\includegraphics[width={0.227\textwidth}]{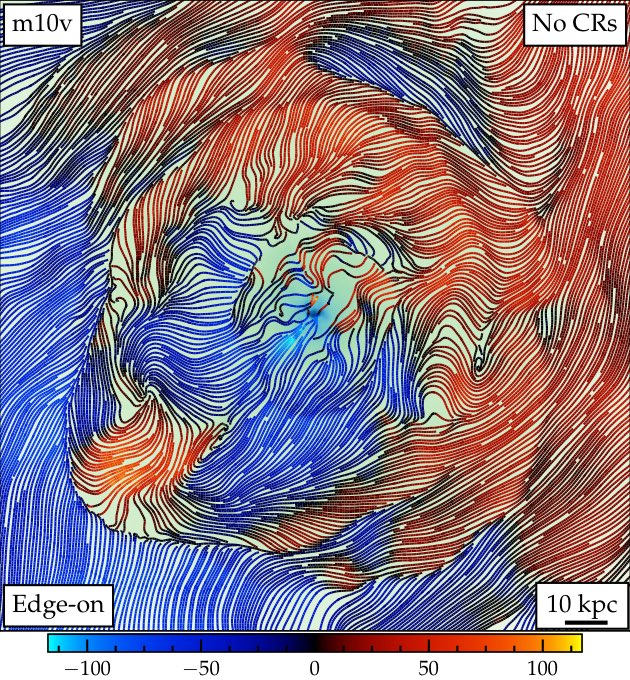}
\includegraphics[width={0.227\textwidth}]{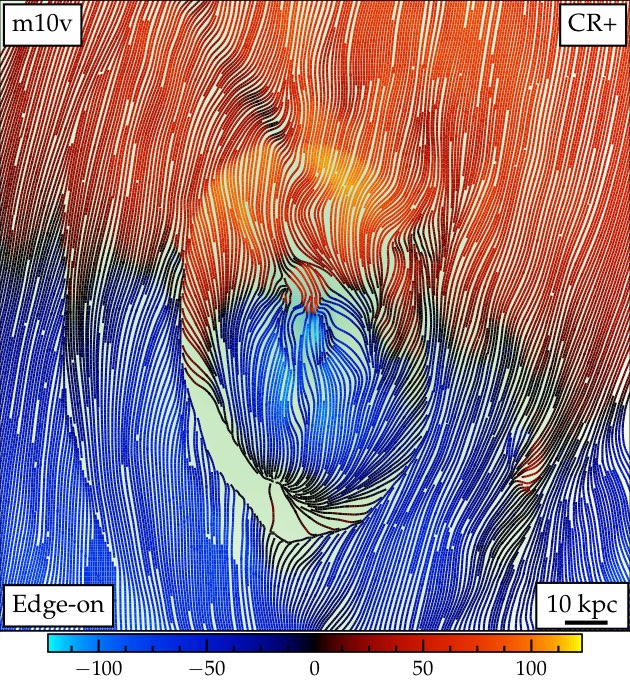} \\
\includegraphics[width={0.227\textwidth}]{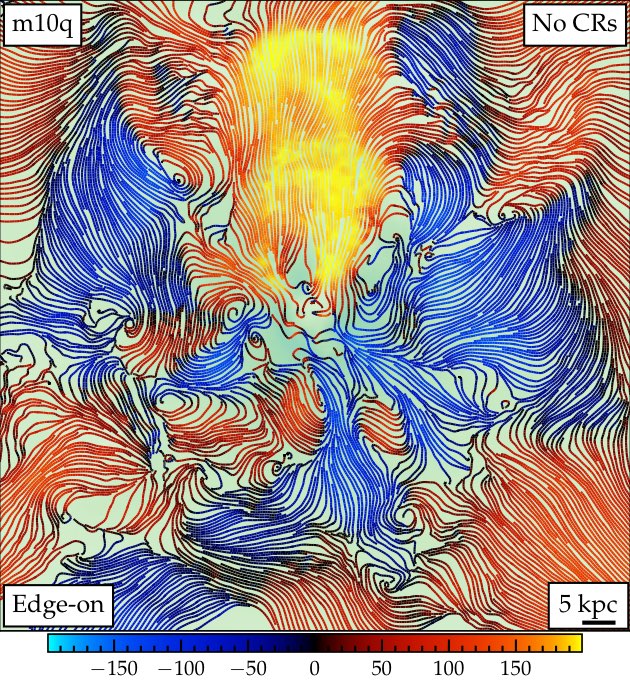}
\includegraphics[width={0.227\textwidth}]{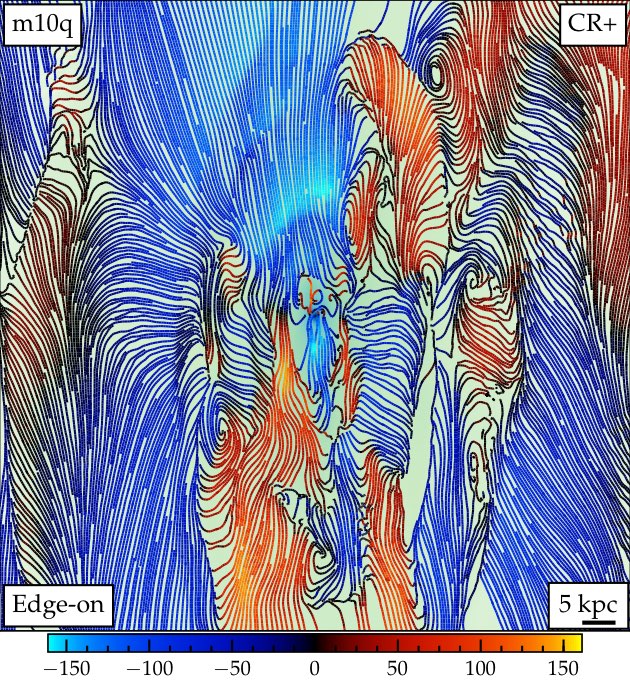} \\
\includegraphics[width={0.227\textwidth}]{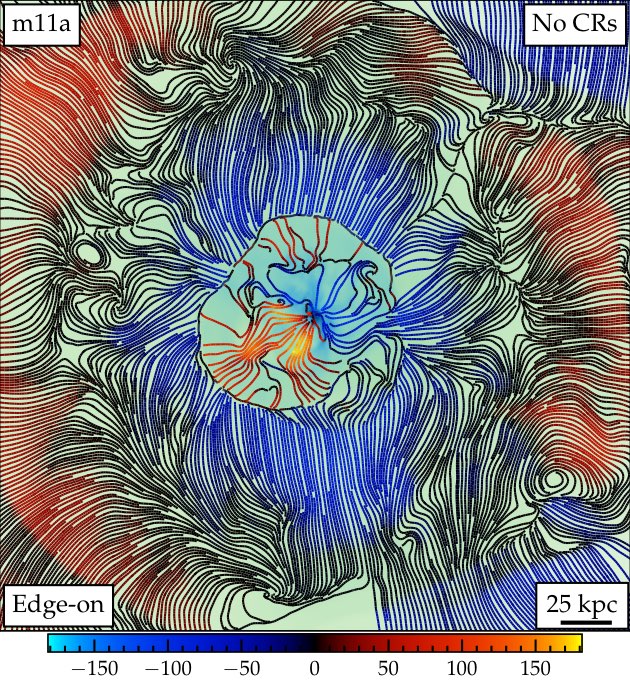}
\includegraphics[width={0.227\textwidth}]{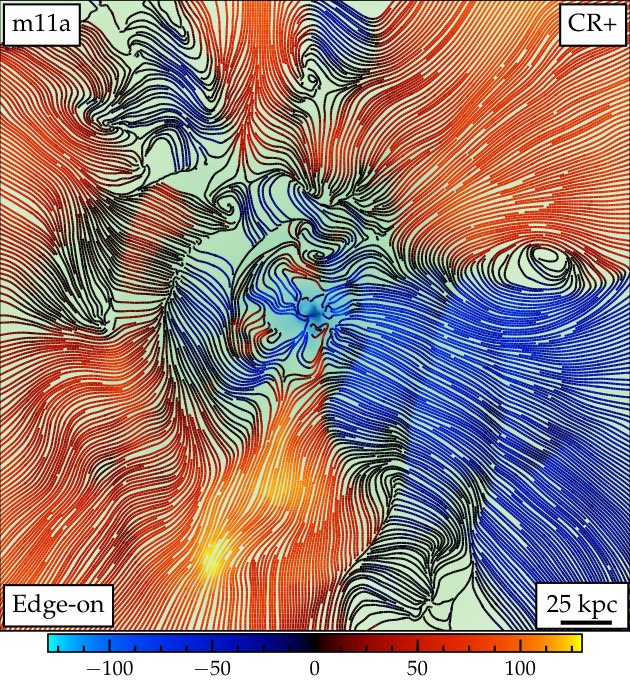}  \\
\includegraphics[width={0.227\textwidth}]{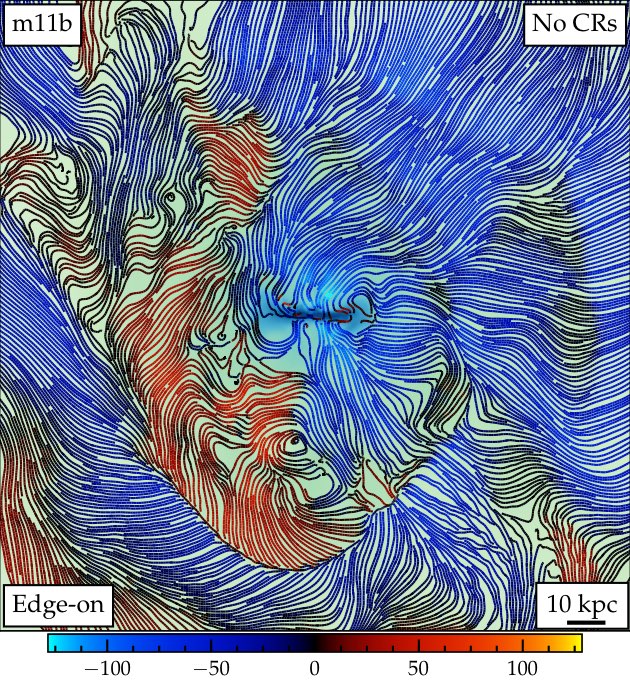} 
\includegraphics[width={0.227\textwidth}]{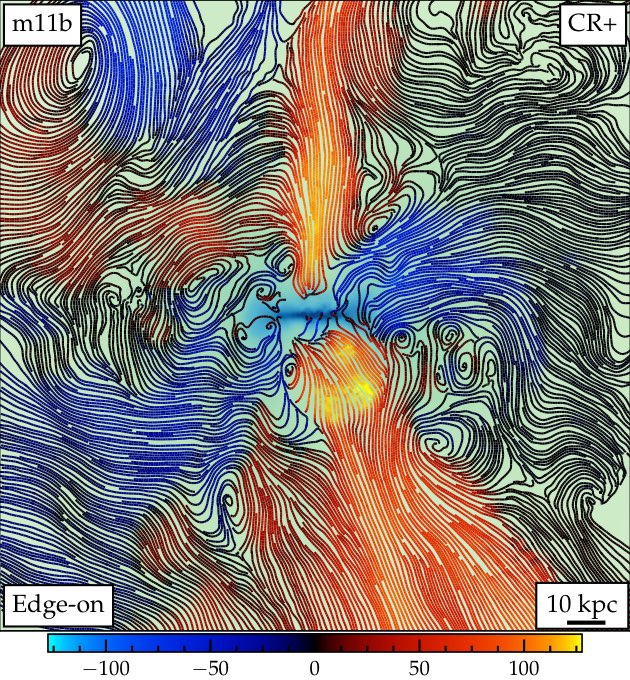} \\
\includegraphics[width={0.227\textwidth}]{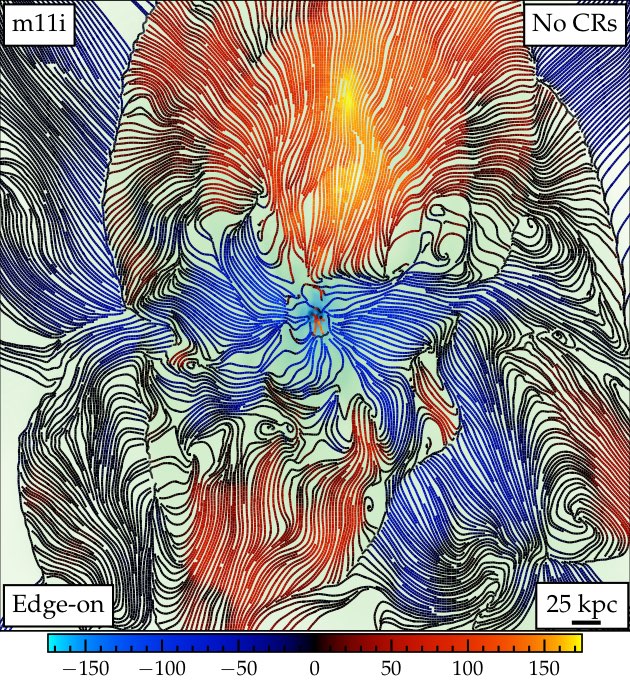}
\includegraphics[width={0.227\textwidth}]{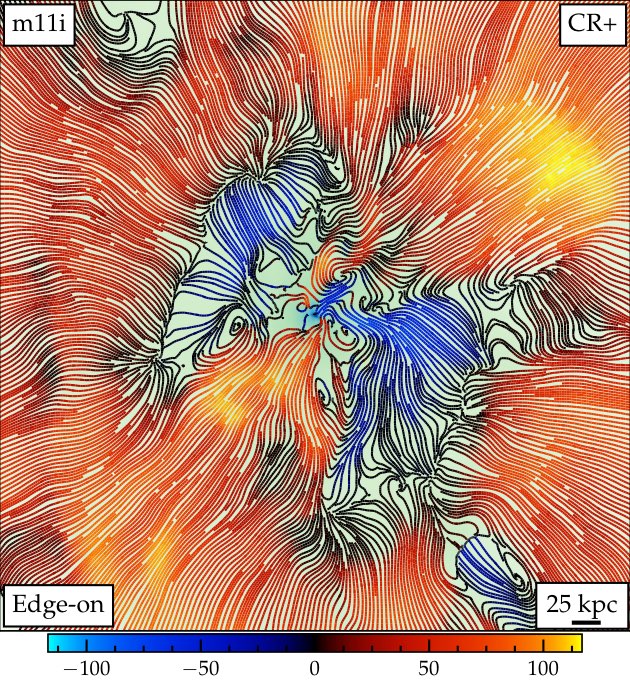} \\
\end{centering}
\vspace{-0.25cm}
\caption{As Fig.~\ref{fig:inflow.outflow.m12i}, showing just edge-on projections, for ``No CRs'' ({\em left}) and ``CR+'' ({\em right}) runs of {\bf m10v}, {\bf m10q}, {\bf m11a}, {\bf m11b}, {\bf m11i} (top-to-bottom, increasing mass). Spatial scale and $v_{r}$ (in ${\rm km\,s^{-1}}$). At the lowest masses, CRs have a weak effect (outflows can be stronger or more polar in ``No CRs'' runs, depending on the recent SF history), while at higher masses, a clear shock appears in ``No CRs'' runs where outflow meets accretion while the ``CR+'' runs begin to develop large-scale bipolar outflow. 
\label{fig:inflow.outflow.dwarfs}}
\end{figure}

\begin{figure}
\begin{centering}
\includegraphics[width={0.227\textwidth}]{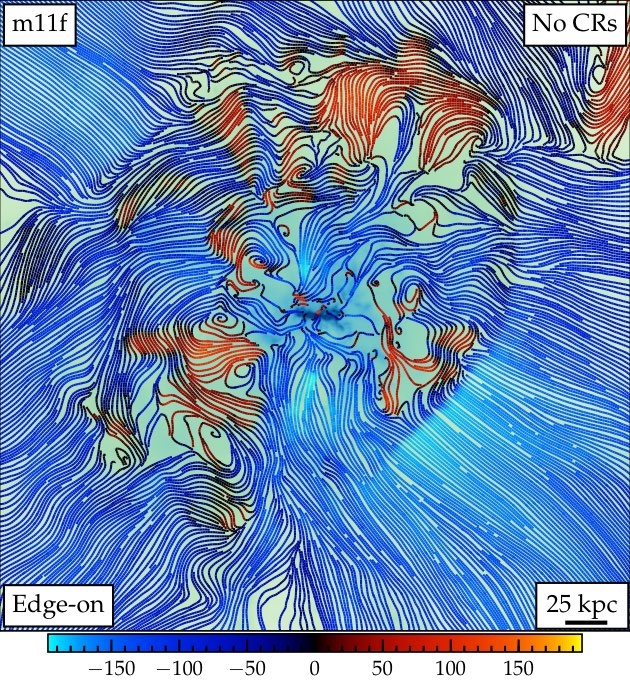}
\includegraphics[width={0.227\textwidth}]{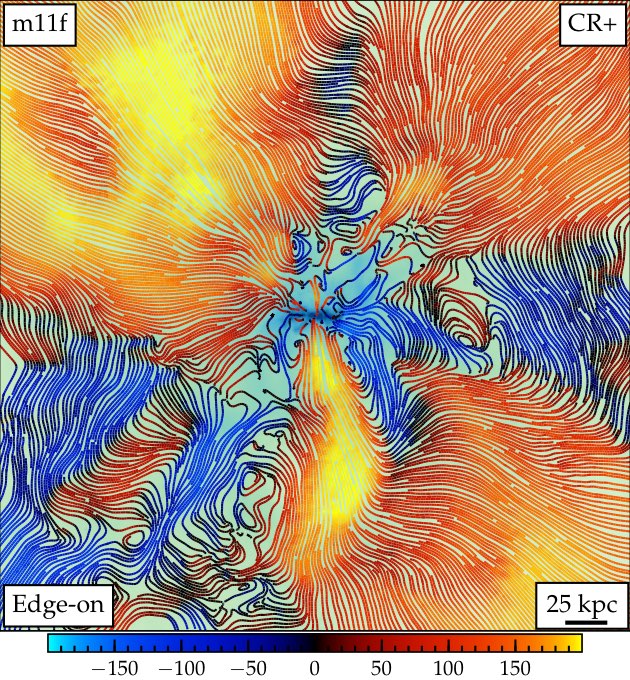} \\
\includegraphics[width={0.227\textwidth}]{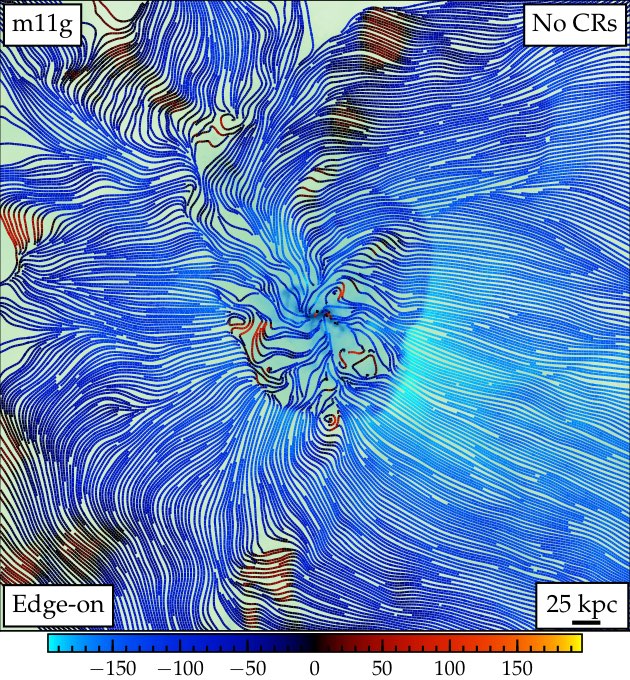}
\includegraphics[width={0.227\textwidth}]{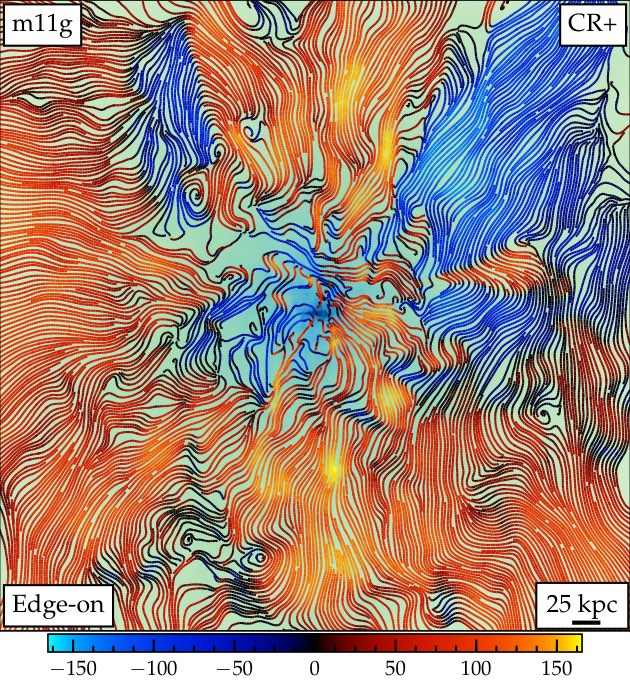} \\
\includegraphics[width={0.227\textwidth}]{figures/m12f_m7/mhdcv/Vmap_vr_R306_bipolar_edgeon}
\includegraphics[width={0.227\textwidth}]{figures/m12f_m7/cr_700/Vmap_vr_R306_bipolar_edgeon} \\
\includegraphics[width={0.227\textwidth}]{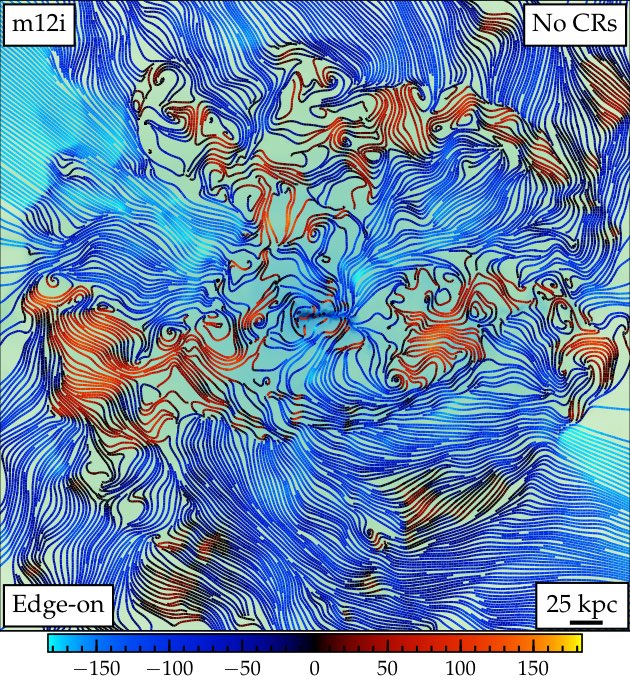}
\includegraphics[width={0.227\textwidth}]{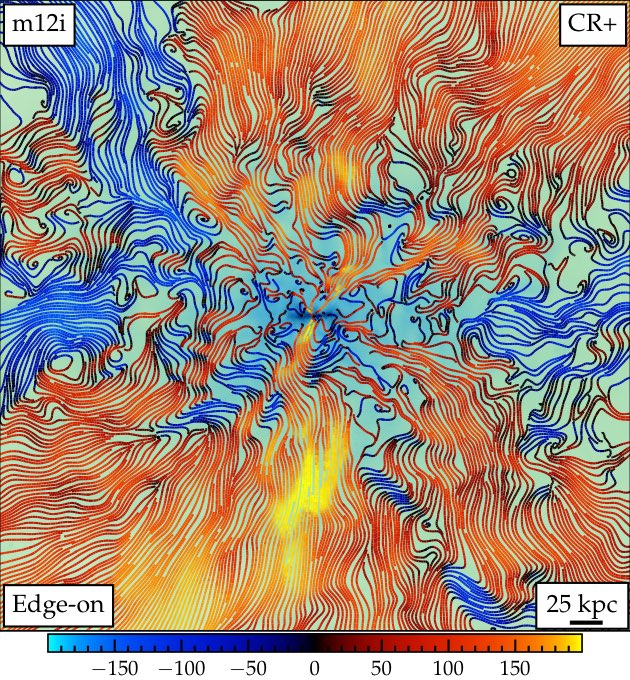} \\
\includegraphics[width={0.227\textwidth}]{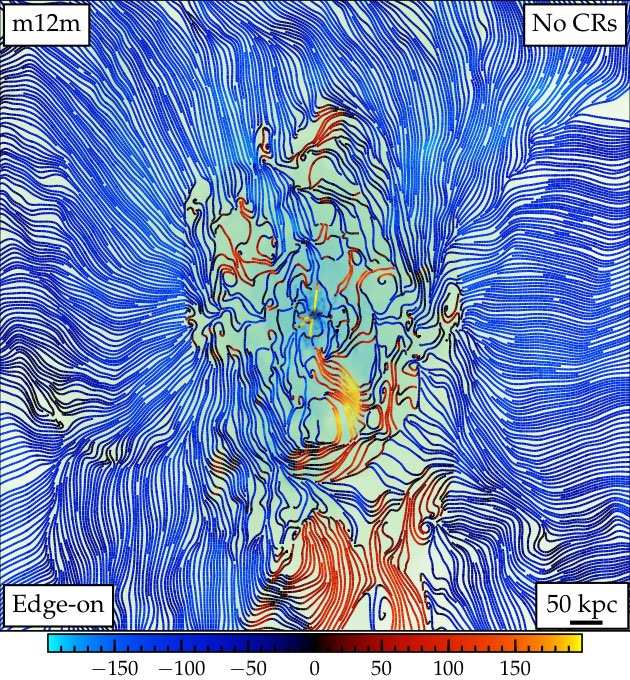}
\includegraphics[width={0.227\textwidth}]{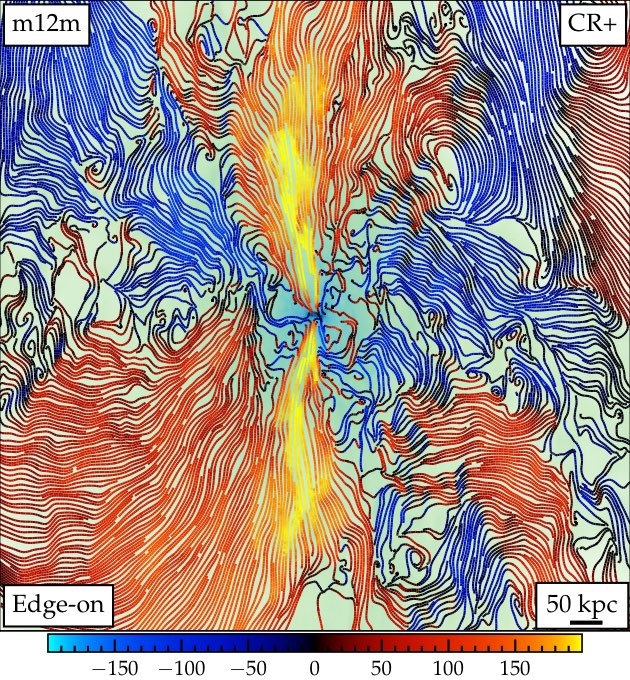} \\
\end{centering}
\vspace{-0.35cm}
\caption{Fig.~\ref{fig:inflow.outflow.dwarfs}, continued to higher masses with {\bf m11f}, {\bf m11g}, {\bf m12f}, {\bf m12i}, {\bf m12m}. The qualitative effect of CRs changing the CGM from turbulent quasi-spherical shock to coherent bipolar inflow-outflow, as in Fig.~\ref{fig:inflow.outflow.m12i}, is similar in each of these halos.
\label{fig:inflow.outflow.massive}}
\end{figure}

\vspace{-0.5cm}
\subsection{Global, Steady-State Wind Solutions}

First for simplicity, consider steady-state (time-independent), global, spherically-symmetric outflow solutions of the Euler-equations, in a CR-dominated halo (so $P = P_{\rm cr}$), with no other forces other than gravity (determined by the dark matter, so ignoring self-gravity of the outflow). From continuity we have $v_{r} = \dot{M}_{\rm out}/(4\pi\,\rho\,r^{2})$.  The momentum equation can then be written $(\dot{M}_{\rm out}/4\pi\,\rho\,r^{2})^{2}\,(2 + d\ln\rho/d\ln r) + \rho^{-1}\,dP_{\rm cr}/d\ln r + V_{c}^{2} = 0$. Inserting $P_{\rm cr} \approx \dot{E}/(12\pi\,\tilde{\kappa}\,r\,[1 + r/r_{\rm stream}])$ (which interpolates between the regimes $r \gg r_{\rm stream}$ and $r\ll r_{\rm stream}$ above), it is easy to verify numerically that this has a continuum of smooth outflow solutions. The resulting density and radial velocity profiles are monotonically decreasing with $r$, positive definite, continuous and infinitely differentiable, 

At small/intermediate radii ($r \lesssim {\rm MIN}(r_{\rm stream},\,r_{s})$), the solutions asymptote to $\rho \rightarrow \rho_{\rm crit}$ and $v_{r} \rightarrow ({G\,M_{\rm  halo}\,\tilde{\kappa}\,\dot{M}_{\rm out}})/({r_{s}^{2}\,\dot{E}_{\rm cr}})\sim $\,constant. Using $\epsilon_{\rm cr}=0.1$, our adopted IMF, the relation between $r_{s}$ and $R_{\rm vir}$ for halos with concentration $\sim 10$, and the fact that $\dot{E}_{\rm cr}=\epsilon_{\rm cr}\,u_{\rm SNe}\, \dot{M}_{\ast}$ (with $\dot{M}_{\ast} = \alpha\,M_{\ast}/t_{\rm Hubble}$), we can re-write this as: $\dot{M}_{\rm out} / \dot{M}_{\ast} \sim 0.6\,(1+z)^{-2}\,(v_{r}/10\,{\rm km\,s^{-1}})\,(\Delta\Omega/4\pi)\,\tilde{\kappa}_{29}^{-1}\,(M_{\rm halo}/10^{12}\,M_{\sun})^{-1/3}$, where $\Delta \Omega$ is the solid angle covered by the outflow (assuming a constant-$\Delta\Omega$ geometry). Outflow solutions exist, and they allow for arbitrarily small outflow velocity (albeit with small associated $\dot{M}_{\rm out} \propto v_{r}$, as well). Along different sightlines/solid angles, one can have different central $v_{r}$ corresponding to different outflow rates along that angle (e.g.\ the bipolar winds we will see below). 
At large radii in the outflow ($r \gtrsim {\rm MAX}(r_{\rm stream},\ r_{s})$), $\rho$ and $v_{r}$ gradually decline as $\sim 1/r$ (decreasing monotonically from their values at small $r$).

\vspace{-0.5cm}
\subsection{Non Steady-State Inflow/Outflow Behavior}

Consider a parcel of gas with initial density $\rho_{i}$ at some radius $r$ in the CR-dominated halo described above. If $\rho_{i} \ne \rho_{\rm crit}$, it experiences a non-zero net acceleration. 

For $\rho_{i} \gg \rho_{\rm crit}$, the dominant term in the equation-of-motion is the gravitational acceleration, so the parcel (at least initially) essentially free-falls onto the galaxy. If we imagine a small (but finite)-sized parcel with all  its mass free-falling on pure-radial trajectories (so the solid  angle subtended by the parcel is conserved), with an initial radial thickness $\Delta r$, then as it falls in it will be compressed in the tangential direction but tidally stretched,\footnote{Of course the ``deformation'' of a parcel and its density evolution will be sensitive to other terms such as magnetic tension and ram pressure. In our simulations these terms are generally small compared to the CR and tidal forces so we neglect them in our simple analytic argument. If we had much lower CR pressure, we of course revert to behavior more akin to classic multi-phase accretion/CGM models relevant in the ``no CRs'' limit. But more detailed study in idealized simulations like those in \citet{2020arXiv200804915B} is warranted.} so $\rho \propto r^{-\alpha}$ with $\alpha\approx 2$ at $r \ll r_{s}$ and $\alpha \approx 3/2$ at $r \gg r_{s}$. At small $r_{i} \ll r_{s}$, since $\rho_{\rm crit} \propto r^{-(1-2)}$ (depending on how $\kappa$ and $v_{\rm st}$ scale), this means the ratio $\rho/\rho_{\rm crit}$ is conserved or increases as the parcel falls towards the galaxy, allowing it to fall ``through'' to the galaxy. But if the parcel is initially moving slowly at large $r$ (where $\rho_{\rm crit} \propto r^{-(3-4)}$) its density increases more slowly than $\rho_{\rm crit}$ as it falls in, so it will eventually de-celerate and halt. 

For $\rho_{i} \ll \rho_{\rm crit}$, the dominant term is outward acceleration by CRs, so the parcel is accelerated into outflow. If we make the same geometric assumptions as above, then the ``tidal'' force from $P_{\rm cr}$ on a parcel or shell is compressive, and $\rho \sim $\,constant as the shell is accelerated at small $r_{i}$ or $\rho \propto r$ at large $r_{i}$. This means the acceleration from CRs becomes weaker as the parcel accelerates and is dominated by the acceleration near $r_{i}$. Eventually, $\rho/\rho_{\rm crit}$ increases and the gravitational force will again  become comparable  to  or dominate  CR  acceleration; however, if the parcel has already been accelerated to very large $v_{r} \gtrsim  V_{c}$, it can travel ``ballistically'' a large distance or even escape. For the assumptions above, the acceleration near $r_{i}$ is quasi-impulsive (relative to the outflow timescale), and accelerates a parcel to a local ``terminal'' $v_{r} \rightarrow \sqrt{2}\,V_{c}(r_{i})\,(\rho_{\rm crit}/\rho_{i} - 1)^{1/2}$ for small $r_{i}$ or $v_{r} \rightarrow \sqrt{2}\,V_{c}(r_{i})\,([\rho_{\rm crit}/\rho_{i} - 1]/3)^{1/2}$ for large $r_{i}$. 

So {\em if} some local process (e.g.\ ejection of low-density wind material from the galactic disk, or rarefactions in a turbulent halo) can generate gas with $\rho_{i} \ll \rho_{\rm crit}$ at a given radius, it will rapidly be accelerated to velocities of order the circular or escape speed, $\sim V_{c}\,(\rho_{i}/\rho_{\rm crit})^{-1/2}$ and can travel to very large radii or be unbound. 

\vspace{-0.5cm}
\subsection{Wind ``Trapping'' or Pressure Confinement}

Above we considered free expanding/contracting solutions. If the galactic disk impulsively ejects some gas, we should also consider the role of the gas column ``above'' it potentially confining these outflows. 

\vspace{-0.5cm}
\subsubsection{Gas Pressure-Dominated Halos (Weak CRs)}

First, consider the case {\em without} CRs. Assume the ``initial'' halo is in hydrostatic equilibrium with gas pressure following a $P = P_{0}\,(\rho/\rho_{0})^{5/3}$ adiabat\footnote{Assuming instead some power-law entropic function $P/\rho^{5/3} \propto r^{n}$ with $n\sim 0-1$, as suggested in e.g.\ \citet{2019MNRAS.488.2549S} for quasi-hydrostatic cooling-flow halos, only changes our argument here by an order-unity coefficient.} inside $R<R_{\rm vir}$ in a \citet{hernquist:profile} profile halo with $c=10$, and that the universal baryon fraction $f_{\rm baryon}\,M_{\rm halo}$ is in gas inside $R_{\rm vir}$ (with $P=({3/16})\,\rho\,V^{2}_{\rm vir}$ just inside $R_{\rm vir}$, appropriate for the post-virial-shock gas). This implies a gas pressure $P=P_{0}\,(\Psi/(1+r/2\,r_{s}))^{5/2}$ where $\Psi \approx 0.2\,{G\,M_{\rm halo}\,\rho_{0}/r_{s}\,P_{0}}$. 
If a spherical outflow moves out of the disk, there is an energetic cost $\Delta E$ associated with the ``PdV'' work of ``lifting'' this column. This can easily be integrated from $r=0$ to $r$, to show at small $r\lesssim a$, $\Delta E  = \int P\,dV \approx P(r=0)\,(4\pi\,r^{3}/3)$, or for a wind expanding at some $v_{\rm wind} = v_{r}$, $\dot{E}_{\rm work} \approx P_{0}\,(2\,\Psi/3)^{5/2}\,8\pi\,r^{2}\,v_{r}$. If we use the normalization conditions above to solve for $P_{0}$ and $\rho_{0}$, and equate this to a constant energy-injection rate $\dot{E}_{\rm wind} \approx (1/2)\,\dot{M}_{\rm out}\,v_{\rm wind}^{2}$, we find that the wind should ``stall''  relatively quickly as the energetic cost of pushing further at $v_{\rm wind}$ becomes larger than the energy injection rate in the wind, with the ``stalling radius'' just $\sim 10\,{\rm kpc}\,(1+z)^{-2}\,M_{\rm halo,\,12}^{-1/3}\,\left( \dot{M}_{\rm out}\,v_{\rm wind} / {\rm M_{\sun}\,yr^{-1}}\,500\,{\rm km\,s^{-1}} \right)^{1/2}$. In other words, even winds launched fairly ``violently'' (with $v \gtrsim 500\,{\rm km\,s^{-1}}$ and $\dot{M}_{\rm out} \gtrsim M_{\sun}\,{\rm yr^{-1}}$) will stall quickly (at $\sim 10\,$kpc), unless they involve extremely large momentum fluxes (usually seen only in AGN-driven winds). 

Of course, it is possible for winds to escape without stalling if they do not entrain the gas but ``punch through'' the hot halo -- e.g.\ if dense clumps or filaments, with small covering factor, are ejected from the disk. These would move ballistically, at least initially, although  secondary Kelvin-Helmholtz or Rayleigh-Taylor instabilities should generally ``shred'' such clouds fair quickly in a hot halo and mix them efficiently.

\vspace{-0.5cm}
\subsubsection{CR-Pressure Dominated Halos}

In the other hand, in the CR-dominated regime, where the pressure primarily comes from CRs, then if the halo gas is primarily sitting at $\rho \approx \rho_{\rm crit}$, and has a much lower temperature set by photo-ionization equilibrium rather than hydrostatic pressure equilibrium, the gas thermal pressure is essentially negligible. If we assume $\rho(r) \sim \rho_{\rm crit}(r) = \dot{E}_{\rm cr}/(12\pi\,V_{c}^{2}\,\tilde{\kappa}\,r)$ and $T\sim 10^{5}$\,K at all radii (using the more exact expression assuming photo-ionization equilibrium makes little difference), then calculate $\dot{E}_{\rm work} \sim P\,dV/dt \sim 4\pi\,P\,r^{2}\,v_{\rm wind}$ and compare this to the energy injection rate $\dot{E}_{\rm wind}\approx (1/2)\,\dot{M}_{\rm out}\,v_{\rm wind}^{2}$, we find that a disk-launch wind would be sufficient to provide the ``PdV'' work required to lift the cool (low-pressure) CR dominated gaseous halo to $r\rightarrow\infty$ for $v_{\rm wind}\,(\dot{M}_{\rm out}/\dot{M}_{\ast}) \gtrsim 30\,{\rm km\,s^{-1}}\,\tilde{\kappa}_{29}^{-1}\,M_{\rm halo,\,12}^{-1/3}$. 

What is important here is that a similar ``PdV'' work is {\em not} required to ``lift'' the CR fluid. Because the CRs are diffusive, with diffusion time on $\sim1-10\,$kpc scales $\sim 1-100\,$Myr much faster than the wind expansion time, the CRs simply diffuse {\em through} the outflowing gas, maintaining the equilibrium CR energy density/pressure profile essentially {\em independent} of the wind. In other words, CRs are not efficiently ``entrained,'' let alone ``compressed'' by these outflows. This means that the gas can gain the benefit of CR acceleration, by feeling the quasi-static CR pressure gradient on large length scales $\sim r$, but does not have to work against the CRs when escaping. 

This means that {\em dense} outflows from the disk will behave more-or-less ballistically, in the halo, and can escape or reach much larger radii before recycling. Moreover, low-density outflows with $\rho_{i} \ll \rho_{\rm crit}$ will be further accelerated in a CR-dominated halo, as described above. In fact, the steady state solutions above show that, for gas with $\rho \lesssim \rho_{\rm crit}$, there is effectively no ``escape velocity'' -- steady-state outflow solutions reaching $ r \rightarrow \infty$ and carrying constant $\dot{M}_{\rm out}$ exist for {\em arbitrarily} low initial wind $v_{\rm wind} = v_{r}$.

\vspace{-0.5cm}
\section{Results}
\label{sec:results}

\subsection{Case Study of a Milky Way-Mass, Low-Redshift Halo}

We first consider a ``case study'' of halo {\bf m12i} at redshift $z\sim 0$. This is instructive because it will allow us to test the analytic theory in \S~\ref{sec:theory}, and demonstrate essentially all of the qualitative features imprinted on outflows in our broader survey. We select halo {\bf m12i}: as a MW-mass halo (and at present-day $z\sim0$), this lies at the ``sweet spot'' where the effects of CRs from SNe are near-maximal on essentially all properties studied here or in \paperone. Here $M_{\ast}/M_{\rm halo}$ is maximized, so the magnitude of CR pressure relative to virial in the halo is largest, but the galaxy still allows most of the CRs to escape the dense star-forming disk gas without losses (see \paperone\ and \citealt{chan:2018.cosmicray.fire.gammaray} for extensive discussion). 

We will compare the two ``baseline'' simulations:  ``No CRs'' or ``MHD+'' from \paperone\ (all physics of star formation, stellar feedback, MHD, conduction, viscosity, but no CRs) and ``CR+'' or ``CR+($\kappa=3e29$)'' from \paperone\ (which has parameters for CRs favored observationally and theoretically, and produces the maximal effect of our CR runs); see \S~\ref{sec:methods}. 
First, we consider several ``snapshots'' of the gas inflow/outflow properties at $z=0$.
\fref{fig:demo.vr.vs.r} shows the distribution of gas inflow and outflow velocities in the disk/galactic fountain regime, CGM, and IGM. Figs.~\ref{fig:inflow.outflow.m12i}-\ref{fig:inflow.outflow.m12i.zoomed} show streamlines of the inflow/outflow on scales $R_{\rm vir}$ and $\gg R_{\rm vir}$. \fref{fig:mdot.inflow.outflow.m12i} shows the total mass inflow and outflow rates at different radii, as well as the flux-weighted mean velocities of the inflowing and outflowing gas. Fig.~\ref{fig:outflow.histories} follows the histories of gas parcels as they are accelerated into outflow, while Fig.~\ref{fig:demo.nh.t.wind} examines the phase structure of the outflows in more detail, specifically correlations between gas density, temperature, and outflow/inflow velocities, in the CGM/IGM gas. Fig.~\ref{fig:angle.outflow} more quantitatively assesses the outflow geometry (polar-angle dependence). 

\vspace{-0.5cm}
\subsubsection{Outflow Kinematics and Acceleration}

From these, we immediately see a number of striking differences with CRs. The CR+ run features a CR pressure-dominated halo, as expected for this mass and redshift range, which (owing to rapid CR diffusion) is well-approximated by a simple spherically-symmetric equilibrium model (see \S~\ref{sec:theory} and \paperone). This, in turn, predicts a simple equilibrium density $\rho_{\rm crit} \equiv |dP_{\rm cr}/d\ln r| / V_{c}^{2}$ at each radius, where CR pressure balances gravity, and indeed where most of the gas appears to reside (see \citealt{ji:fire.cr.cgm}). Lower-density material has a net outward acceleration ($a_{\rm cr} \propto \rho^{-1}\,dP_{\rm cr}/dr$ outward, larger than the gravitational $a_{\rm grav} \propto V_{c}^{2}/r$ inward), and appears to be re-accelerated at each $r$ up to the expected terminal velocity for its density (\S~\ref{sec:theory}), giving rise to both ``fast'' and ``slow'' outflows at large $R \sim 0.5-5\,R_{\rm vir}$.

Moreover, if anything, Figs.~\ref{fig:demo.vr.vs.r} \&\ \ref{fig:mdot.inflow.outflow.m12i} show that without CRs, outflows {\em within and around} the disk ($r\lesssim 30\,$kpc) actually have slightly {\em larger} mass-loading $\dot{M}_{\rm out}/\dot{M}_{\ast}$ (and recall, the run without CRs has a $\sim 2.5\times$ higher SFR, so the absolute $\dot{M}_{\rm out}$ is correspondingly larger), and significantly larger velocities ($> 100\,{\rm km\,s^{-1}}$). But without CRs the outflow rate drops precipitously at larger radii, while with CRs, we see $\dot{M}_{\rm out}$ fall then rise to dominate over inflow $\dot{M}_{\rm in}$ over $>$\,Mpc scales, with a mean velocity which increases to $\sim 70-80\,{\rm km\,s^{-1}}$ out to $\sim 1.5-2\,R_{\rm vir}$.

\fref{fig:outflow.histories} goes a step further and follows the time-history of Lagrangian gas elements in the CR+ simulation. The fast outflows at large-$r$ are accelerated over a broad range of radii in good agreement with our simple analytic expectations for constant $\rho/\rho_{\rm crit}$, without necessarily being heated to (or spending much time at) ``hot'' temperatures $T\gtrsim 10^{6}\,$K where cooling is inefficient. In fact, the outflows tend to begin ``cold'' (at typical ISM temperatures $\sim 10^{4}\,$K) and are mildly photo-heated as they expand (roughly tracing photo-ionization equilibrium with the UV background, with $T \propto \rho^{-0.2}$ or so, see \citealt{ji:fire.cr.cgm}). For the kinematics, we specifically compare the predicted $v_{r}$ of $r$ obtained from the model in \S~\ref{sec:theory} if we take the actual $V_{c}(r)$ from the simulation and assume a fixed ratio $\rho \approx \rho_{\rm crit}/2$ (approximately what we see for the fast winds in \fref{fig:demo.nh.t.wind}), for material starting at $r\approx 10\,$kpc. This provides a remarkably good description of the fast outflows. 

Fig.~\ref{fig:demo.nh.t.wind} shows the outflow velocities, densities, and temperatures at a {\em fixed} time in a specific radial annulus, to show that even at a fixed time, and at a given radius, the CR+ simulation outflow velocities trace our simple analytic expectation for the ``terminal'' velocity at a given density given acceleration by the CR pressure gradient, while remaining at the relatively cool/warm temperatures given by photo-ionization equilibrium. In contrast, in the ``No CRs'' runs, the outflows are strongly associated with gas whose thermal temperature exceeds the virial temperature, suggestive of traditional hydrodynamic pressure-driven outflows. 

All of these behaviors are clear demonstrations of CR acceleration in the ``CR+'' run. In fact the acceleration we see in the CR+ runs is not generically possible for an energy {\em or} momentum-conserving hydrodynamic wind. Consider: over much of the range of spatial scales where we see the outflows, the gas follows a density profile of approximately $\rho \propto r^{-1}$ (which follows from $\rho \sim \rho_{\rm crit} \propto 1/(r\,V_{c}^{2}) \sim 1/r$ over the range where $V_{c}\sim$\,constant, approximately $\sim 20-100\,$kpc here). In this regime, the velocities of a hydrodynamic wind with constant energy input ($\dot{E}_{\rm wind}$) or momentum input ($\dot{P}_{\rm wind}$) rates, or a conserved/constant initial/impulsive energy or momentum ($E_{\rm wind}$ or $P_{\rm wind}$) would necessarily decrease with $r$ (following a given fluid element).

\vspace{-0.5cm}
\subsubsection{Outflow Phases and Metallicities}

As noted above, in the CR+ run, the outflows lie preferentially at somewhat lower densities (as this material is efficiently accelerated outward by CR pressure gradients) and lower temperatures (as the acceleration is non-thermal and the densities low, most of this gas is simply at the equilibrium temperature for photo-ionization by the UV background), compare to our non-CR simulations. This is part of a broader trend examined in detail in \citet{ji:fire.cr.cgm}, wherein the {\em entire} CGM (both inflow and outflow) is shifted in density and (more dramatically) temperature when the halo is dominated by CR pressure (so gas can remain in the halo at thermal pressures well below virial). 

We also see in \fref{fig:outflow.histories} that the ``slow'' outflows at larger radii contain a mix of some de-celerated fast material, but also material which is accelerated {\em in situ} at large $r \gtrsim R_{\rm vir}$ -- material which {\em never} enters the galaxy. This also means that the metallicity of the outflows does not necessarily trace that of the galaxy, depending on where the material is ``swept up.'' We see this in Fig.~\ref{fig:mdot.inflow.outflow.m12i}, where the metallicity of the outflows steadily decreases with galacto-centric distance, clearly indicating ``new'' (lower-metallicity material which was residing in the halo) gas is swept up (either directly entrained or accelerated in-situ) to join this outflow. This occurs to some extent as well in our non-CR runs (and is discussed in detail in \citealt{muratov:2015.fire.winds} and \citealt{ma:2015.fire.mass.metallicity}), but the trend is much more clear and monotonic in the CR runs. This has a very important consequence: depending on where one defines or measures the outflow, it's metallicity can be much lower than the ISM metallicity (which is roughly the wind metallicity at the base of the disk), and so its ``metal-loading'' factor can be much lower than its mass-loading factor.

\vspace{-0.5cm}
\subsubsection{Outflow Geometry and Morphology}

Figs.~\ref{fig:inflow.outflow.m12i}-\ref{fig:inflow.outflow.m12i.zoomed} clearly illustrate a dramatic change in the morphology and geometry of inflows and outflows between our No CRs and CR+ runs. Absent CRs, the gas forms a quasi-spherical virial shock (at a radius of $\approx 1\,R_{\rm vir}$). External to the virial shock, the gas is in spherical inflow, while internal, it is largely turbulent, with outflows from the galaxy ``stalling'' and driving strong mixing with the CGM gas as they recycle. With CRs, the virial shock is hardly evident (the actual changes to the virial shock structure will be studied in future work), and the outflow and inflow assume a clear biconical structure which is also bipolar (approximately aligned with the angular momentum vector of the galactic disk). Inflow proceeds in the dense midplane sheets and filaments and joins smoothly onto the rotating disk at $\sim 20\,$kpc. Outflow extends outwards biconically with a widening outflow angle at increasing distance, filling the majority of the volume of the CGM and IGM out to $\sim\,$Mpc distances.

Note that the alignment of the outflows and the inner ($\sim 10\,$kpc) disk axis is not perfect in Fig.~\ref{fig:inflow.outflow.m12i.zoomed}.  Moreover we show below some example galaxies where the bipolar outflow ``base'' is well above the disk (at $\sim 30-50\,$kpc above the disk) as opposed to joining onto the disk. We also find that the disk orientation can sometimes vary (owing to e.g.\ minor mergers) on relatively short ($\lesssim\,$Gyr) timescales, but the outflow geometry on large scales remains stable. We also argued above that much of the outflow is accelerated well above the disk in the CGM. And we see the outflow is not particularly biconical, even very close to the disk, in our ``No CRs'' runs. All of this strongly argues that the bipolar morphology of the CR-driven outflows is {\em not} a result of ``shaping'' by the disk (as occurs for e.g.\ nuclear pressure-driven outflows emerging from a disk). Rather, the direction and morphology of {\em both} the outflows and the disk follow from a common cause -- the geometry of the large-scale IGM accreting onto the halo. The dense planar/filamentary structures accreting onto the halo define the preferred angular momentum axis and hence, disk direction, but they also define the directions where CR pressure will not overcome inflow ram pressure. Instead CR pressure will drive outflows in the remaining, lower-density volume which has lower ram pressure. 

Fig.~\ref{fig:angle.outflow} quantifies the angular dependence of outflow/CGM properties in more detail. We see that the mean density profile of the CGM is relatively weakly modified by CRs. This is consistent with the conclusions in \citet{ji:fire.cr.cgm}, who showed the effects on CGM temperatures were much larger. There is a strong radial density gradient (as expected), and a weaker trend at all radii towards lower density in the poles -- this occurs even outside $R_{\rm vir}$ in the ``No CRs'' runs, indicating it follows from large-scale structure. With CRs, there is a more well-ordered metallicity trend (discussed above), with higher metallicity in the polar direction (owing to outflow). Together these offsetting effects mean that the metal columns are not wildly different in the polar and planar directions. The velocity, as expected from Figs.~\ref{fig:inflow.outflow.m12i}-\ref{fig:inflow.outflow.m12i.zoomed}, shows the most clear difference between no-CR and CR+ runs, shifting from essentially no angular dependence at any radius in ``No CRs'' to clearly polar structure in ``CR+.''

\vspace{-0.5cm}
\subsection{Scaling With Halo Mass and Redshift}

Having considered a detailed case study of one galaxy ({\bf m12i}) above, we now use our larger sample of simulations to explore how the effects of CR-driven winds scale as a function of galaxy mass. Recall, \paperone\ showed that the effects of CRs (from SNe) on galaxy properties dropped off steeply at halo masses $M_{\rm  halo} \lesssim 10^{11}\,M_{\sun}$ or redshifts $z \gtrsim 2$. \citet{ji:fire.cr.cgm} showed the same for the effects of CRs on CGM phase structure. This is also naturally predicted by the simple analytic scalings in \S~\ref{sec:theory}: at low $M_{\rm halo}$, the ratio $M_{\ast}/M_{\rm halo} \propto \dot{M}_{\ast} / M_{\rm halo} \propto \dot{E}_{\rm cr}/M_{\rm halo}$ drops precipitously -- there simply isn't enough energy in CRs to compete with other forces (moreover, mechanical input from SNe becomes more efficient, further limiting the relative contribution of CRs). And at high-$z$, high densities within the galaxy deplete CR energy collisionally while dense halos contribute greater pressure than CRs can support. 

\fref{fig:profile.pressure} illustrates this showing the CGM pressure support from CRs, now comparing some $\sim 10^{10}$, $10^{11}$, and $10^{12}\,M_{\sun}$ halos. We see that by $<10^{11}\,M_{\sun}$, CR pressure is not sufficient to provide hydrostatic equilibrium support (we show in \paperone\ that at all radii in these halos, the gas thermal pressure is larger than the CR pressure). By $\sim 10^{10}\,M_{\sun}$ the CR pressure is more than an order-of-magnitude sub-dominant at all radii. We therefore expect the effects seen in {\bf m12i} ($\sim 10^{12}\,M_{\sun}$) to drop off rapidly in our less massive halos below $\sim 10^{11}\,M_{\sun}$. \fref{fig:mdot.inflow.outflow.others} shows the inflow/outflow rates of gas and SFRs at several halos across this mass range: indeed, the effects on galaxy SFRs (as shown in detail in \paperone) drop off rapidly and become second-order or negligible below $M_{\rm halo} \lesssim 10^{11}\,M_{\sun}$, and we see the same in the outflow rates and velocities at any radii. By the lowest-mass ($10^{10}\,M_{\sun}$) halos the outflows become strongly dominated by the effects of the most recent large ``burst'' of SF.

Figs.~\ref{fig:masssurvey.vr.vs.r}, \ref{fig:masssurvey.vr.vs.nh}, and \ref{fig:masssurvey.vr.vs.t} repeat our study of the outflow/inflow velocity distribution vs.\ radius, density, and temperature. For the halos with $M_{\rm halo} \gtrsim 10^{11}\,M_{\sun}$, we find in every case qualitatively similar conclusions to our {\bf m12i} case study, with the effects generally becoming stronger in the more massive halos, as predicted, while they drop off at lower-mass halos. A couple of the ``marginal'' cases are interesting: the effects of CRs on the CGM and outflows of e.g.\ {\bf m11q} are stronger here than than they are on its stellar mass and CGM properties (see \paperone), but this halo lies exactly at $\sim 10^{11}\,M_{\sun}$ (the border between weak/strong CR effects), so this is perhaps not surprising. Where the CR pressure is weak (in lower-mass halos), we see our analytic prediction for the CR-pressure driven outflow velocities falls well short of the actual velocities: this is just the statement that the outflows are not CR-driven.

Closely related, Fig.~\ref{fig:outflow.vdist} shows the outflow velocity distributions in the CGM and CGM+ISM of the galaxies. Again at low masses there is no significant difference. At high masses, this quantifies again the extent to which ISM outflows are pressure-confined in the massive halos without CRs, while high-velocity material primarily resides at large radii in the CR-dominated halos.

Figs.~\ref{fig:inflow.outflow.m12f}, \ref{fig:inflow.outflow.m11f}, \ref{fig:inflow.outflow.m11b}, \ref{fig:inflow.outflow.m11a}, \ref{fig:inflow.outflow.m10q} repeat the morphological comparison of inflows and outflows on CGM/IGM scales. Again in more massive halos the qualitative conclusions match {\bf m12i}. The radii and mass range where the outflow/inflow morphology is strongly altered corresponds to those where CRs strongly alter the total pressure balance, in \fref{fig:profile.pressure}. In low-mass halos the morphology is dominated by the shocks from outflowing gas, and turbulence, but is not particularly sensitive to the presence of CRs.

Note that in many of these plots, in the low-mass systems (e.g.\ {\bf m10q}, {\bf m10v}) we  can clearly see in both the CR+ and non-CR runs, the impact of successive ``bursts'' of SF on the CGM/IGM: they produce successive ``spikes'' of outflowing material (``shells'' with $v_{r} \propto r$ over a small range, consistent with a burst of material launched at the same time), and a series of concentric shocks visible in the inflow/outflow morphology.

\fref{fig:profile.pressure} also shows the effects of changing redshift. Again in \paperone, we showed in detail that the effects of CR pressure in the CGM decrease with increasing $z$ (at fixed $M_{\rm halo}$ or $M_{\ast}$), even more rapidly than the effects drop off with lower halo mass -- by $z \gtrsim 1-2$ (depending on the halo mass) the effects of CRs are completely negligible in the CGM pressure. We confirm this here. As a result, the effects of CRs ``shaping'' the outflows are weak at high-redshift, and we cannot identify any obvious morphological or quantitative differences at $z\gtrsim 2$ (but because high-redshift galaxies are quite ``bursty'' and often undergoing mergers, side-by-side morphological comparisons such as those above tend to be dominated by chaotic differences in timing of bursts, etc.). So we do not explore this further here.

It is worth noting, as discussed further in \paperone, that some previous studies \citep[e.g.][]{Boot13,Simp16} found stronger effects of CRs in dwarfs than we see here; however that particular study included very weak mechanical feedback from SNe (using a ``pure thermal energy deposition'' scheme that the authors noted tends to over-cool), and a lower diffusivity by a factor of $\sim 100$. Thus CRs are more strongly trapped in the ISM (see \paperone), and SNe are relatively weak, so the relative effect of CRs is more prominent. Per \paperone, with higher diffusivity CRs escape dwarfs more efficiently \citep[as observed; see e.g.][]{lopez:2018.smc.below.calorimetric.crs}, but more importantly with more careful coupling of mechanical SNe feedback and better resolution of super-bubble cooling radii \citep[see][]{hopkins:sne.methods}, the SNe mechanical energy (which is an order-of-magnitude larger than CR injection energy, by definition) simply dominates in dwarfs where cooling is often inefficient.

\vspace{-0.5cm}
\section{Discussion \&\ Conclusions}
\label{sec:conclusions}

We study the properties of galactic outflows in a large survey of high-resolution cosmological FIRE-2 simulations, with explicit treatment of mechanical and radiative stellar feedback (SNe Types Ia \&\ II, O/B \&\ AGB mass-loss, photo-ionization and photo-electric heating and radiation pressure), magnetic fields, anisotropic conduction and viscosity, and cosmic rays injected by SNe (with anisotropic streaming and diffusion; advection and adiabatic interactions; hadronic, Coulomb and streaming losses). Previous work has extensively explored how mechanical and radiative feedback influence outflows, and has also shown that their properties (at least insofar as relevant for bulk galaxy/ISM/CGM/IGM predictions) are not particularly sensitive to magnetic fields, conduction, and viscosity. We therefore focus on the role of CRs.

\vspace{-0.5cm}
\subsection{Key Conclusions}

In previous studies (see \paperone), we have shown that the effect of CRs (from SNe) on galaxy properties is maximized in intermediate and massive ($M_{\rm halo} \sim 10^{11-12}\,M_{\sun}$) halos at relatively low redshifts $z\lesssim 1-2$. This is where, for physically-reasonable and observationally-allowed CR parameters (from $\gamma$-ray, grammage, CR energy density, synchrotron, and other constraints; see \papertwo), CR pressure can dominate over gas thermal pressure in the halo. Not surprisingly, we find the same for outflows. Specifically, the effects of CRs on outflows are strongly correlated with their relative prominence in the CGM (e.g.\ ratio of CR to thermal gas pressure, at $\sim 0.1-2\,R_{\rm vir}$). Where this ratio is large, CRs dramatically alter outflows; where the ratio is small, they have small effects. 

In ``CR-dominated'' halos (near this ``sweet spot'' in mass and redshift, where CRs dominate the CGM pressure), we find the following effects on outflows and inflows:

\begin{itemize}

\item The morphology of outflows and inflows is drastically re-shaped. In CR-dominated halos, outflows are coherently bi-conical from disk-through-IGM  ($\gtrsim\,$Mpc) scales, with most of the {\em volume} at large radii in outflow, and inflow confined to relatively small covering-angle, dense planar/filamentary structures (Trapp et al., in prep.). Some collimation occurs even when CRs do not fully-dominate, if they still contribute an order-unity fraction of the CGM pressure. Absent CRs, the bi-conical morphology is largely destroyed and what little outflow remains in $\sim L_{\ast}$ halos is confined to the near-region around the disk, and essentially all directions show inflow onto the halo outside $R_{\rm vir}$.

\item The total or gross cosmological inflow rates onto halos (at large radii) are not dramatically altered, although inflows appear to be slower (more ``gently'' decelerated as they enter the CGM and approach the galaxy) and the virial shock is much less pronounced (this will be studied in future work; Ji et al., in prep). However, in CR-dominated halos the inflow is primarily confined to dense inflow structures (e.g.\ filaments) which carry most of the mass (primarily in the plane of the disk), but represent little volume or covering factor. Absent CRs, nearly all gas outside $R_{\rm vir}$ is inflowing.

\item Outflow rates {\em within and near the disk} ($r\lesssim 30\,$kpc, in MW-mass systems) both absolute and per-unit-star formation, are comparable or larger {\em without} CRs, and the mean velocities and temperatures of outflowing material are also larger ($v_{r} \gtrsim V_{c}$, and $k\,T \gtrsim \mu\,V_{c}^{2}$) without CRs. However, in massive halos absent CRs, this outflow is strongly confined by the very large thermal gas pressure of the over-lying halo, and so outflows decelerate and ``halt'' rapidly outside the galaxy (re-cycling quickly and stirring the central regions; see \citealt{muratov:2015.fire.winds,muratov:2016.fire.metal.outflow.loading}). Where CRs dominate the pressure, outflows from the center can escape, owing to the much lower thermal gas pressure of the halo (owing to rapid diffusion, the CR pressure does not ``resist'' outflow expansion), and propagate effectively to extremely large radii $\gtrsim\,$Mpc.

\item In CR-dominated regimes, gas is also accelerated ``in situ'' in the halo to large velocities by the large-scale CR pressure gradient (on $\sim 10-10^{3}\,$kpc scales). There is a critical density $\rho_{\rm crit}$ in the CGM, as shown in \paperone\ and \citet{ji:fire.cr.cgm}, where CR pressure balances gravity; less dense material is accelerated rapidly to $v_{r} \sim V_{c}(r)$ by CR pressure gradients. We show this ``in situ'' acceleration provides a good explanation of the wind dynamics, both for (i) material escaping the central $\sim 10-30\,$kpc and reaching large radii as ``fast'' outflows (because $V_{c}$ where it is accelerated is large), and also (ii) for accreted material which never reaches the galaxy and is ``turned around'' or accelerated ``in situ'' at large $r \gtrsim R_{\rm vir}$ into ``slow'' outflows ($v_{r} \lesssim V_{\rm vir}$). 

\item These effects (in the CR-dominated regime) directly supply CR-driven outflows at large radii, so the outflow rate or mass-loading actually {\em increases} further away from the disk, reaching and sustaining {\em net} outflow (compared to cosmological accretion rates) at essentially all radii $\gtrsim 100\,$kpc. Thus CRs act primarily as a ``preventive'' feedback mechanism: they suppress inflow rates into the galaxy and inner CGM, and extend the recycling times of gas which has already been blown out of a galaxy.

\item The less-efficient ``trapping'' in CR-dominated cases means that outflows are not recycled nearly as rapidly as they are in runs without CRs. It is therefore import to revisit previous calculations based on following fluid elements over time, which argued that galactic outflows may be recycled many times at low redshifts in the CGM of massive galaxies.

\item Because the acceleration by CRs does not depend (directly) on gas thermal energy, and more so because much of the material at large radius is never shock-heated (even by the virial shock), the outflows in CR-dominated cases are primarily ``cool/warm'' ($\sim 10^{5}\,$K). This is a broader consequence, however, of the fact that the CGM as a whole is more dominated by gas which can be supported by CR pressure so does not need to have large temperatures to be in (initial) virial equilibrium \citep[see][]{ji:fire.cr.cgm}.

\end{itemize}

We show that all of the above phenomena can be predicted (with surprising accuracy) by simple equilibrium analytic scalings, derived in \S~\ref{sec:theory}. These can also provide useful ``fitting functions'' to predict CR pressure-driven outflow rates and velocities, where relevant.

Although there has been considerable study in the literature of CR-driven outflows (see references in \S~\ref{sec:intro}), most of this has focused either (a) on driving in or very near the disk (e.g.\ around the CR scale-height above the disk plane), or (b) around extremely massive halos (e.g.\ clusters) where the CRs are likely sourced by AGN. Although some of our conclusions are similar (e.g.\ CRs can re-accelerate pressure-confined winds and enhance the fraction of ``cool'' gas in the outflows), what is truly remarkable here is the enormous spatial scale over which the CRs have a dramatic effect. In fact, we find that almost all of the most dramatic effects only occur at radii $\gtrsim 30\,$kpc, and extend to radii $\gtrsim\,$Mpc. Obviously, exploring these far-field CGM/IGM scales requires cosmological simulations. Moreover, doing so self-consistently, and capturing the effects of different phases of gas launched out of the disk initially, requires simulations that not only model CR transport and coupling but also magnetic fields, a multi-phase ISM, star formation, and stellar feedback processes (e.g.\ mechanical and radiative feedback). This explains why it has not been seen in most previous studies.

\vspace{-0.5cm}
\subsection{Observational Implications}

These modifications to outflows have a number of observational ramification, which fall into three categories:
\begin{itemize}

\item {\bf Outflow Phase Structure:} As noted above, the CRs modify the phase structure (temperatures and densities) of outflows, as part of a general shift in the CGM phase structure when it is CR-pressure supported. A more detailed study of the observational effect of this change in CGM phase structure (including these outflows), and its consequences for UV and X-ray observations of CGM warm/hot gas, is the subject of \citet{ji:fire.cr.cgm}. Briefly, the overall lower temperatures, and higher gas densities in most of the CGM lead to an increase in the columns of warm absorbers (e.g.\ OVI) and decrease in hot absorbers (NeVIII) in MW-mass halos. However, given the large scatter observed in the strength of these absorbers, it is difficult to un-ambiguously rule out either model at present, but this may be possible with larger statistical samples of simulations and observations.

\item {\bf Outflow Kinematics:} A number of observations have suggested that there is a correlation between outflow velocity and galacto-centric radius or impact parameter, of the form $v_{r} \propto r^{0.2-0.8}$ at radii $\sim 2-200$\,kpc \citep[e.g.][]{steidel:2010.outflow.kinematics}, which is suggestive of continuous outflow acceleration. Indeed this is strikingly similar to the trend in the ``upper envelope'' of $v_{r}$ versus $r$ seen in our Fig.~\ref{fig:demo.vr.vs.r} for our CR simulation. However, (a) the observational result remains controversial, (b) the {\em mean} $v_{r}$ in the same simulation from Fig.~\ref{fig:demo.vr.vs.r} does not actually increase with radius in the same manner (see Fig.~\ref{fig:mdot.inflow.outflow.m12i}), despite acceleration of individual parcels (as they can reach lower terminal velocities at larger radii; see Fig.~\ref{fig:outflow.histories}) so this clearly depends on how the ``velocity'' measured is defined or weighted, and (c) a similar trend of increasing $v_{r}$ with $r$ can emerge naturally from thermal pressure-driven or even ballistic (decelerating) outflows, simply owing to the fact that if outflows are launched with a range of $v_{r}$, those with larger $v_{r}$ reach larger $r$ and do so more rapidly \citep{hopkins:2013.merger.sb.fb.winds}. It is also the case that many observations have suggested a large fraction of outflow mass may be in ``slow'' outflows \citep{2015ApJ...809..147H}, qualitatively similar to the predictions in the CR-dominated models here. However, (a) absolute outflow rates across different gas phases are notoriously difficult to robustly measure and compare, (b) ``slow'' outflows in particular are difficult to distinguish from turbulent or fountain motion within the halo \citep{muratov:2016.fire.metal.outflow.loading}, and (c) once again, this is not a unique signature of CRs, as the same effect can arise from e.g.\ winds preferentially driven by stellar radiation pressure instead of mechanical (SNe) stellar feedback \citep{hopkins:stellar.fb.winds,zhang:dusty.cloud.acceleration}.

\item {\bf Outflow Morphology:} Perhaps the most direct and robust testable prediction here is the effect of CRs on outflow morphology. A number of observational studies (primarily of Mg II and Fe II absorbers) have suggested outflows around $\sim L_{\ast}$ galaxies at $z\lesssim 2$ are preferentially bipolar and biconical \citep{2011MNRAS.416.3118K,2012ApJ...760L...7K,2012ApJ...758..135K,2012MNRAS.426..801B,2014ApJ...794..156R}, in agreement with the CR-dominated halo predictions (Fig.~\ref{fig:inflow.outflow.m12i}). Even when non-CR feedback mechanisms can drive strong outflows in our simulations (e.g.\ in dwarfs, or high-redshift galaxies, or starburst systems), they are generally {\em not} strongly bipolar (and even when they are collimated they are not particularly well-aligned with galactic disks; see \citealt{faucher-giguere:2014.fire.neutral.hydrogen.absorption,muratov:2015.fire.winds,hafen:2016.lyman.limit.absorbers}). However, some care is still needed: because the CR-driven outflows are ``slow'' and the density in outflowing gas is relatively low, it is not obvious if the clear bipolar outflow structure seen in e.g.\ Fig.~\ref{fig:inflow.outflow.m12i} actually translates to a clear observable trend of absorber equivalent width or velocity width as a function of polar angle. In subsequent work, we will forward-model the absorption-line profiles of MgII absorbers to quantitatively compare with these observations.

\end{itemize}

\vspace{-0.5cm}
\subsection{Caveats \&\ Future Work}

There are many interesting aspects of these simulations which remain to be explored. In future work, we plan to examine more detailed wind diagnostics, e.g.\ their phase, column density, and ionization-state distributions, which will allow us to directly compare to observational constraints on galactic outflows (and to make predictions for e.g.\ which observable phases/diagnostics should represent the bulk of the outflow material). 

Given the deep physical uncertainties in CR transport in the CGM, this may represent the best path forward to constrain these models. Those uncertainties in CR physics will also be explored. We wish to strongly emphasize that we chose as our ``default'' CR model here the implementation which had a near-maximal effect in \paperone. There and in \citet{chan:2018.cosmicray.fire.gammaray}, we showed that a much lower CR diffusivity $\kappa$ essentially eliminates all the effects here, as CRs are trapped too close to the galaxy (where their pressure is less important, and they lose energy rapidly to hadronic and Coulomb collisions), and leads to excessive $\gamma$-ray production (compared to observations). But if the diffusivity increases too rapidly in the CGM, or CRs de-couple from the stress tensor (e.g.\ ``slip'' and stream out), then they will essentially do nothing to CGM gas (and such a scenario is theoretically at least plausible, and observationally allowed). So clearly it is important to develop and test more sophisticated CR transport models beyond the streaming+diffusion approximation here. Indeed, preliminary comparison with simulations including more complicated CR transport parameterizations motivated by extrinsic turbulence or self-confinement theories in \citet{hopkins:cr.transport.constraints.from.galaxies,2020arXiv200402897H} suggests that while the CRs probably do not completely ``de-couple,'' they do stream more rapidly at large galacto-centric radii, weaking some of their effects in the CGM.

We have also emphasized above that the CR-driven outflows can reach enormous ($>$Mpc) scales: they may in fact go further and pollute a substantial fraction of the IGM, but we cannot continue our analysis much further before we start to approach the boundaries of the high-resolution ``zoom-in'' region of these simulations (a few Mpc,  at most). Large-volume simulations  are clearly required to explore the potentially radical effects on even larger scales. 

Finally, we have also focused our analysis on halos with $z\sim 0$ masses $M_{\rm halo} \lesssim 10^{12}-10^{13}\,M_{\sun}$. In \paperone, \citet{su:2018.stellar.fb.fails.to.solve.cooling.flow}, and herein, we showed that in more massive halos, or (equivalently) halos which reach $\gtrsim 10^{12}\,M_{\sun}$ at high redshifts $z\gtrsim 2$, the effects of CRs from SNe are weak, owing to a combination of lower $M_{\ast}/M_{\rm halo}$, higher halo/CGM gas pressure and column densities, and higher gas densities in the galaxy. However, such massive halos have an additional, obvious source of CR feedback, namely AGN, which we have not included. This will be explored in companion papers such as \citet{su:turb.crs.quench} as well as future work focused on the broader question of the role of AGN feedback in such very massive halos.

\vspace{-0.7cm}
\datastatement{The data supporting the plots within this article are available on reasonable request to the corresponding author. A public version of the GIZMO code is available at \gizmourl. Additional data including simulation snapshots, initial conditions, and derived data products are available at \FIREurl.}

\acknowledgments 
Support for PFH and co-authors was provided by an Alfred P. Sloan Research Fellowship, NSF Collaborative Research Grant \#1715847 and CAREER grant \#1455342, and NASA grants NNX15AT06G, JPL 1589742, 17-ATP17-0214. 
CAFG was supported by NSF through grants AST-1517491, AST-1715216, and CAREER award AST-1652522, by NASA through grant 17-ATP17-0067, by STScI through grants HST-GO-14681.011, HST-GO-14268.022-A, and HST-AR-14293.001-A, and by a Cottrell Scholar Award from the Research Corporation for Science Advancement. DK was supported by NSF grant AST-1715101 and the Cottrell Scholar Award from the Research Corporation for Science Advancement. Numerical calculations were run on the Caltech compute cluster ``Wheeler,'' allocations FTA-Hopkins supported by the NSF and TACC, and NASA HEC SMD-16-7592.\\

\vspace{-0.2cm}
\bibliography{ms_extracted}

\begin{appendix}

\section{Additional Velocity Field Images}

Here we include some additional detailed images of the galaxy velocity fields, in the style of Fig.~\ref{fig:inflow.outflow.m12i}.

\begin{figure*}
\begin{centering}
\includegraphics[width={0.24\textwidth}]{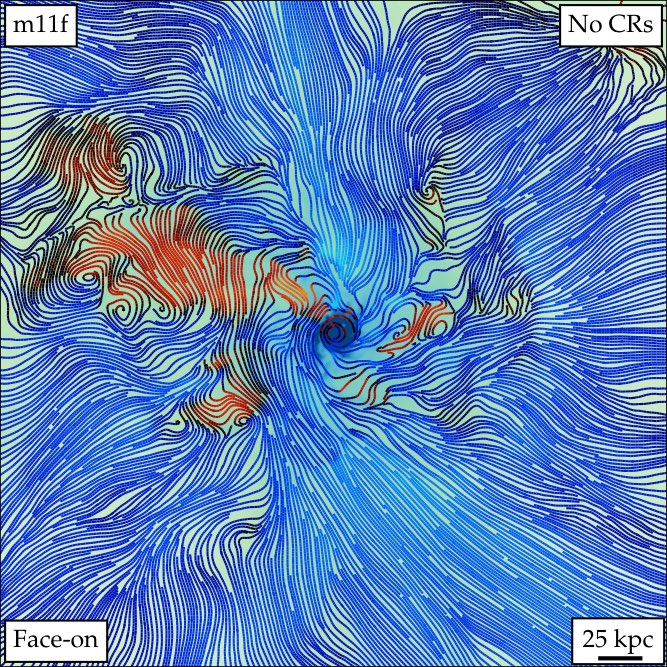}
\includegraphics[width={0.24\textwidth}]{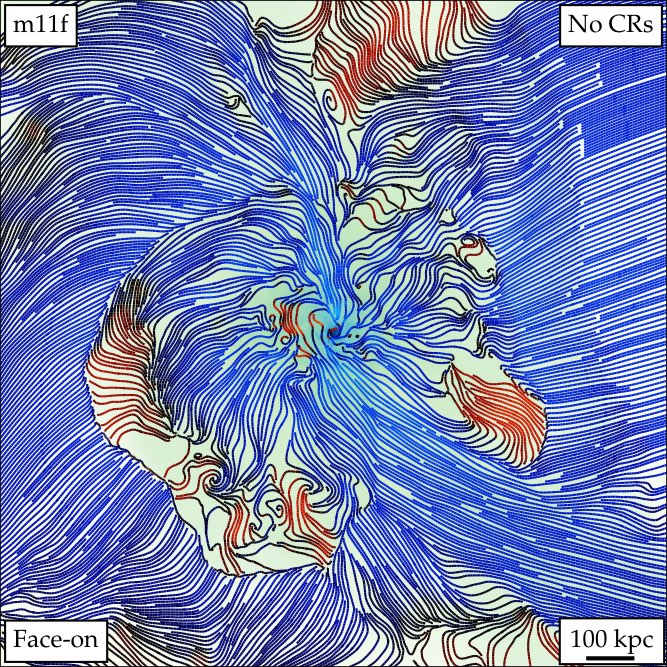}
\hspace{0.4cm}
\includegraphics[width={0.24\textwidth}]{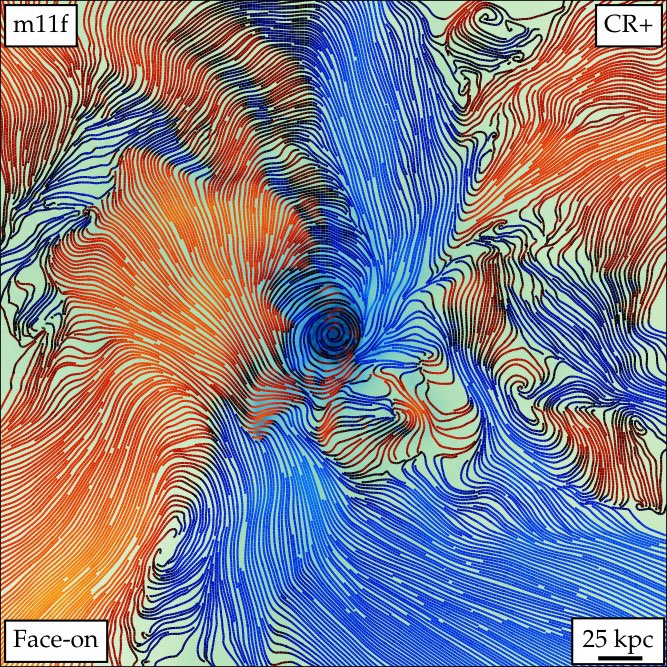}
\includegraphics[width={0.24\textwidth}]{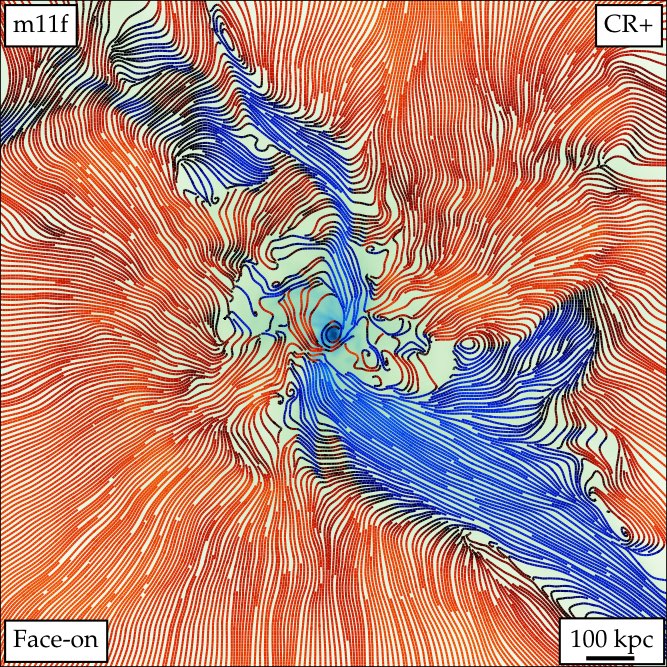} \\
\includegraphics[width={0.24\textwidth}]{figures/m11f/mhdcv/Vmap_vr_R208_bipolar_edgeon}
\includegraphics[width={0.24\textwidth}]{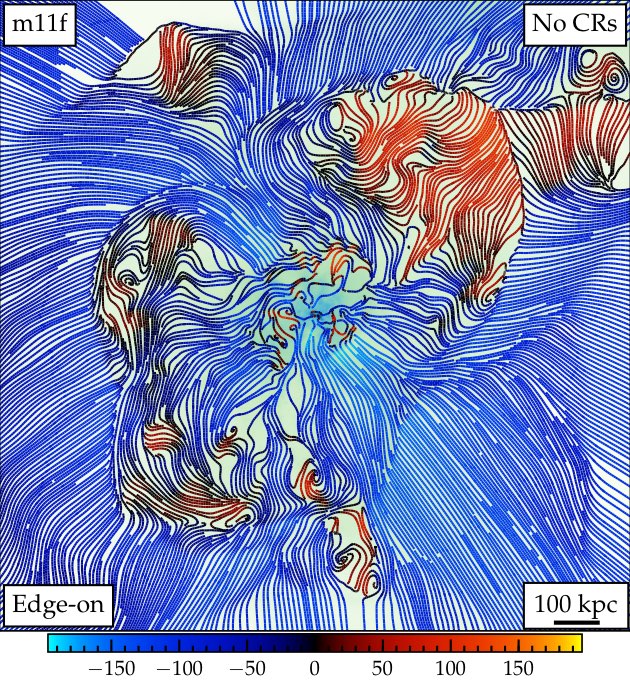}
\hspace{0.4cm}
\includegraphics[width={0.24\textwidth}]{figures/m11f/cr_700/Vmap_vr_R208_bipolar_edgeon}
\includegraphics[width={0.24\textwidth}]{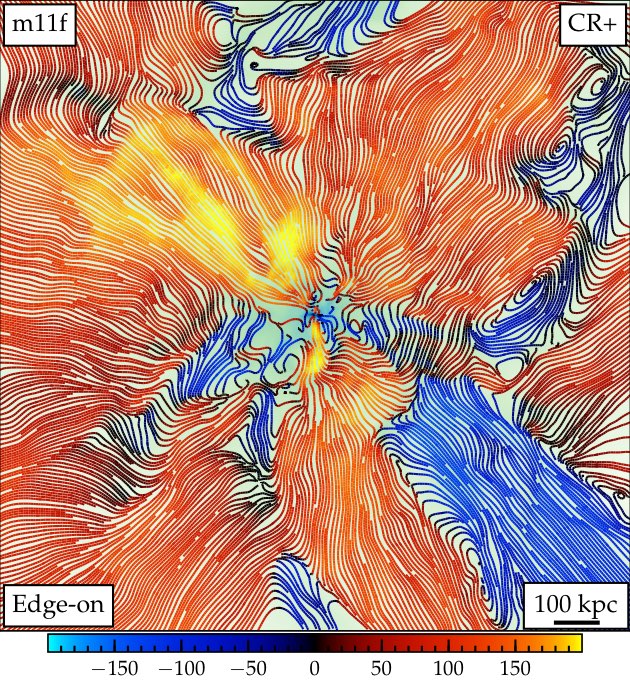} \\
\end{centering}
\caption{As Fig.~\ref{fig:inflow.outflow.m12f}, for {\bf m11f}. The galaxy is somewhat less-massive, but still has a SFR strongly suppressed by CRs and a CR-dominated halo in the ``CR+'' run. The lower halo mass (more rapid cooling) means the virial shock is somewhat less sharp in the ``No CRs'' run (compared to the {\bf m12} runs), but we still see a similar qualitative change in behavior out to $\gg R_{\rm vir}$. 
\label{fig:inflow.outflow.m11f}}
\end{figure*}

\begin{figure*}
\begin{centering}
\includegraphics[width={0.24\textwidth}]{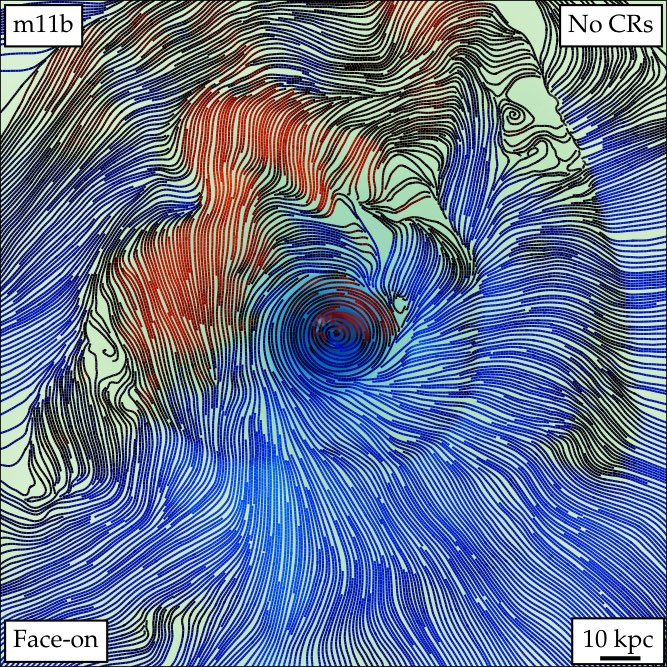}
\includegraphics[width={0.24\textwidth}]{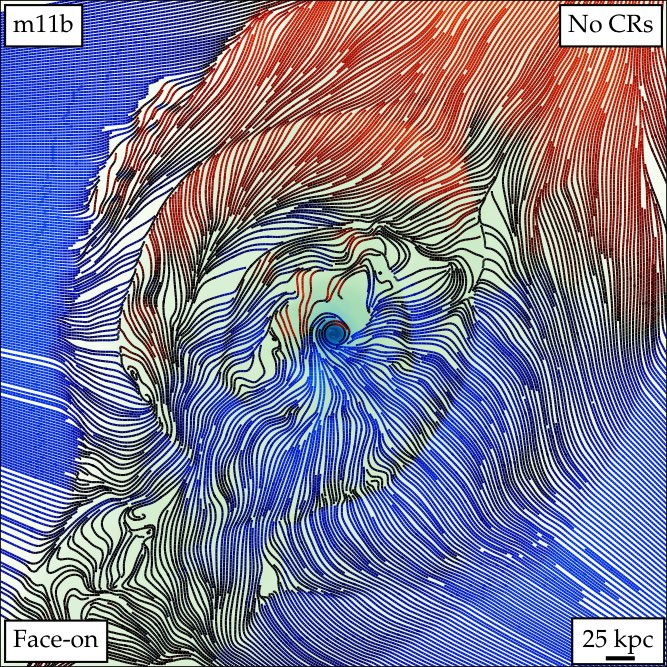}
\hspace{0.4cm}
\includegraphics[width={0.24\textwidth}]{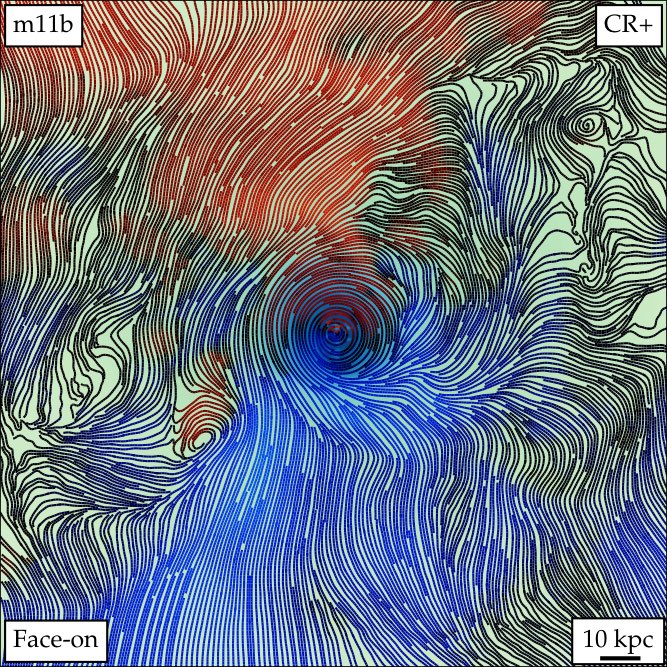}
\includegraphics[width={0.24\textwidth}]{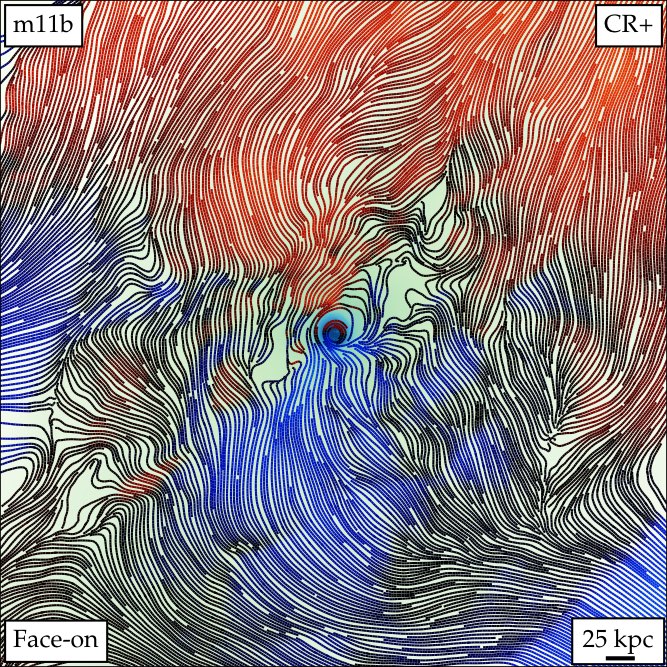}\\
\includegraphics[width={0.24\textwidth}]{figures/m11b/mhdcv/Vmap_vr_R92_bipolar_edgeon}
\includegraphics[width={0.24\textwidth}]{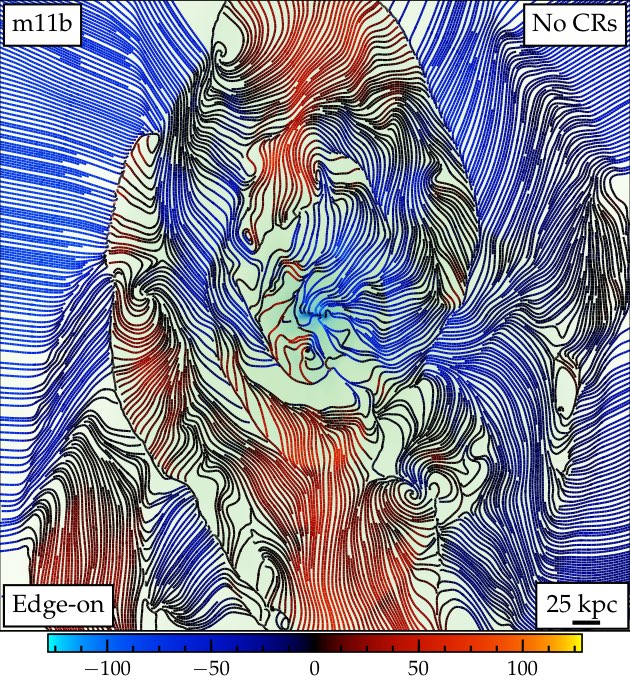}
\hspace{0.4cm}
\includegraphics[width={0.24\textwidth}]{figures/m11b/cr_700/Vmap_vr_R92_bipolar_edgeon}
\includegraphics[width={0.24\textwidth}]{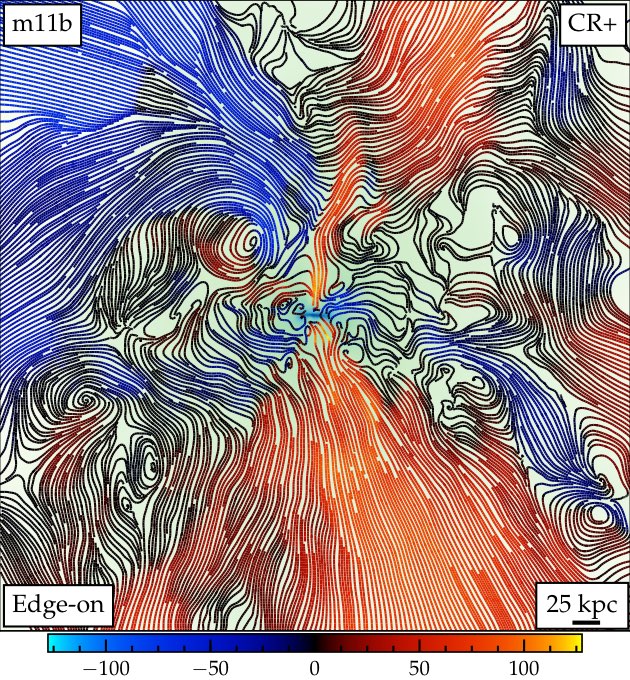}\\
\end{centering}
\caption{As Fig.~\ref{fig:inflow.outflow.m12f}, for {\bf m11b}. This galaxy is low enough in mass ($M_{\ast} \sim 10^{8}\,M_{\sun}$, $M_{\rm halo} \sim 4\times10^{10}\,M_{\sun}$) such that the effect of CRs on galaxy properties is significantly weaker. However, significant effects of CRs on CGM/IGM scales $\sim 1-4\,R_{\rm vir}$ are still apparent. Edge-on, the ``No CRs'' run features shocks and a quasi-spherical/isotropic \&\ turbulent flow structure, while the ``CR+'' run exhibits a bipolar structure (albeit with less volume-filling outflow). Face-on, the ``No CRs'' run exhibits a clear series of concentric shocks towards $R_{\rm vir}$, but these are absent in the CR run.
\label{fig:inflow.outflow.m11b}}
\end{figure*}

\begin{figure*}
\begin{centering}
\includegraphics[width={0.24\textwidth}]{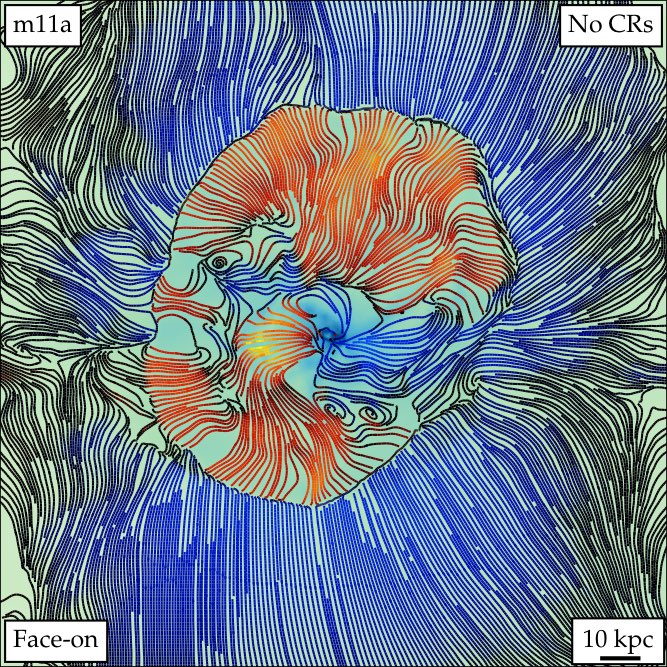}
\includegraphics[width={0.24\textwidth}]{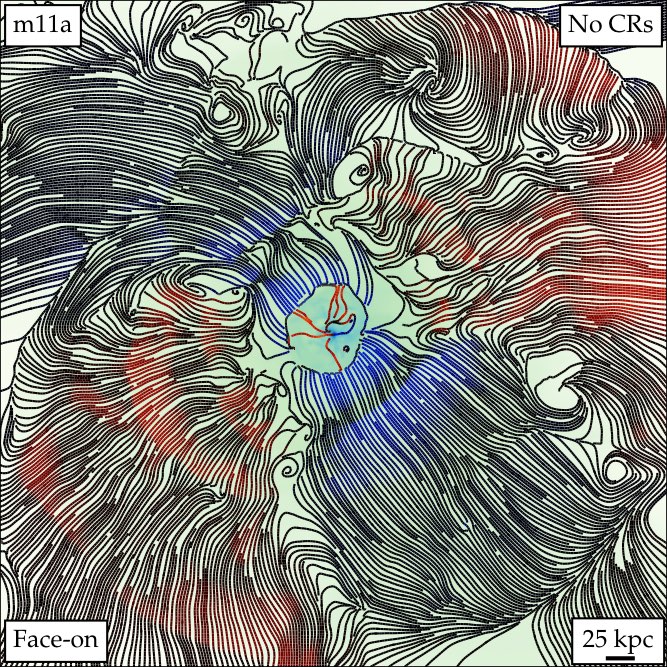}
\hspace{0.4cm}
\includegraphics[width={0.24\textwidth}]{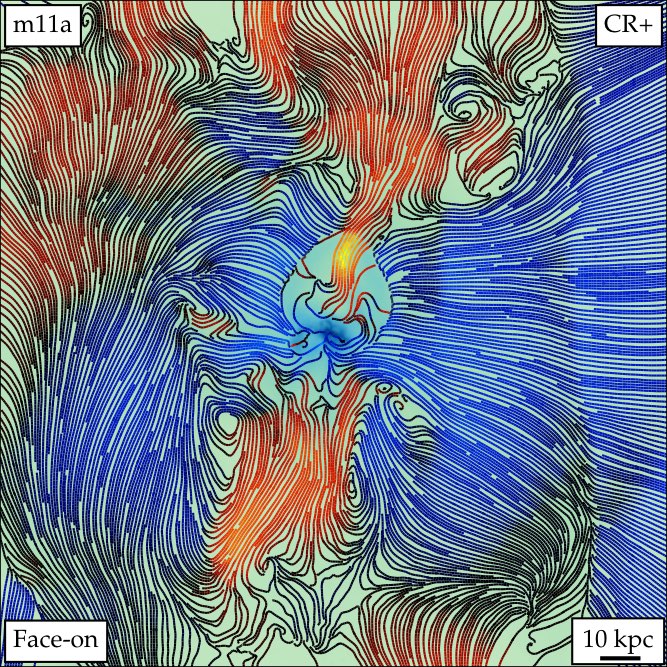}
\includegraphics[width={0.24\textwidth}]{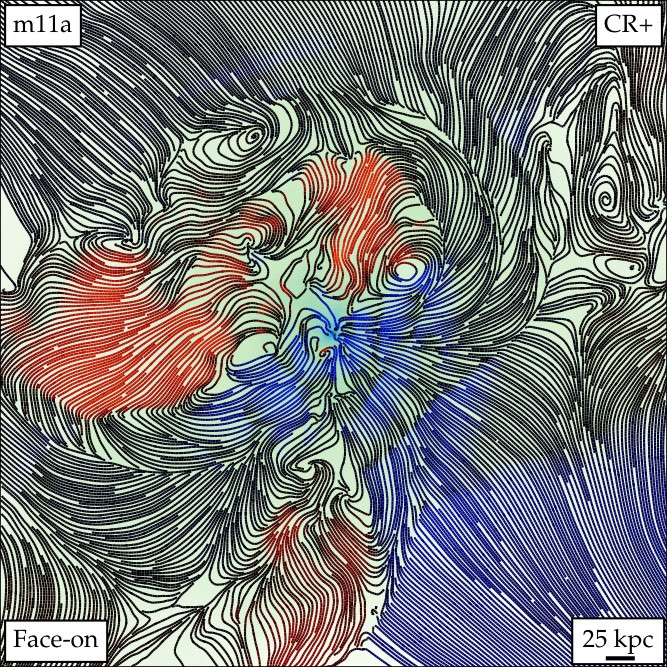}\\
\includegraphics[width={0.24\textwidth}]{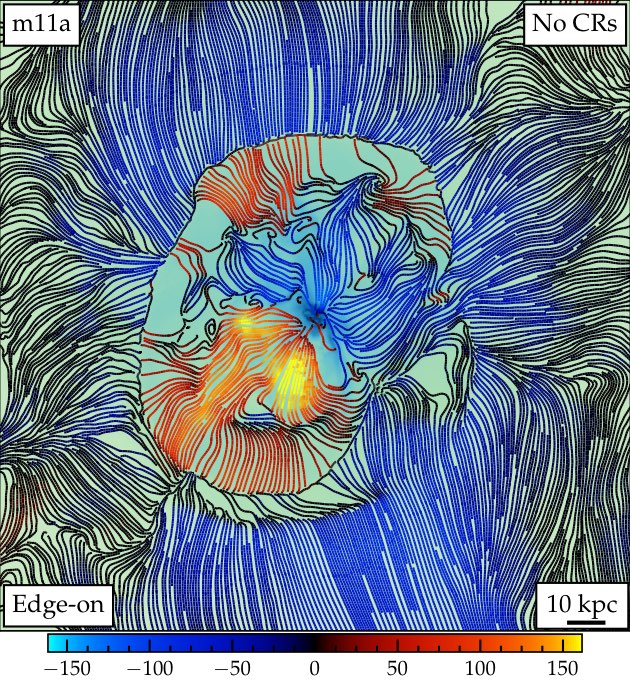}
\includegraphics[width={0.24\textwidth}]{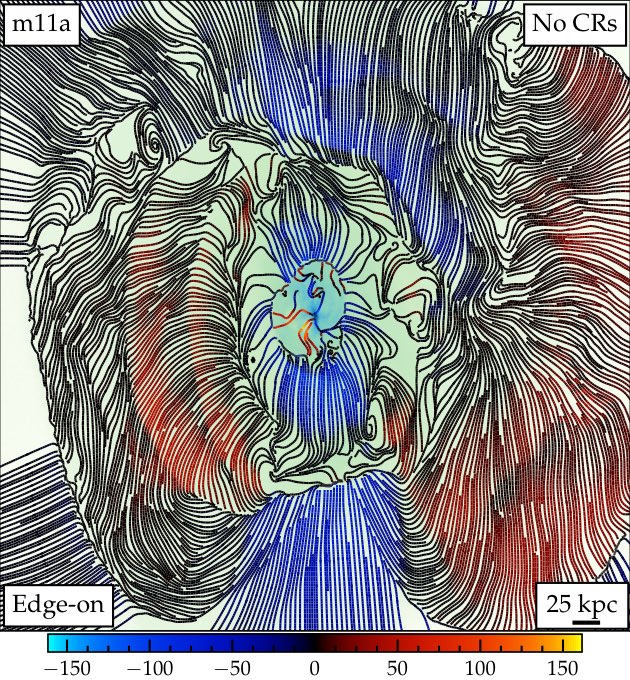}
\hspace{0.4cm}
\includegraphics[width={0.24\textwidth}]{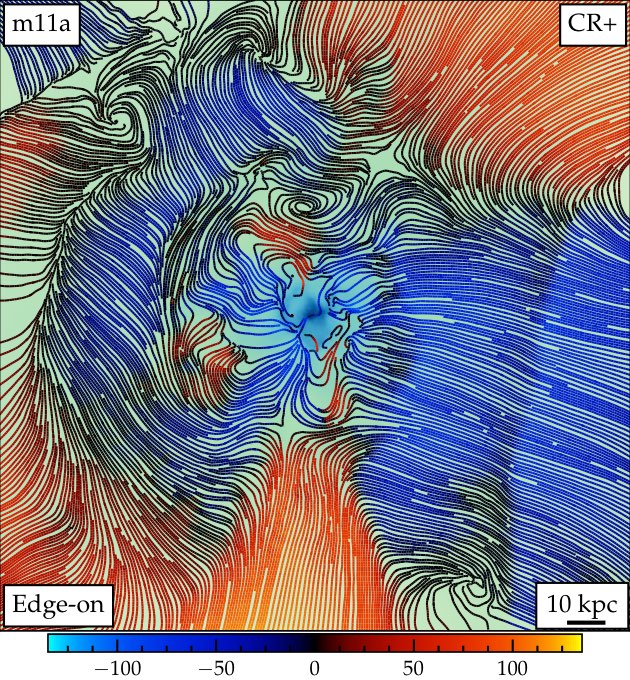}
\includegraphics[width={0.24\textwidth}]{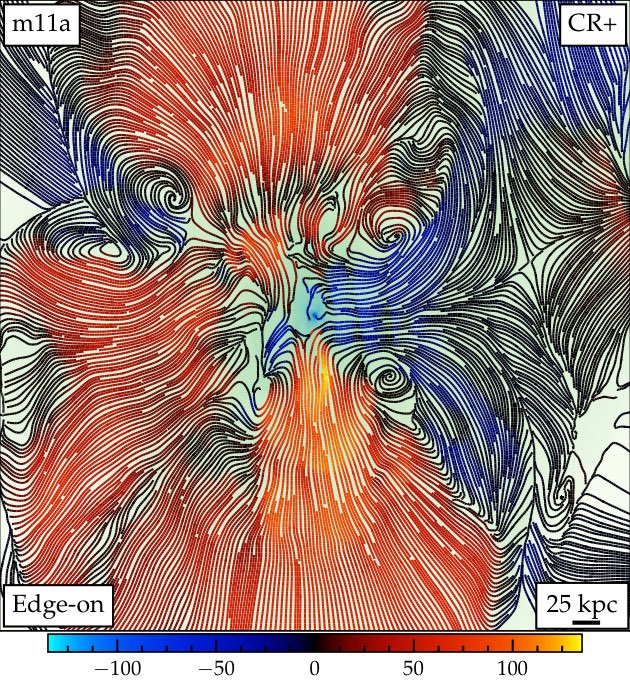}\\
\end{centering}
\caption{As Fig.~\ref{fig:inflow.outflow.m12f}, for {\bf m11a}. Like {\bf m11b}, the low mass ($M_{\ast} \sim 5\times 10^{7}\,M_{\sun}$, $M_{\rm halo} \sim 4\times10^{10}\,M_{\sun}$) means CRs have relatively modest effects, but some differences on scales $\sim 1-4\,R_{\rm vir}$ are still evident. The ``No CRs'' run is close to spherically-symmetric (with a clear shock where disk outflows meet accretion at $\sim 0.5\,R_{\rm vir}$), and weak inflow shocks at larger radii. The ``CR+'' run shows a large-scale bipolar outflow which does not extend to the disk but has a ``base'' at $\sim 30-50\,$kpc from the disk, reflecting collimation and acceleration by the CGM.
\label{fig:inflow.outflow.m11a}}
\end{figure*}

\begin{figure*}
\begin{centering}
\includegraphics[width={0.24\textwidth}]{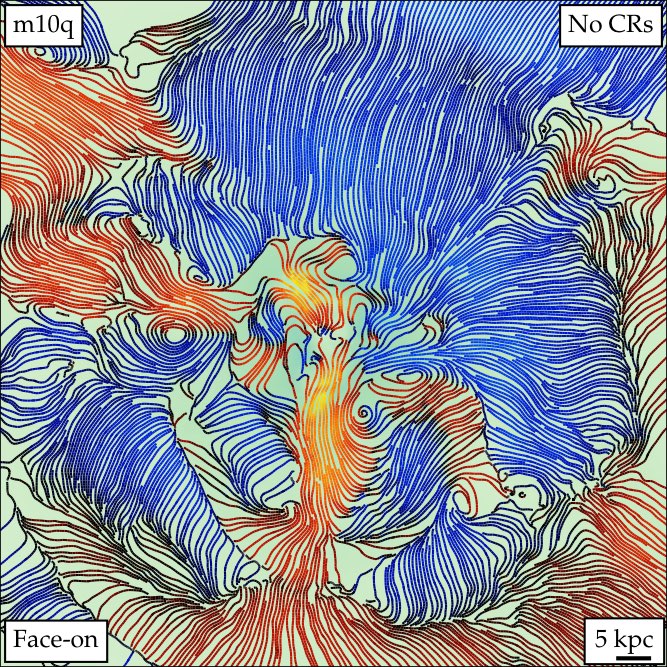}
\includegraphics[width={0.24\textwidth}]{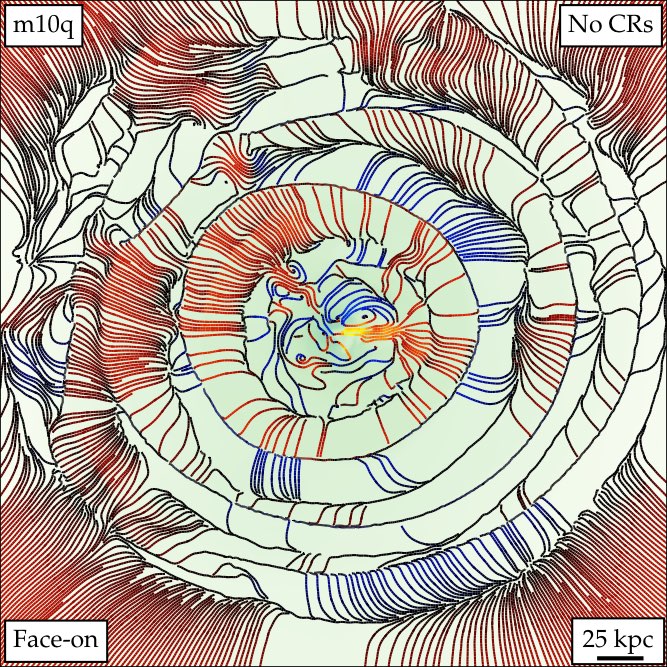}
\hspace{0.4cm}
\includegraphics[width={0.24\textwidth}]{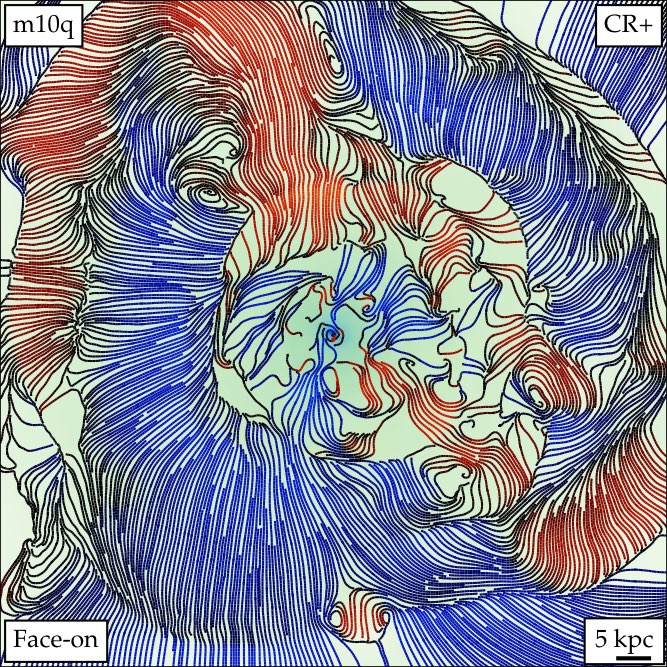}
\includegraphics[width={0.24\textwidth}]{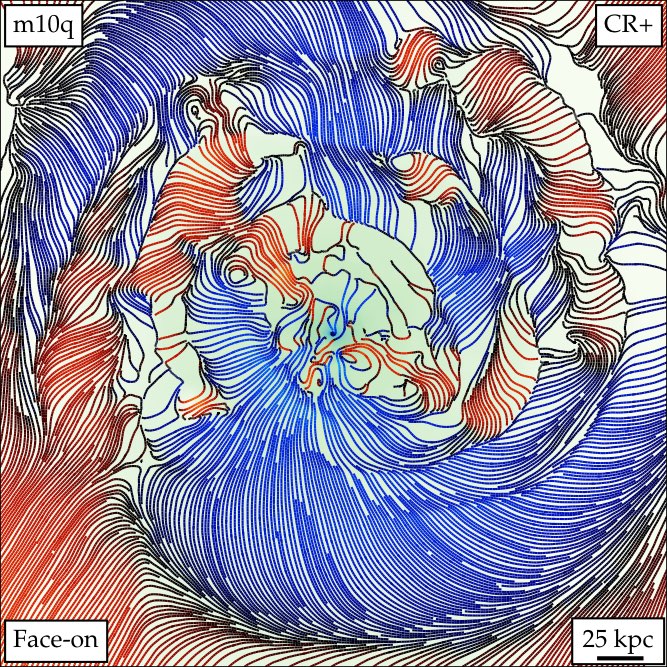}\\
\includegraphics[width={0.24\textwidth}]{figures/m10q/mhdcv/Vmap_vr_R54_bipolar_edgeon}
\includegraphics[width={0.24\textwidth}]{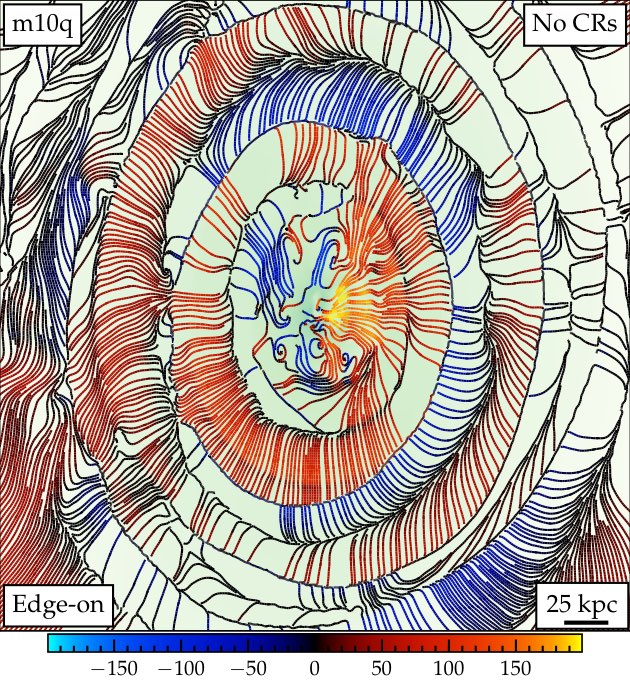}
\hspace{0.4cm}
\includegraphics[width={0.24\textwidth}]{figures/m10q/cr_700/Vmap_vr_R54_bipolar_edgeon}
\includegraphics[width={0.24\textwidth}]{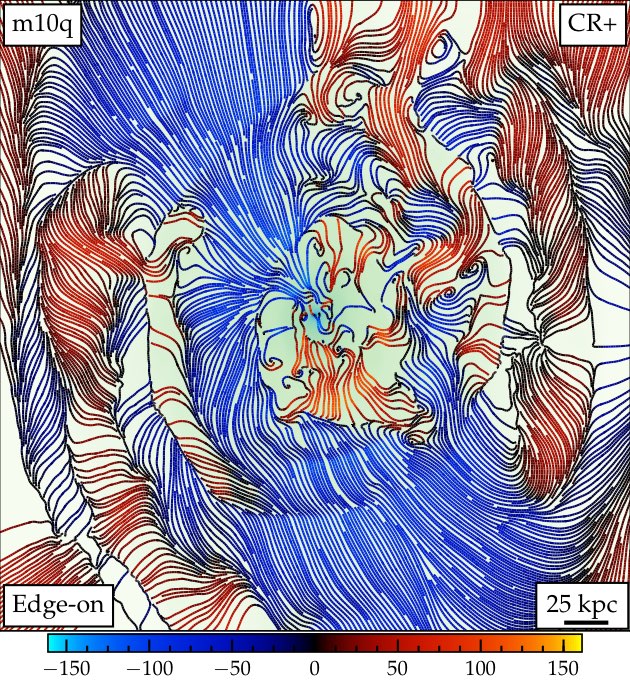}\\
\end{centering}
\caption{As Fig.~\ref{fig:inflow.outflow.m12f}, for {\bf m10q}. By this mass ($M_{\rm halo} \sim 10^{10}\,M_{\sun}$), the contribution of CRs to the halo pressure is quite weak (Fig.~\ref{fig:profile.pressure}), and it is difficult to discern an obvious systematic change to the inflow/outflow structure around the galaxy - the ``No CRs'' run actually has a more obvious strong outflow at this particular time, owing to a recent burst of star formation.
\label{fig:inflow.outflow.m10q}}
\end{figure*}

\end{appendix}

\end{document}